\documentclass[journal,compsoc]{IEEEtran}
\usepackage[pdftex]{graphicx}
\usepackage{url}
\usepackage[center]{caption}
\usepackage{todonotes}
\usepackage{algpseudocode}
\usepackage{subfig}
\usepackage{multirow}
\usepackage{xcolor}
\usepackage{stackengine}
\usepackage{tikz}
\makeatletter
\tikzset{noclip/.code={
    \tikzoption{clip}[]{\pgf@relevantforpicturesizefalse}
    \tikzoption{use as bounding box}[]{\pgf@relevantforpicturesizefalse}
    }
}
\makeatother

\def\techreport{1}
\newcommand{\papertechreport}{technical report}

\begin{document}

\presetkeys{todonotes}{inline}{}

\author{Jens Van den Broeck, Bart Coppens, Bjorn De Sutter
\thanks{The authors thank the Agency for Innovation by Science and Technology in Flanders (IWT) for supporting Jens. Part of this research was conducted in the EU FP7 project ASPIRE, which has received funding from the European Union Seventh Framework Programme (FP7/2007-2013) under grant agreement number 609734.}
}

\title{Extended Report on the Obfuscated Integration of Software Protections}
\IEEEtitleabstractindextext{
\begin{abstract}
To counter man-at-the-end attacks such as reverse engineering and tampering, software is often protected with techniques that require support modules to be linked into the application. It is well-known, however, that attackers can exploit the modular nature of applications and their protections to speed up the identification and comprehension process of the relevant code, the assets, and the applied protections.
To counter that exploitation of modularity at different levels of granularity, the boundaries between the modules in the program need to be obfuscated.
We propose to do so by combining three cross-boundary protection techniques that thwart the disassembly process and in particular the reconstruction of functions: code layout randomization, interprocedurally coupled opaque predicates, and code factoring with intraprocedural control flow idioms.
\if\techreport0
By means of an experimental evaluation on realistic use cases and state-of-the-art tools, we demonstrate our technique's potency and resilience to advanced attacks.
\else
By means of an elaborate experimental evaluation and an extensive sensitivity analysis on realistic use cases and state-of-the-art tools, we demonstrate our technique's potency and resilience to advanced attacks.
\fi
All relevant code is publicly available online.

\end{abstract}
}

\maketitle

\IEEEdisplaynontitleabstractindextext

\section{Introduction}
Software protection techniques such as code obfuscation and remote attestation
aim to mitigate man-at-the-end (MATE) attacks that target software assets that
come with confidentiality and integrity requirements.

The protections typically do not aim to prevent attacks completely. Because
MATE attackers have white-box access to the software in their labs,
the protections aim to raise the costs of (i) identifying
successful attack vectors in the attacker's lab, and (ii) scaling up the attacks
to exploit them outside the lab. The protections are in many cases best-effort
rather than providing well-defined security, and part of their protection comes
from security through obscurity. In practice, their effectiveness decreases when
attackers gain more knowledge about their inner workings.

To be effective, protections should provide resistance against many of the
possible methods with which they can be overcome, worked around, bypassed, and undone~\cite{emse2019}. Multiple protections defending against different attack
methods hence need to be layered upon each other, ideally to the point where
attackers consider the attack path of least resistance not profitable enough to
attack the software.

Advanced protections, such as code mobility~\cite{cabutto2015software},
ba\-rri\-er-slicing with server-side execution~\cite{ceccato2007barrier}, remote
attestation~\cite{viticchie2016reactive}, anti-debugging by
self-debugging~\cite{abrath2016tightly}, and instruction set
randomization~\cite{custom_instruction}, are deployed by means of two forms of
adaptions to that software. First, components implementing functionality of the
protections are linked into the software. Secondly, the original code is
transformed.

To delay an attacker in overcoming protections, it is useful to embed the
linked-in protection components stealthily, meaning hard to identify. For that
reason, protections components are always linked statically into native software
to be protected, whether that software is itself a main binary or a dynamically
linked library. Static linking does not offer very strong protection,
however. In experiments with both professional penetration testers and amateur
hackers~\cite{emse2019}, we have observed that once attackers identify a small
part of a statically linked-in protection, they can all too easily expand their
reverse-engineering to the most vulnerable program points.
Beyond static linking, a number of design obfuscations (such as function
merging, inlining, and outlining) are available to obfuscate the design of the
interfaces between the application and the protection components, but those can
only be deployed when all source code is available. In practice, this is not the
case: vendors of protection tools do not make the source code of their
protections available to their customers, because of the practical
security-by-obscurity reasons. There hence exist few practical techniques to
obfuscate how protection components are integrated into the software they help
to protect. While some security vendors have post-processing tools to secure
their components after they are integrated into their customers' software, both
the inner workings of those tools and their effectiveness are tightly protected
secrets.

\emph{In this \papertechreport{}, we present novel techniques and combine them with
  adaptations of existing techniques to hide the location and boundaries of
  software components that are linked together, including linked-in protection
  components, with the goal of hampering MATE attacks.}

All our techniques are based on post link-time binary rewriting.
It hence does not suffer from module boundaries and separate compilation the way compile-time or source code
techniques do. %
It offers an additional advantage over
source-to-source code rewriting, as our %
techniques
are not limited to the expressiveness of the used source
language(s).

Our combined techniques are (1) whole-program code layout
randomization, (2) insertion of fake direct control flow transfers between
procedures, and (3) factorization of code fragments common to multiple
 components without embedding them in separate functions.
Together, they make it much harder to for attackers and their tools to
identify and structure the relevant code and the control flow in the
program. Moreover, as our evaluation will demonstrate, the combination of
techniques is resilient against a number of commonly used and academic
state-of-the-art manual and automated deobfuscation techniques.

This \papertechreport{} offers the following main contributions:
\begin{itemize}
\item We present new forms of code factoring to serve as module boundary obfuscations.
\item We discuss how to combine them with code layout randomization and opaque
  predicates to resist automated and manual attacks.
\item We present an open-source tool chain that implements the presented techniques.
\item We analyse and evaluate the presented techniques on use cases of
  real-world complexity.
\if\techreport1
\item We perform a sensitivity analysis on the obfuscator's most relevant parameters.
\fi
\end{itemize}

This \papertechreport{} is structured as follows: Section~\ref{sec:attack_model} discusses our attack model. Sections~\ref{sec:randomization}--\ref{sec:factoring} discuss the three forms of obfuscations we combine.
\if\techreport1
Section~\ref{sec:evaluation} presents an elaborate quantitative experimental evaluation including an extensive sensitivity analysis, after which Section~\ref{sec:related} discusses related work and Section~\ref{sec:conclusions} draws conclusions and looks forward.
\else
Section~\ref{sec:evaluation} presents a quantitative experimental evaluation, after which Section~\ref{sec:related} discusses related work and Section~\ref{sec:conclusions} draws conclusions and looks forward.
\fi

\section{Attack Model}
\label{sec:attack_model}

We protect native software from man-at-the-end (MATE) attacks. MATE attackers have full access to, and full control over, the software under attack
and over the end systems on which the software runs. They can use static
analysis tools, emulators, debuggers, and all kinds of other hacking tools. The
attacks are looking to break integrity and confidentiality requirements of
assets embedded in the software, e.g., to steal keys or IP, or to break license
checks and anti-copy protections. They do so mainly by means of reverse
engineering and by tampering with the code and its execution.

MATE protections mostly aim at economically driven
attackers~\cite{collbergbook}.
They are considered effective when the provider's cost of
deploying the protections is compensated by a resulting reduction in the loss of
income due to successful attacks. This reduction can result simply from delaying
attacks. The protection is maximally effective if it stops attackers before they
reach their goal, or even before they start an attack, e.g., because the
(supposed or observed) presence of protection lowers the attackers' perceived
return-on-investment to the extent that they give up.

MATE attackers execute an attack strategy in which they execute a series of
attack steps. The strategy is adapted on the fly, based on the results obtained with previous attack steps. These include the testing of hypotheses regarding assets and protections.
We refer to literature for more information on and
models of MATE attack processes on protected software~\cite{emse2019}.

To be effective, protections deployed on software and assets should cover as
many as possible relevant attack paths, i.e., paths that might be
paths-of-least-resistance for certain attackers.
It is commonly accepted that this can only be achieved by combining
many protections in a layered fashion. The deployed protections then become
assets themselves, that protect each other just like they protect the original
assets.

In this section, we focus on the attack processes and attack
activities that are impacted by the protections presented in this
\papertechreport{}. These are the essential processes of:
\begin{itemize}
\item identifying and structuring the code components and their functionality at different levels of granularity and abstraction;
\item identifying relevant relations between components;
\item determining their features based on the relations;
\item browsing through those elements to locate and identify the relevant fragments on which to execute additional attack
  steps.
\end{itemize}

Attackers use tools, techniques, and heuristics to build structured
program representations such as control flow graphs (CFGs), call graphs, execution
traces, data dependency graphs, etc.\ of disassembled binary code. 

They then build mental models of the software in which they assign meaning
(i.e., some higher-level semantics) to the different components and derive
relevant features thereof. They do so in terms of all the concepts they know as
relevant from past experience~\cite{emse2019}.

This assignment process and the derivation of features is typically an iterative
process that starts from easily identified elements such as API calls and system
calls, XOR-operations, references to strings, known patterns or fingerprints of
certain algorithms, etc. That leads the attacker towards the specific
components of interest, such as the data or code he wants to lift from the
software, or those parts of protections he wants to tamper with to overcome the
protections. Table~\ref{table:relations} lists some of the relations between
components that attackers exploit.

\begin{table*}
\centering
\begin{tabular}{|p{8.5cm}|p{8cm}|}
\hline
Relation exploited by attackers & Examples of exploitation in concrete attack\\
\hline
\hline
Control flow transfers %
& Disassemblers such as IDA Pro deploy recursive descent algorithms to identify code bytes to be disassembled and to be partitioned into functions.\\
\hline 
Data flow dependencies %
& If an attacker has observed that values are XOR'ed before they are output, he often assumes they are being encrypted. The code that produces the mask used in the XOR then draws the attention of the attacker if he is after the embedded encryption key. \\
\hline 
Spatial proximity of code fragments in code sections & If an attacker has identified a code guard function, e.g., because it reads from the code sections as it hashes the code bytes, he looks in the proximity of that function for other functionality related to tamper detection, such as the functions that check the final hash value. This is based on the assumption that related functionality is linked into the program together.\\
\hline 
 Temporal proximity of code fragments in an execution trace & When an attacker tampers with the code of a program, and as a reaction the program halts almost immediately, the attacker will focus on the code executed right before the halting to find the code fragment that checks the integrity of the code. \\
\hline 
Spatial proximity of data stored in memory & When attackers know that structs on the heap hold values with known patterns as well as unknown values they want to steal, they search for the known patterns to find the locations of the values to steal.\\
\hline
\end{tabular}
\caption{Relations between structured software components and uses thereof by attackers.}
\label{table:relations}
\end{table*}

It is clear that if we can prevent tools from correctly identifying the relevant
relations and structures, we can make the attacks harder to execute. From
conversations with professional reverse engineers
at Dagstuhl Seminar 17281 in July 2017, we also learned that if tools
present incorrect relations and structure, this hampers attackers even more
because they then waste additional time performing activities based on incorrect
assumptions and data.

From the field of software engineering, we know that code comprehension benefits
from well-structuredness of the code~\cite{cognitive_size,Woodward} and
a separation of concerns, with each fragment having a single responsibility.
It then follows that attackers have a harder time comprehending code that does
not adhere to structures and concepts they are familiar with. In this \papertechreport{}, we
build on the hypothesis that attackers have a harder time handling code
fragments that each individually implement multiple parts of multiple, unrelated
high-level functions in a program, in particular when those code fragments are
not structured correspondingly.

Concretely, consider the procedures in a program. Attackers recognize procedures
by their prologues and epilogues, and by the fact that they are invoked
through function calls. It is a natural assumption that procedures can be
invoked from within different contexts.
As long as the semantics of the function in the
multiple contexts are somewhat related, i.e., it performs roughly the same
functionality in those contexts, the process of assigning a meaning to the function can require little effort.

It becomes much harder, however, to comprehend code if a function implements
multiple completely unrelated functionalities, depending on the context from
which they are called. This is exploited by obfuscation techniques called
function merging and fusion~\cite{collbergbook}. The fused function is then invoked
from completely unrelated contexts, to perform completely unrelated
computations, i.e., to implement very different semantics.

Comprehending the code becomes even harder if the code fragment that implements
those different functionalities in different contexts is not even recognizable
as such, i.e., if it does not look like a procedure in the first place. In
software obfuscation, it is also a well known technique to hide calls, returns,
epilogues and prologues by replacing their standard assembler idioms by
alternative instruction sequences with the same semantics but with different
looks~\cite{linn2003obfuscation}. This thwarts disassemblers that do not
recognize the replacements, and it slows down human reverse
engineers.%

The obfuscations proposed in this work explicitly build on this observation about
the challenges that human attackers face when they try to attack and reverse
engineer software. The obfuscations do so by factoring out code (outlining
code) from unrelated contexts without putting the factored code in separate
procedures, instead using control flow idioms typically used for
intraprocedural transfers.

Not only humans are challenged when facing such factored out code
fragments. Automated attack steps, such as de-obfuscating transformations and
data flow analysis on which attackers rely, are also hampered.

First, it is well known that many data flow analyses return more precise results
when their sensitivity is improved. Higher sensitivity, e.g., in the form of
flow sensitivity, path sensitivity, or context sensitivity comes at the cost of
rapidly increasing running times and resource consumption, however, so attackers
need to compromise between more precision and faster analyses. While context
sensitivity has been shown to be both useful and practical in the context of
multiple whole-program binary analyses such as liveness analysis and constant
propagation~\cite{diablo2005,debray2000compiler,muchnick1997advanced} we know of
no path sensitive variants that are practical. As context-sensitive analyses do
not consider separate contexts for factored code fragments that do not look like
procedures, they are of little help to attackers that aim to recover the same
information they would on unprotected code.

Secondly, powerful, automated de-obfuscation approaches are available that build
on the detection of quasi-invariant behavior in obfuscated code. In essence,
those techniques iteratively filter out and simplify instructions that are observed to behave
quasi-invariantly (i.e., instructions that produce the same result every time
they are executed on some selected program inputs), as well as code that does
not contribute to the software semantics (i.e., to the input-output relation the
software displays for the selected inputs).
This de-obfuscation approach has been shown to succeed in undoing obfuscations
ranging from opaque predicates (with corresponding conditional branches that are
either always or never taken) to the use of packers (because the unpacking of
program code does not depend on program inputs). Based on our experience with
human attackers, this form of de-obfuscation is also performed mentally by
attackers that analyze code manually, i.e., when attackers derive properties
from program behavior observed, e.g., with debuggers. Although such derivations
are often unsound, MATE attackers only care about the result, not about
soundness.

By factoring out code fragments from multiple, unrelated contexts, we aim to prevent
that the fragments and the surrounding control flow behave invariantly, and
hence that they fall victim to the generic deobfuscation approach.

In summary, with the protections presented in this \papertechreport{}, we aim to protect
against MATE attacks on binary code, and we specifically aim for hampering
a number of relevant manual and automated MATE attack steps.

We concede that this attack model is fuzzy, rather than well-defined. To the
best of our knowledge, in the domain of practical software protection against
MATE attacks, there is no alternative, however.

\section{Code Layout Randomization}
\label{sec:randomization}
Attack heuristics include spatial proximity.
Each source code file typically contains code fragments that are closely
related. Software libraries to link programs against are also structured along
related functionality.

Compilers and linkers typically do not mix the binary code
generated for different functions in a source code file. Whole
function bodies are typically placed one after another in the text sections of object
files, and text sections of object files are placed one after the other in linked
applications or libraries, in which they are largely grouped by the archive from
which they were linked in.
Unless countermeasures are taken, related code fragments are hence more likely
located close to each other in binaries. Attackers hence sometimes use proximity
as a guide during their hunt for code to attack. In other words, they sometimes
browse the code linearly.

Taking countermeasures in a link-time rewriter like Diablo~\cite{diablo2005} is
trivial, as already demonstrated in the context of software
diversification~\cite{coppens2013feedback}. Mixing unrelated code can be done at
any level of granularity, because all code is represented in one big
CFG~\cite{diablo2005}, from which binary code in virtually any (randomized) order can be
generated.

The level of granularity at which the code layout is randomized has to be
considered carefully. At the coarsest level, we can simply leave function bodies
intact, but randomize their order throughout a whole program or library, as
previously proposed to prevent memory exploits~\cite{kil2006address}. This
already breaks proximity assumptions regarding the archive and compilation unit
levels. By mixing protection and application functionality, we can already
improve the stealthiness of protection components. For example, identifying one
function as one code guard computation then no longer automatically leads the
attacker to the related functionality in related functions.

We can also randomize the order of instructions and basic blocks, and mix
instructions from all function bodies. At the abstraction level of instructions,
code is most often represented and analyzed in CFGs constructed by disassemblers
such as IDA Pro, Binary Ninja, GHIDRA, or DynInst. Those tools deploy recursive
descent strategies to disassemble the binary code and to partition it into
function bodies. In other words, they assign code to functions and reconstruct
their CFGs based on the observed direct control flow transfers between
them. Fine-grained code layout randomization does not hamper
that partitioning at all. Moreover, the extra branches and possibly
worse instruction cache behavior following from fine-grained layout
randomization can severely impact performance.
When applied in isolation, fine-grained randomization below the function level
is therefore costly but hardly useful.

When the randomization is combined with obfuscations that break the
recursive descent strategy of the disassembler, more fine-grained randomization
can still be useful, however. In that case, splitting up function bodies and
placing the parts in a randomized order prevents the tools from deploying linear
sweep strategies to make up for the then defunct recursive descent strategy. How
to do so is precisely the aim of the obfuscations discussed next.

\section{Interprocedural Opaque Predicates}
\label{sec:opaque}

\subsection{Disassembler Function Reconstruction Thwarting}
To thwart the strategy of partitioning disassembled instruction sequences into
functions based on direct control flow transfers, we have two options. First, we
can replace direct transfers with indirect ones, such as branch
functions~\cite{linn2003obfuscation}, to prevent that the disassembler infers
that two code fragments relate and belong in the same function. Note that this
goal of thwarting the disassembler's function CFG reconstruction after bytes
have already been disassembled into instructions is complementary to the
original goal of branch functions, which was to thwart the disassemblers'
ability to identify the locations of instruction bytes in the executables, which is known to be a difficult task~\cite{Meng2016binary}.

Secondly, we can add ``fake'' direct transfers that trigger incorrect
assignments of basic blocks to functions. Such transfers can be added easily by
means of opaque predicates and corresponding conditional branches. If we choose
the predicate of the conditional branch to opaquely evaluate to false, implying
that the branch will never be taken, we can simply choose any point in the
program as the target of the conditional branch, thus injecting branch-taken CFG
edges between completely unrelated code fragments. If we choose the predicate to
opaquely evaluate to true, we can inject fall-trough CFG edges between code
fragments from completely unrelated functions. This is trivial with the already
existing support for code layout randomization.

Importantly, whereas choosing the targets of the fake edges is to be done at
link-time when all linked-in code is available, the actual injection of opaque
predicates does not necessarily need to occur at link time. Source-level
obfuscators or obfuscating compilers can be used for the latter as well. They
can typically inject more complex opaque predicates, which are then integrated
in the original code more stealthily as they are compiled together with the
original source code as long as they can inform the link-time rewriter about the
location of the opaque predicates in the code. Obfuscating compilers can do so
by adding comments and mapping symbols to the generated assembly code or object
code, source-level obfuscators can do so by describing the locations of inserted
opaque predicate code in terms of source line numbers. By means of debug
information in the object files, a link-time rewriter can then translate the
source line numbers to object code addresses, thus identifying the locations
where fake edges can be redirected to unrelated fragments at link time.

Fake CFG edges confuse the disassembler tools' CFG construction algorithms
because their recursive descent strategies are implemented greedily: starting
from function entry points (identified through symbol information or pattern
matching), they traverse the code and greedily assign traversed fragments to
functions. During the traversal, they treat idioms for intraprocedural control
flow, such as conditional branches, as precisely that: intraprocedural control
flow. For unobfuscated compiled code, this works fine, because few if any source
languages feature interprocedural gotos, and standard compilers don't insert
interprocedural branches (with the rare exception of tail call optimization).

But without more complex data flow analysis or other mechanisms to distinguish
real from fake direct edges out of conditional branches, the greedy strategies
fail. Depending on whether a basic block is first reached through a fake or a
true edge, it will be assigned to the correct or incorrect function body. This
implies that we can try to steer the tools towards incorrect function
partitioning and CFG reconstruction by inserting fake edges in a controlled
manner, but it also implies that the result of the reconstruction and
partitioning will depend on the order in which basic blocks are traversed by the
tools. In that regard, we observed that tools like IDA Pro and Binary Ninja
tend to give fall-through paths precedence over branch-taken paths.

It is important to note that tools like IDA Pro offer different views on the
CFGs to human attackers on the one hand, and to analysis tools on the other. In
CFGs stored in a database in support of plugins and external analysis tools, IDA
Pro stores all direct CFG edges it has discovered during the disassembly
process. This includes all edges from direct transfers such as both paths out of
conditional branches. This database hence includes the mentioned fake edges,
which can be considered false positives (FPs). The IDA Pro GUI, which is
typically used by humans to study code, however, does not display all such
edges. Instead, it omits such edges if they are interprocedural according to IDA
Pro, meaning that they connect basic blocks IDA Pro has put in different
functions. So attackers manually browsing through CFGs in the tool's GUI don't
get to see them. When fake edges are (accidentally) omitted that way, we can
consider them as semi-true negatives (STNs): They are FPs in the
database view, but true negatives (TNs) in the GUI view. When true edges are
omitted as a result in the GUI, they correspond to semi-false negatives (SFNs):
they are TPs in the database, but false negatives (FNs) in the GUI view.

SFN and STN CFG edges hamper manual code
comprehension and code browsing activities on the GUI, as they result in code
from different components, such as protections and original application code,
being presented as if it is part of the same functions, and code originating
from the same functions not being displayed as such. By inserting such edges, we
can contribute to a much more stealthy integration of protection components.

\if false
----------------

TEXT JENS:

Compilers are designed to compile source code in such a way that the compilation time and/or execution time is optimized, depending on the compilation parameters. In both cases, compilers normally don't add unnecessary control and data flow to a program, so only relevant code and data is generated. Additionally, compilers don't add direct links between functions and libraries, but rather execute code from these different components using function calls. So, the program consists of multiple, decoupled components,

Attackers construct control and data flow graphs of a program to analyse its inner workings. To achieve this, they have to assume that, at least initially, all the code and data in a program is relevant. Moreover, they have to assume that all conditional jumps can be taken in both directions, as the code that calculates the condition is considered to be relevant. Finally, attackers prefer to split up the control flow graph of a program into smaller chunks (i.e., functions using partitioning heuristics) to make their analyses faster and more accurate.

Based on these observations, it follows that attackers use control and data flow graphs to comprehend the program. Moreover, the control flow graph can easily be split up into smaller chunks because of the way compilers work, which makes control and data flow analysis easier to calculate.

Again, taking countermeasures in a link-time rewriter like Diablo is trivial, by taking advantage of the whole-program overview. The reconstruction of correct control and data flow graphs can be countered by introducing bogus flow in the program. One well-documented way to do this is by using opaque predicates to insert bogus data flow, and, based on the (opaque) result, bogus control flow using conditional jumps~\cite{collberg1998manufacturing}. This bogus control flow can then be used to link unrelated components in the program, which is important when thwarting the recursive descent algorithms and function partitioning heuristics.

We observed that tools like IDA Pro, Binary Ninja and DynInst use rather simple function partitioning heuristics. When following a control flow edge, the sink, in case it has not been associated with a function yet, is added to the function containing the source: most of the time, edges are considered to be intraprocedural. In the case of conditional jumps, we have observed that IDA Pro has a tendency to prioritize fallthrough edges over other edge types: the sink of a fallthrough edge is mostly added to the same function as the one the source is in\footnote{As we are not allowed to reverse engineer the algorithms implemented in IDA Pro, these observations are solely based on results, rather than the concrete implementations of the algorithms.}. Based on these observations, we propose to counter the partitioning heuristics by deliberately inserting fake fallthrough edges in the program, i.e., by ensuring that opaque predicates can end in a fake outgoing fallthrough path.
\fi
\subsection{Resilience against Counterattacks}
\label{sec:resilience}
So far, we only discussed the potency of code layout randomization and
interprocedural opaque predicates to confuse attackers and tools. Another
important aspect is resilience to attacks, because attackers can of course still
deploy all kinds of automated attacks to make up for the deficiencies of the
existing, basic CFG partitioning strategy. These include static attacks such as
pattern matching~\cite{ngo2007detecting}, abstract interpretation~\cite{dalla2006opaque}, and symbolic
execution~\cite{yadegari2015symbolic} to detect opaque predicates, and dynamic attacks such as
generic deobfuscation~\cite{yadegari} and fuzzing~\cite{Madou06applicationsecurity}. The dynamic ones are not
sound, but that typically does not hamper attackers.

A first, critical point to make is that none of the mentioned static techniques have
been scientifically validated as successfully breaking complex forms of opaque
predicates (such as the graph-based ones from Collberg et al.~\cite{collberg98}) on
software of real-world complexity. Symbolic execution, for example, was only
tested on programs of at most two functions~\cite{yadegari2015symbolic}. Abstract interpretation was
only evaluated on opaque predicates of which the program slice (i.e., the code
computing the predicate) consisted of a tiny fragment immediately preceding the
conditional branch~\cite{dalla2006opaque}.

In practice, we have observed that both pattern matching and local symbolic
execution are effective attack techniques~\cite{emse2019}. In both cases, small slices
of the predicates used in conditional branches are then analyzed to determine
whether or not they (likely or definitely) correspond to opaque
predicates. Depending on the size of the software under attack and the
immediate availability of a working attack tool box, attackers perform this
analysis manually or by means of tool plug-ins that automate the analysis. Less
skilled attackers reuse existing plug-ins as is, expert attackers can also
customize plug-ins. On small software, attackers prefer manual analysis
when they assess that the cost of setting up and customizing the tools will not
be worthwhile. As completely manual analysis does not scale to larger software
with many predicate instances to analyze, automation is typically preferred for
attacks on larger software. That automation also often requires manual
effort, however, if only because the customization of plug-ins requires the
attacker to first determine which forms of opaque predicates are useful to
search for, i.e., which code patterns to try to support.

At first sight, the attacker's ability to perform these attacks seems not
hampered by the interprocedural nature of the opaque predicates we propose to
inject. After all, the interprocedural aspect only directly impacts the control
flow from the conditional branch on, not the code computing the predicate
leading up to the conditional branch.

However, by carefully choosing the targets of the fake edges, we can directly
impact the code slices of the opaque predicates, or at least the perception
thereof by the attacker. We can in fact do so trivially by interrupting a slice
of one opaque predicate by means of a fake edge coming in from another one. In
the best case, this results in the assembler mistakenly assigning the
instructions computing the predicate to multiple functions. In that case, the
GUI will not show all relevant instructions in one
function CFG. This will certainly hamper all manual activities of the attacker
as discussed above. But even if the whole slice is assigned to the same function
and hence shown on screen with the correct control flow between the relevant
instructions, the attacker will still to some extent be confused when the fake
edge is drawn as well.

\begin{figure}
  \centering
  \includegraphics[width=0.4\columnwidth]{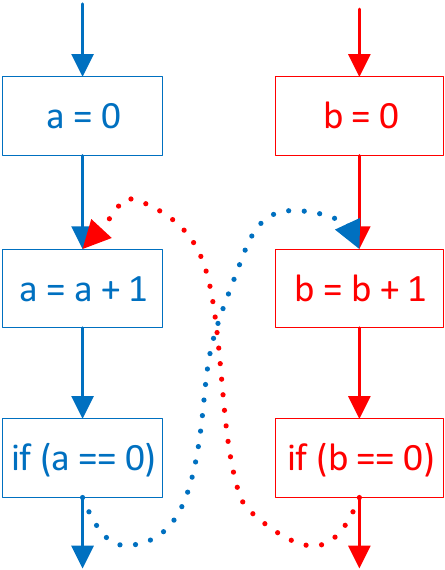}
  \caption{Example of coupled opaque predicates}
  \label{fig:coupling}
\end{figure}

To overcome this confusion, how small or big it may be in practice, the attacker
has to consider multiple instances of opaque predicates together.  Consider the
example in Figure~\ref{fig:coupling} with predicates of contrived
simplicity. Fake edges are drawn dotted, but at first, the attacker does not
know they are fake. To learn that they are truly fake, the attacker needs to
consider both fragments.
In practice, we are not limited to coupling pairs of opaque predicates mutually,
we can easily couple more in larger cycles. A local code comprehension task
for the attacker then becomes a global one; the effort needed to undo the
protection grows.

A similar reasoning holds for fully automated analyses. Had the opaque
predicates not been mutually coupled in the example of
Figure~\ref{fig:coupling}, a \emph{simple constant propagation}, applied locally
and iteratively with unreachable code elimination, would have sufficed to detect
them. In the coupled case, simple constant propagation no longer
suffices. Instead, a more complex \emph{conditional constant
  propagation}~\cite{wegman1991constant} is now required. In general, the (mutual) coupling of
opaque predicates by letting fake edges interrupt slices implies that
path-sensitive versions of analyses are needed. If those are applied locally,
i.e., one slice at a time, they can suffice to identify likely opaque
predicates, i.e., predicates that evaluate to constants on some execution
paths. In that case, the necessary increase in complexity of the attack step is
rather limited. If the attacker wants to deploy a sound(ish) analysis, however,
to get a degree of certainty about the opaque predicates, the analysis has to be
performed on all mutually coupled fragments together. This implies a
considerable increase in complexity.

In summary, we can conclude that the resilience of code layout randomization and
interprocedural opaque predicates, i.e., the effort needed to minimize their
potency, with respect to attacks of which we know they are used in practice, is
improved by coupling them in the proposed way.

Admittedly, this security analysis is fuzzy rather than well-defined. We
consider a formal %
analysis out of reach at this point in time,
not only for the protections against MATE attacks presented in this \papertechreport{}, but
for most if not all MATE protections.
In Section~\ref{sec:evaluation}, we will perform a quantitative evaluation of a prototype
implementation in which we mimick some real-life attacks.
\if\techreport1
We will also provide an extensive sensitivity analysis on the parameters of our obfuscator tool that steer how the proposed protections are applied.
\fi

Finally, we acknowledge that because it only injects invariant behavior into the
software, the proposed protection via mutually coupled opaque predicates and
code layout randomization does not protect in any way against dynamic attacks
such as the tracing-based ge\-ne\-ric deobfuscation. While we deem this acceptable,
as other protections can be used to shield of dynamic attacks, such as
anti-debugging, anti-emulation, and anti-taint protections, we will still build
on the protections presented so far in the next section to also make some
dynamic attacks less effective.

\if false

The novelty of our use of opaque predicates as presented so far, i.e., linking
fragments from unrelated functions via the never-taken execution paths, by
itself does not make those deobfuscation techiques less or more effective or
efficient. We do note, however, that in literature some techniques, including
some forms of symbolic execution (that execute the whole program) and generic
deobfuscation, have only been demonstrated to be effective and sufficiently
efficient on tiny programs, in particular on programs with very few (to the best
of our knowledge, 1 or 2) functions. There is reason to doubt their practical
value to attackers on complex, protected applications like the ones we
target.

For other techniques that operate locally, i.e., on the program slices of
conditional branche conditions, such as some forms of symbolic execution,
pattern matching, and abstract interpretation, there is no reason why they would
not scale (with linearly increasing running times) to larger programs. So
against such techniques, our novelty as introduced above provides little added
value by itself.

In addition, the attackers can try to repartition the code into functions to
improve upon the original partitioning and CFG visualization delivered by the
basic tools. Such automated attacks start from the database information that,
e.g., IDA Pro can export. This includes all true direct CFG edges plus the fake
edges, and a more or less incorrect partitioning of the basic blocks into
functions.

\textbf{TODO: decide what to write here about such attacks. The text below from Jens is somewhat problematic: attack 1 assumes the additional incoming edge that splits up a predicate computation, but the above text has not yet mentioned that we want to choose the targets of never-taken branches in that way; attack 2 and the example in figure 2 are strange because without attack 1, why would a block like B have been assigned to the wrong function?}

\fi

\if false

\subsection{Chaining Opaque Predicates}
To counter local code analysis attacks that aim for detecting opaque predicates
by analyzing static slices of conditional branch conditions, we propose to chain
opaque predicates such that they ``protect'' each other. Concretely, suppose we
have inserted $n$ opaque predicates $p_1,...,p_n$. Those consist of computations
$c_1,...c_n$ of the predicates, i.e., instruction sequences, and corresponding
conditional branches $b_1,...,b_n$. For each $b_i$, we then choose the sink of
the never-taken CFG edge at $b_i$ to be a program point somewhere on the path
from the first instruction in $c_j$, with $i\neq j$, to the last instruction therein. Consider
the example in Figure~\ref{}, in which the predicates are not really opaque, but
simple constants one. Fake edges are drawn dotted. With simple constant
propagation, those constants cannot be deduced. Had the opaque predicate
computations not be chained, simple constant propagation would have sufficed to
detect the opaque predicates.

Obviously, this example is of contrived simplicity. Breaking it does not require
a lot of additional complexity: conditional constant propagation starting from
the two top nodes in the CFGs would already suffice. Alternatively, an attacker
might use pattern matching or other heuristics to guess that the two fake edges
might well be fake. If he then performs a simple constant propagation on the CFG
minus those fake edges, the analysis will validate his guesses.

In real deployments of opaque predicates, the computations of the predicates are
more complex. Resolving them therefore requires more complex analysis, for which
symbolic execution and abstract interpretation have been proposed in
literature~\cite{yadegari2015symbolic,dalla2006opaque}. While the latter is mostly an academic approach, we
know of professional penetration testers that use local symbolic execution
techniques rather than pattern matching. Both the abstract interpretation
techniques presented in literature and those local symbolic execution techniques
we have observed in practice are deployed on the static slices of predicates
used in conditional branches to assess whether those predicates unconditionally
evaluate to true or false. In practice, those slices are taken from the CFGs
produced by tools like IDA Pro. While the initial learning curve for using and
tuning a symbolic execution engine is steeper than that of pattern matching, in
the long run it require less effort, because it does not need to be retuned
whenever the attacker faces new variations (i.e., new patterns) of previously
attacked opaque predicate constructs.

Furthermore, in real deployments of opaque predicates, the predicate
computations can and will be spread over larger program fragments, i.e., larger
slices, which can easily include multiple CFG paths to the conditional
branch. The fake edges can enter those slices and paths at any point. This
implies that the original abstract interpretation or symbolic execution tool
that would have sufficed for identifying non-chained forms of opaque predicates
has to be made path-sensitive in one way or another. Whether that is easy or not
to implement likely depends on the existing implementation and on the forms of
opaque predicates being attacked. In any case, it is safe to assume that the
path-sensitive versions will not run faster. More likely, they will feature much
longer running times.

Moreover, if the opaque predicate chains are circular, it is no longer possible
to attack the opaque predicates one after another. Instead, all of them have to
be attacked together, in the sense that the validation of guessed never-taken
paths ...

\fi

\section{Code factoring}
\label{sec:factoring}
To prevent that some of the stronger attacks can reconstruct the CFGs of a
program's functions completely by identifying fake edges that can never be
executed, we need to insert control flow transfers with more than one true
outgoing edge. In line with what we discussed in Section~\ref{sec:attack_model},
those true outgoing edges should look like intraprocedural edges. In other
words, intraprocedural control flow transfer idioms should be used in general. In
order to thwart the partitioning of code into functions, however, the edges
should be interprocedural, connecting code from diffferent functions.

\begin{figure}[t]
  \centering
  \includegraphics[width=\linewidth]{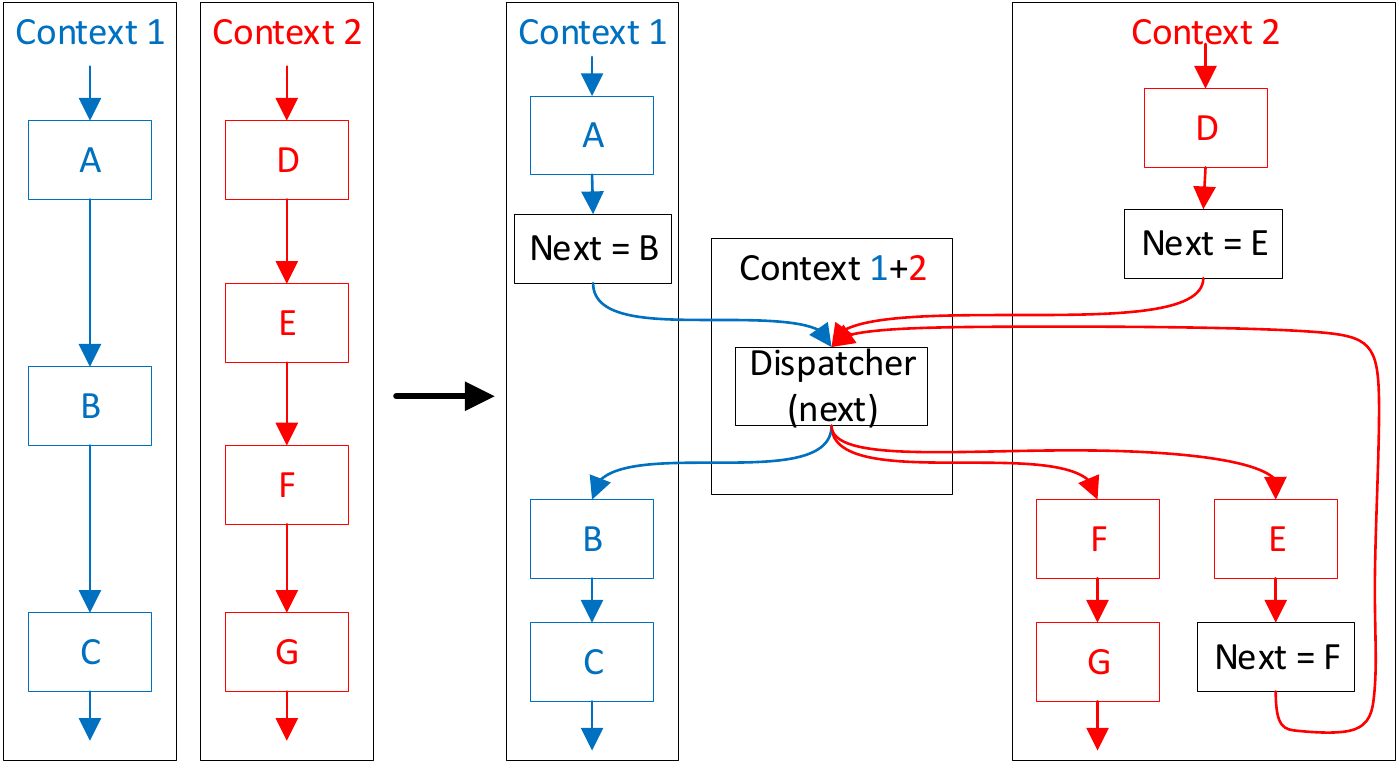}
  \caption{Control flow flattening}
  \label{fig:flattening}
\end{figure}

\begin{figure}[t]
  \centering
  \includegraphics[width=\linewidth]{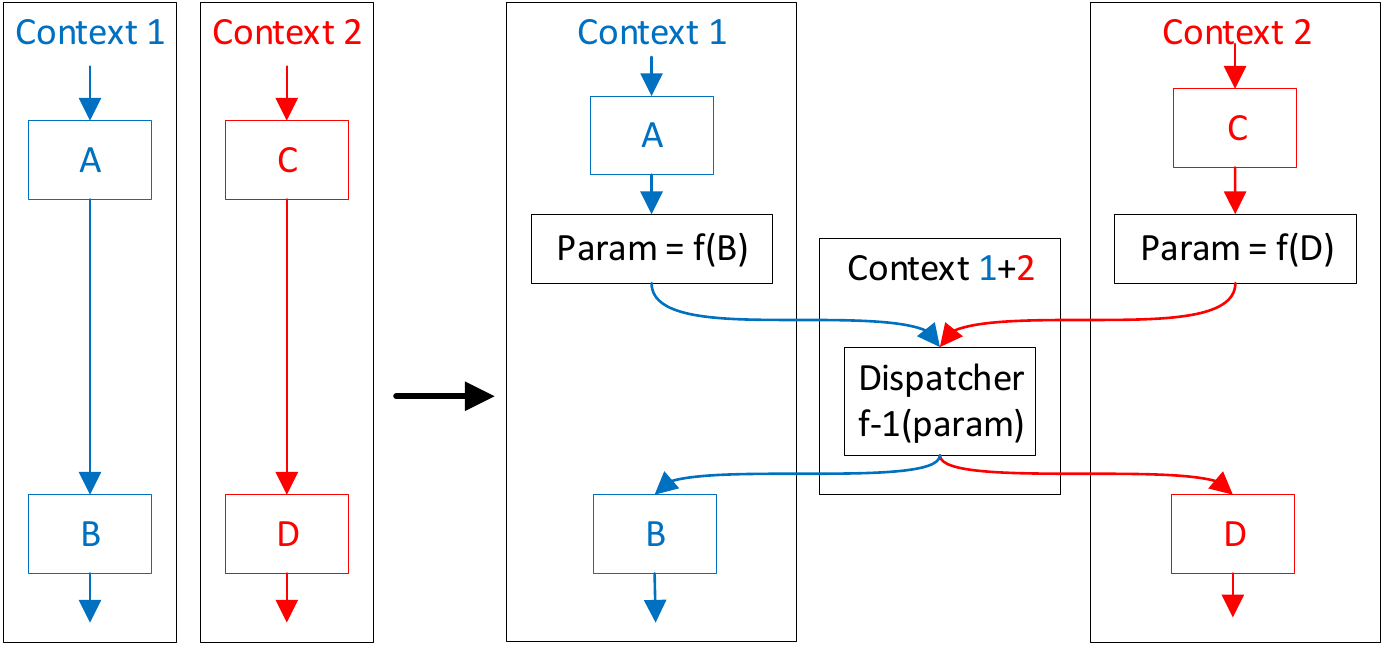}
  \caption{Branch function}
  \label{fig:branchfunction}
\end{figure}

We can meet these requirements by deploying control flow
flattening~\cite{wang2000software} and branch
functions~\cite{linn2003obfuscation} across multiple functions. Both control
flow obfuscations can be implemented with many forms of intraprocedural looking
control flow transfers such as conditional branches, switch tables, and computed
jumps. Some simple examples are depicted in Figure~\ref{fig:flattening} and
Figure~\ref{fig:branchfunction}. However, in that case the transfers can still
be observed to be semantically irrelevant: In a program trace, their executions
will never depend on actual input values, only on constants such as those
assigned to \texttt{next} in Figure~\ref{fig:flattening} and \texttt{param} in
Figure~\ref{fig:branchfunction}. Furthermore, apart from steering control to the
appropriate continuation points depending on how they are reached, the injected
code fragments then do not contribute to the output of the program. For both
reasons, these fragments will get de-obfuscated by the approach of Yadegari et
al.~\cite{yadegari}.

\begin{figure}[t]
  \centering
  \includegraphics[width=\linewidth]{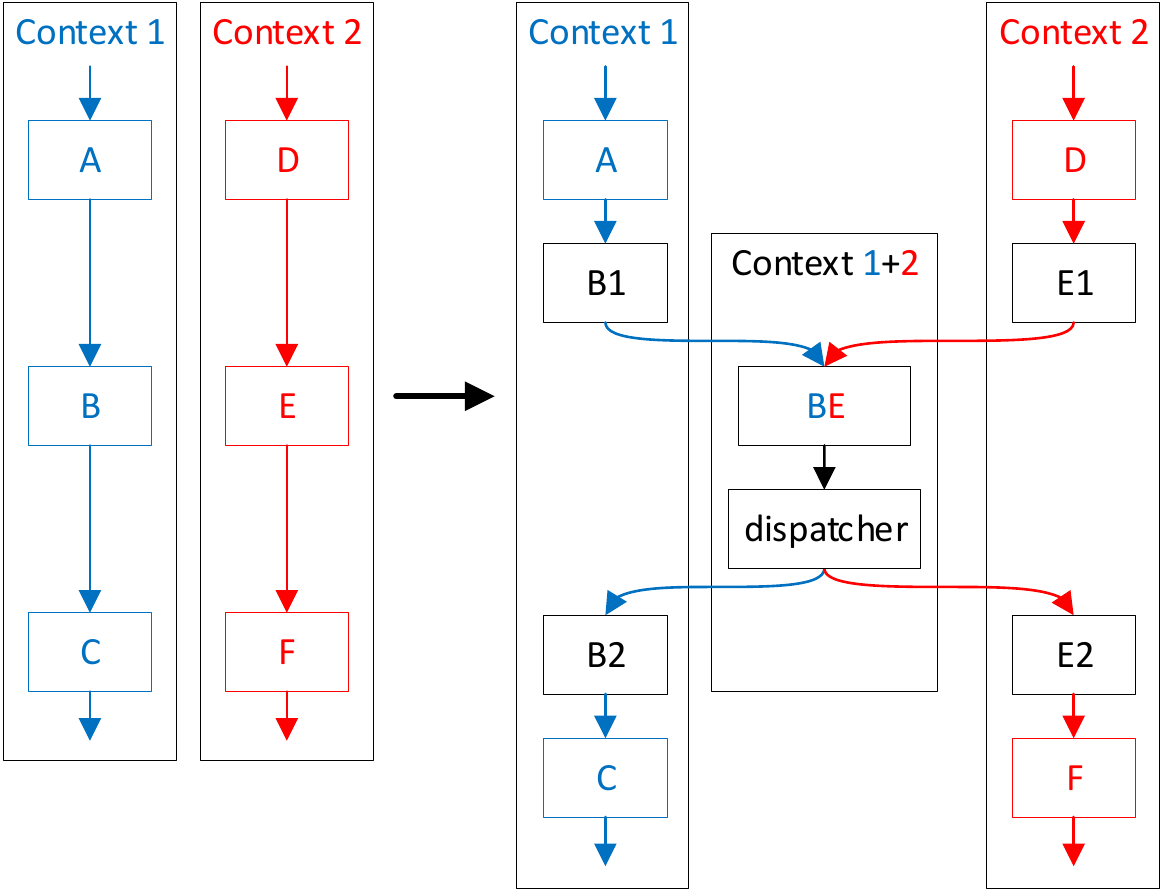}
  \caption{Code factoring for obfuscation}
  \label{fig:factoring_concept}
\end{figure}

To counter this, we propose to combine the mentioned obfuscations with code
factoring, as illustrated in Figure~\ref{fig:factoring_concept}. Blocks B and E
are identical in the original code. If both of them can actually be executed in
the original program, both edges coming out of the factored block BE will be
executable in the transformed program. So the transfer at the end of block BE
will show variable behavior. Moreover, the code in BE will be executed on data
from two different contexts, and hence also display variable behavior. Moreover,
as the original fragments B and E mattered for the original program, we can
assume the factored block BE to be semantically relevant in the transformed
program. The generic de-obfuscation approach of Yadegari et al.\ will therefore
fail.

Code factoring is not new. Several forms have been proposed in the past to
compact programs~\cite{debray2000compiler}.
Our deployment of factoring serves the purpose of obfuscation, however, so it
differs in two significant ways from previous deployments. First, we do not
factor code into new functions that get called and end with return
instructions. Instead, we use idioms of intraprocedural control flow, such as
conditional branches, switch tables, and computed jumps. Secondly, we do not
strive for more compact code. This implies that we can transform non-identical
code fragments to make them identical, even when that involves prepending or
appending extra instructions to the original fragments %
that move values between registers.

With these different requirements, we developed a significantly different code
factoring technique. The most relevant aspects are a fast preliminary
identification of potential fragments to be factored, the identification of
actual factoring candidates, the order in which those are selected
for transformation, the preparation of the selected ones, and the actual
factoring transformations themselves.

\subsection{Potential Factoring Candidates}
\label{sec:potential_candidates}
To factor code, identical code fragments need to be identified or
created. Existing factoring
techniques~\cite{debray1999compiler,de2002sifting,diablo2005} pre-partition code
fragments using fingerprints. The fingerprinting functions are simple and strike
a balance between recall and precision: They are defined such that code
fragments that are ``similar enough'' to be likely candidates for factoring are
mapped onto the same fingerprint. Much more complex and time-consuming
precise checks of factoring pre-conditions, which also analyse the fragments'
surroudings, are only performed on
sets of fragments within the same partition, i.e., with the same fingerprint.

In existing code factoring techniques focussing on compaction, ``similar
enough'' is defined as ``nearly identical'', i.e., having identical instruction
schedules, and (almost) identical register allocations. The underlying
assumption is that less similar fragments might well be factorable, but likely
glue code will have to be injected around them before they can be factored,
which will likely undo the compaction gains.
Furthermore, to further limit the search space by focusing on worthwhile cases,
existing techniques typically consider fragments consisting of one or more basic
blocks, such as whole single basic blocks, single-entry CFG subgraphs of
multiple blocks, and whole functions/methods~\cite{microsofticf,debray2000compiler,siftingmud,edler2014exploiting,rocha2019function}. An underlying assumption
is that it is much less likely to find nearly identical, worthwhile fragments
inside single basic blocks if the containing blocks are not nearly identical as
a whole.

For our obfuscation purpose, the size of the glue code is only a secondary
concern. We hence have to strike a balance differently. We opted to do so by not
factoring fragments consisting of one or more whole basic blocks. Instead we
focus on slices (as defined by Horwitz~\cite{slices}) and instruction sequences
that are limited to, i.e., originate from within, single basic blocks. We only
consider slices and sequences that exclude control flow transfer instructions.

The \emph{slices} we consider as candidates for factoring are directed acyclic graphs
(DAGs) with a single sink node. The DAGs' nodes are instructions and their edges
are data dependencies. Instructions can define multiple slices, ranging from
the single-instruction slice consisting of only the instruction itself, to the
largest possible incoming data-dependency DAG within the instruction's basic
block. Besides in the slices they define themselves, instructions can also show
up in the slices defined on instructions further down in their basic blocks.
The \emph{sequences} we consider are sequences of instructions in the order in
which they occur in the basic blocks. All subsequences of the instruction
sequence constituting the block are considered. %
In the remainder of this \papertechreport{}, we use the term \emph{fragments} to denote both
slice and sequences. They are treated mostly identically in our factoring
approach.

The only point where their treatment differs is in the computation of
fingerprints. For sequences, we iterate over the instructions in their order in
the original program. For slices, we iterate over the instructions in a
canonical order that abstracts from the precise order in which the instructions
occur in the program. This canonicalization is useful because nodes in a DAG
are only partially ordered, and compilers generate different instruction orders
for the same DAGs depending on the other instructions mixed in between them.

The fingerprints consist of the concatenation of at most four instructions' opcode
(e.g., ADD, MOV, ...), their operand types (e.g., two registers, one register and
an immediate, ...) and some of their flags (e.g., pre- or post indexed).
We found that including only four instructions in the fingerprint strikes a good
balance between precision, recall, and memory consumption.

It can also be useful to consider the hotness of code fragments, i.e., their
contribution to the total execution time of a program as determined with
profiling. Excluding the hottest fragments helps to reduce the performance
overhead.

\subsection{Actual Factoring Candidates}
\label{sec:actual}
Being nearly identical does not suffice for actual factoring. For sets of nearly identical fragments, we also need (i) to extract the fragments from their basic blocks; (ii) to make fragments truly identical by reallocating registers and by replacing non-identical immediate operands by constants stored in registers; (iii) to add a dispatcher to ``return'' from the factored fragment and to feed that dispatcher with the necessary inputs at each ``call site''. The latter two result in increased register pressure. Our binary rewriter does not convert the higher-level executable code to a higher level IR. Hence we need to transform the code and handle the register pressure locally. Concretely, this means we have to inject glue code in the form of register transfer instructions such as move, copy, swap, and spills to memory around the fragments. Foremost, we need to check whether we can actually perform the required rewriting within the capabilities (available transformations and analysis precision) of the link-time rewriter. As different dispatchers come with slightly different requirements, we also need to check which dispatchers can be used for which sets.

\begin{figure}[t]
  \centering
  \includegraphics[width=0.6\linewidth]{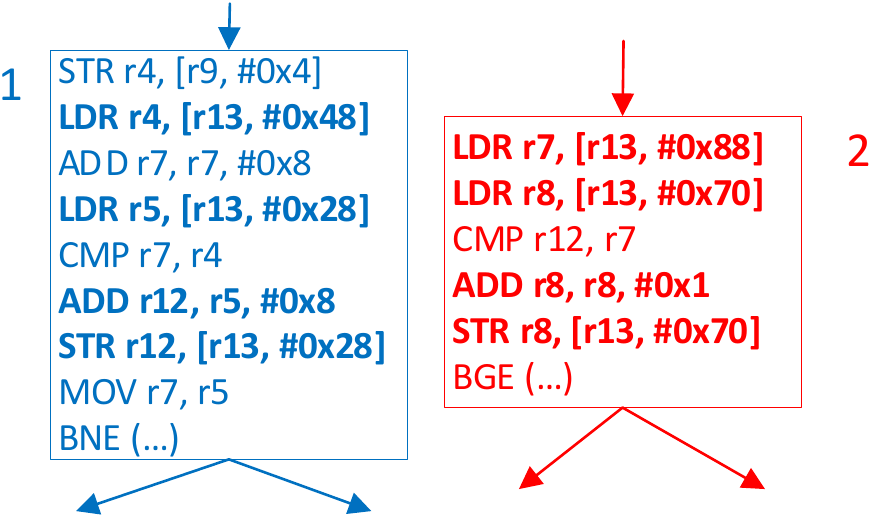}
  \caption{Factoring candidate slices in bold in their respective basic blocks}
  \label{fig:fragments}
\end{figure}

\begin{figure}[t]
  \centering
  \includegraphics[width=0.7\linewidth]{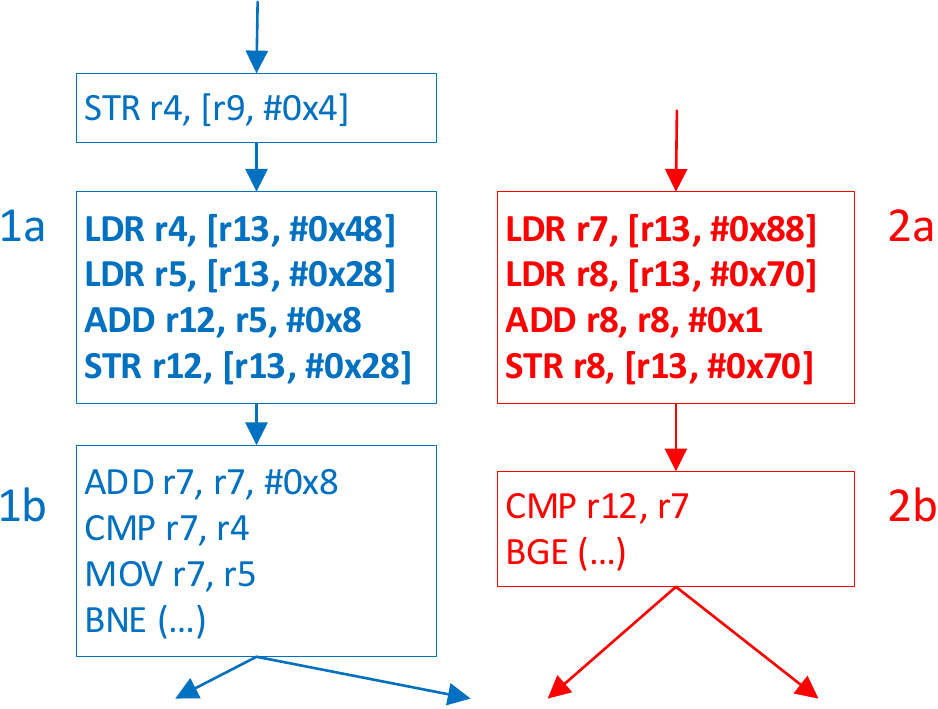}
  \caption{Split factoring candidate slices}
  \label{fig:fragments_split}
\end{figure}

\begin{figure}[t]
  \centering
  \includegraphics[width=0.7\linewidth]{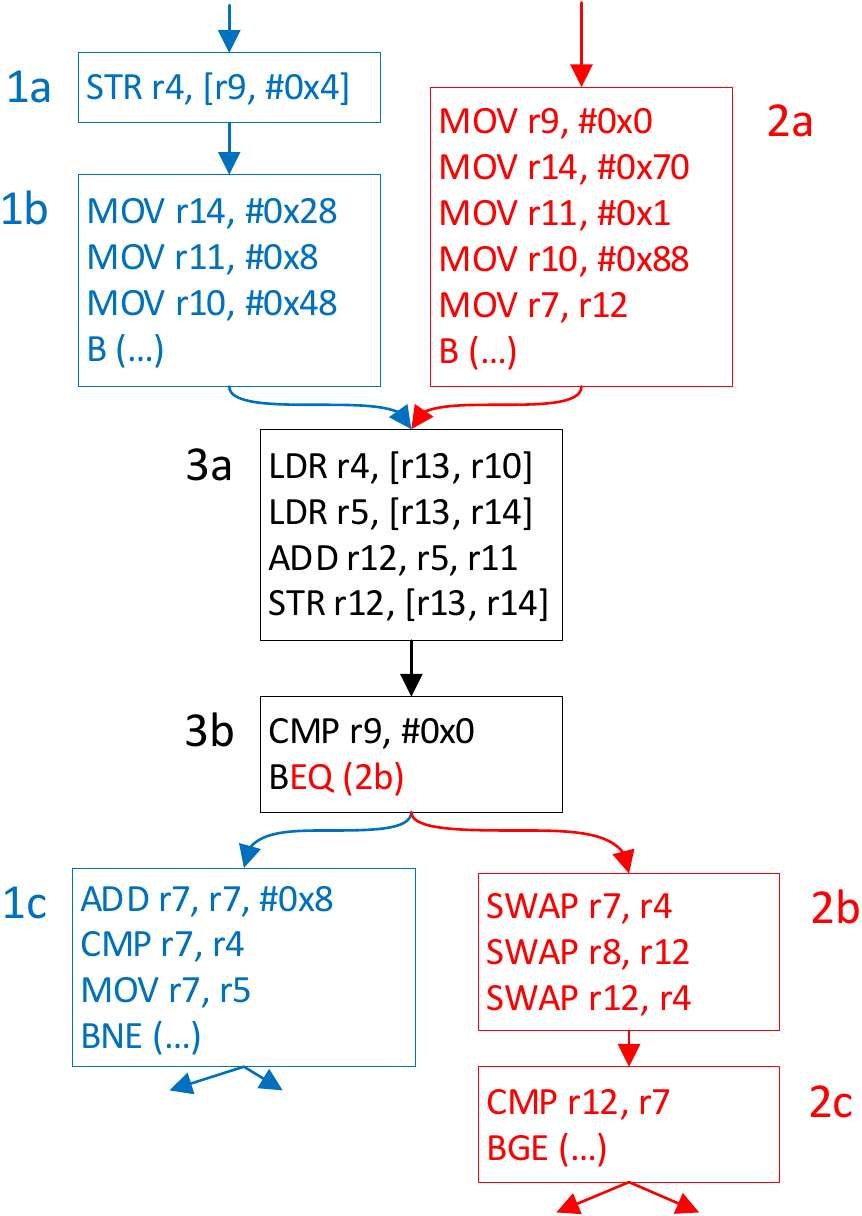}
  \caption{Factored slices}
  \label{fig:fragments_factored}
\end{figure}

\if false
To illustrate the required glue code, consider the concrete example in Figure~\ref{fig:fragments}, where the selected slices are marked in \textbf{bold}. To extract the slices, it suffices to reschedule the code and split basic blocks, as shown in Figure~\ref{fig:fragments_split}. Differences in immediates (0x48--0x88, 0x28-0x70, 0x8-0x1) are compensated with the moves of constants into registers r10, r11, and r14, as shown in the code after factorization in Figure~\ref{fig:fragments_factored}. Registers are reallocated in slice 2 to become identical to those in slice 1. These fragments consume no live-in registers besides the identically used r13 (the stack pointer), but in slice 1 live-out registers r4 and r12 are overwritten, which happen to be live throughout slice 2, and the corresponding registers r7 and r8 respectively in slice 2 are live throughout slice 1. To reallocate the registers in slice 2, one move (r12 to r7) is prepended, and three swap instructions are appended.\footnote{Note that these moves and swaps might get eliminated afterwards when the binary rewriter performs a copy propagation optimization post-pass.} Finally, in this factoring of only two slices, a simple conditional branch preceded by a compare-to-zero instruction is used as dispatcher. This dispatcher is controlled via register r9. The value zero is moved into r9 in the glue code preceding slice 2. For slice 1, no value is moved into r9, however. Instead, the fact that r9 is used as a base address in the store preceding slice 1 is relied upon: as user applications have no data mapped onto the lowest page in virtual memory, we can assume that r9 will be non-zero in the code following the store with a small immediate offset, for if the base address were zero it would have resulted in an exception. Likewise, we can assume that any registers later used as a base address in a memory access with a small immediate will not be zero, for if they will, the program will crash shortly after the fragment.\footnote{Note that relying on these assumptions is optional: Users of our tool can disable them, e.g., when protecting system software for which the assumption might not hold, or in software that relies on exception handlers to handle base addresses being zero, or if the crashing behavior should, for whatever reason, and despite it not being part of the, e.g., the C specification, be maintained.}
\else
Figures~\ref{fig:fragments},~\ref{fig:fragments_split}, and~\ref{fig:fragments_factored} illustrate the required transformations with 32-bit ARMv8 code. The selected slices are marked in bold in Figure~\ref{fig:fragments}. They have been rescheduled into separate blocks in Figure~\ref{fig:fragments_split}. To enable the factoring already applied in Figure~\ref{fig:fragments_factored}, the differences in immediate operands and register allocations have been overcome by inserting a number of move and swap operations in blocks 1b, 2a, and 2b. The dispatcher in block 3b is a simple conditional branch. In the first instruction of block 2a, the controlling register r9 is set to zero, to control and enable the execution path 2a-3a-3b-2b. For controlling and enabling the path 1a-1b-3a-3b-1c, register r9 does not need to be set to a specific value. Instead, the fact that r9 is used as a base address in the store preceding slice 1 is relied upon: as user applications have no data mapped onto the lowest page in virtual memory, we can assume that r9 will be non-zero in the code following the store. This assumption is optional and can easily omitted in scenarios where it would not hold, such as kernel code.
\fi

To test whether sufficient glue code can be generated to make a fragment set actually factorable, we use a bi-directional, context-sensitive interprocedural liveness analysis~\cite{JSAliveness}. To identify already available constants as input to dispatchers, we perform a flow-sensitive, context-sensitive (k-depth with k=1) constant propagation analysis~\cite{TOPLAS00}. On top, we developed a simple flow-sensitive, context-sensitive (k-depth with k=1), bidirectional, interprocedural non-zero analysis that tracks which registers hold values that are definitely non-zero. %
As these data flow analyses operate at the level of executable code, where useful alias information is sparse~\cite{alias_executable_muth}, they only analyze data in registers.%

The constant analysis and the non-zero analysis allow us to reuse values that already have semantic relevance in the original program to control the dispatcher. If, for some factored fragment, this is the case for more than one of the contexts from which the factored fragment was extracted, the dispatcher is then controlled by semantically relevant data originating from more than one execution context. The invariants that held in those original contexts in isolation likely do not hold in the merged context after factoring. We conjecture that this makes code comprehension harder. It also ensures that de-obfuscation techniques based on (quasi-)\-invariants will not work on the factored code.

In the example, slice 2's registers were renamed to those of slice 1. In many cases, candidate sets consist of more than 2 slices. Trying out all possible register renamings to select the best one would increase the code analysis time significantly, so instead we use a simple heuristic to select one of the slices as reference slice to which the others are renamed. This simple heuristic in practice also favors more likely successful renamings over less likely successful ones. In slice 1 of the example, the value loaded into r5 by the second load is live-out. In slice 2, the value loaded into r8 by the corresponding load is overwritten by the add. So an allocation like that of slice 2 cannot replace that of slice 1. In our simple heuristic, we count the number of different registers occurring in the original fragments, and we pick the one with the highest number as reference fragment. %
In case the heuristic does not favor one fragment over the others, and when (optional) profiling information is available, we pick the fragment with the highest execution count as reference fragment.%
While these simple heuristics are clearly not optimal, they provide a good balance between analysis time, performance and size overhead, and success ratio of the transformations.

\subsection{Selection Order}
\label{sec:selection_order}

Instructions can be present in multiple factoring candidate sets, but each instruction can only be factored once. Furthermore, factoring a set of fragments changes the data flow properties in the surrounding code, e.g., by making previously dead registers containing non-zero or constant data live, so one factoring can impact the potential of another candidate one. The order in which we select and apply actual factorings is therefore important.

The selection order also needs to strike a balance between the level of protection and obfuscation speed. The former requires a global optimization and decision process that considers all potential candidate sets. However, that would require too much computation time. The potential candidate sets can be very large, up to hundreds of fragments, especially for small fragments of one or two instructions. The larger subsets thereof are typically not actual factoring candidates because our local register renaming technique is not powerful enough to overcome the differences in data flow properties of all the fragments surroundings. For smaller candidate subsets, the renaming is much more likely to succeed. Our approach hence starts from small candidate sets, that we expand as much as possible, i.e., as long as the estimated protection value increases.

\subsubsection{Priority Function}

To order and compare candidate sets in terms of protection value, we need to consider measurable features (i.e., metrics) that contribute to the potency, resilience, and stealth of factoring them. We propose the following ones:
\begin{enumerate}
  \item the fragment size as their number of instructions;
  \item the numbers of archives, object files, and functions from which the fragments come;
  \item the numbers of archives, object files, and functions in which fragments were observed to be executed for at least one input, as determined by (optionally) profiling or fuzzing;
  \item the possible dispatchers, and, if applicable, the already available constants or non-zero values.
\end{enumerate}

The first metric prioritizes larger code fragments over smaller ones. We conjecture this is useful because factoring larger fragments results in more semantics being merged from different contexts, thus increasing the potency of a factoring transformation. It can also be useful for stealth, as it allows for better mixing of the injected dispatcher code with the factored code. Finally, it can contribute to the resilience against certain attacks. For example, undoing a factoring transformation by statically rewriting the code is more difficult when more instructions need to be re-inserted in the contexts from which they were factored.

The second metric, which actually consists of three metrics, contributes to potency. Assigning higher value to factorings of unrelated fragments originating from multiple object archives, object files, or functions, allows us to prioritize candidate sets that break proximity-based attack heuristics and that obfuscate component boundaries.

The third metric, again a set of three metrics, relates to resilience against dynamic attacks that build on observations of executions of the software under attack. These metrics allow us to prioritize candidate sets of which the effect of factoring them on the reconstructed CFGs cannot be undone by omitting edges and nodes that the attacker cannot trigger during dynamic attacks and by then simplifying the remaining code, as is done in the generic de-obfuscation attack by Yadegari et al.~\cite{yadegari}. %

The fourth metric allows to consider the potency, resilience, and stealth of the different types of dispatchers: some are harder to analyze but not very stealthy (e.g., dynamic switch dispatchers), others are stealthy in the sense that they ressemble already occuring fragments in the original programs (e.g., conditional jumps). Some are more resilient to automatic de-obfuscation, others are less so. The different dispatchers are discussed in Section~\ref{sec:transformation}.

The metrics can be combined in a priority function in various ways: in weighted sums, in decision trees, etc. They can also be combined with profile information to give lower priority to fragments on frequently executed code paths to minimize the performance impact of the factorings. The definition of the best priority function is out of the scope of this \papertechreport{}. Importantly, a user of our protection tool chain can customize it depending on his use case at hand, taking into account the security requirements of the software assets at hand (confidentiallity, integrity, ...), a risk assessment of different attack scenarios, and the performance budget.

\subsubsection{Selection and Actual Factoring}

Our factoring algorithm consists of two phases.%

At the start of the \emph{selection phase}, we perform the already mentioned data flow analyses. Then a list of \emph{initial factoring candidates} is assembled, ordered by their protection value. This list includes sets of fragments that are actual factoring candidates in the untransformed program. In other words, the data flow properties of the original program meet the necessary pre-conditions to apply the factoring transformations. No factorings are applied yet, however.

To decide on the initial candidate sets to add to the list in the selection phase, we implemented an iterative algorithm that is applied to each of the potential candidate sets. For each such set, the algorithm starts by marking \emph{pairs} of fragments that can be factored, i.e., pairs for which register renaming can be performed and at least one dispatcher can be generated. Using the priority function to sort all possible pairs in terms of protection value, we select the best starting pair as the seed set. Next, we iteratively try to expand the seed set. In each iteration, we add the one fragment from the potential candidate set that results in the biggest increase in protection value. This continues as long as the protection value increases. The final expanded set is then added to the list of actual factoring candidates, in which we also keep track of the possible dispatchers, available constants or non-zero values, and other useful information to steer the dispatcher. The fragments in the expanded set are removed from the potential candidate set, and the whole process is repeated with other seeds until no sufficiently valuable seed sets can be found anymore.

In the \emph{factoring phase}, we iterate over the ordered list of actual factoring candidate sets in decreaseing priority. We factor each set if the necessary pre-conditions have not been invalidated by a previously applied factoring. Our prototype implementation can be configured on how to choose specific dispatchers from the available ones for each factoring, such as randomly or giving priority to specific forms. After each factoring, we update data flow information by means of incremental versions of the mentioned analyses to propagate the impact of the performed factoring on available registers, constants, and non-zero values to the necessary program locations.

\subsection{Dispatchers}
\label{sec:transformation}
Many different dispatchers can be designed. We developed support for four types.%

\subsubsection{Conditional jump dispatcher}
\label{sec:dispconditional}
For sets of two fragments, a simple conditional branch can serve as dispatcher, as in Figure~\ref{fig:fragments_factored}. A branch condition like equal-to-zero can be steered with a zero and an unknown non-zero value that already has a semantic role in the original program. %
If no constants or non-zero values are available at the program locations of the original fragments, glue code is injected to produce them, possibly in an obfuscated manner
and hoisted in the code such that a local static analysis does not suffice to detect it. We will come back to this in Section~\ref{sec:factoring_integration}. Moreover, there is no need to keep it in a register, it can also be stored in memory. %
All kinds of schemes can be imagined that opaquely produce or load specific constant values or other values, always negative or always positive values, etc.

These dispatchers offer the major advantage that disassemblers like IDA Pro will recognize them as intraprocedural control flow, and thus we can rely on them to steer IDA Pro towards incorrect partitioning of code into functions. %

In terms of preconditions, it is important to note that this type of dispatcher sets the processor's status flags. If those were live-out in the original fragments, it means the status bits have to be saved somehow, either in registers or by spilling them to memory. Saving and spilling status flags is rarely done in compiler-generated code, however, so when it occurs, it makes the code immediately suspicious in the eyes of attackers. For that reason, we opted not to use this type of dispatcher when the status flags are live-out in any of the involved fragments. Whether or not this is the best choice under all circumstances %
admittedly is open for debate. %

For sets of more fragments, trees of multiple conditional branches can be used, but our prototype implementation is currently limited to single branches that are fed data (zeroes and non-zero values) directly through registers.

\subsubsection{Indirect branch dispatcher}
\label{sec:dispindirect}
For larger fragment sets, we can use branch-to-register dispatchers, similar to the branch functions of Linn et al~\cite{linn2003obfuscation}. In the simplest implementation, the exact addresses of the destination blocks are produced in the glue code preceding the extracted fragments, but less manifest schemes can easily be constructed.%

Very simple schemes in which addresses are produced directly and locally, i.e., in glue code immediately preceding the transfer to the factored fragment, are not resilient to even relatively simple static analysis. For example, %
IDA Pro out-of-the-box identifies directly produced addresses during its recursive disassembly process and continues disassembling at those addresses. If the bytes at those addresses correspond to valid instruction encodings, IDA Pro adds the code at those addresses to CFGs, albeit in separate functions to which it does not create edges from the dispatcher. Complex schemes in which addresses are computed right before the branch-to-register instruction can be made completely resilient against static analysis and even the generic de-obfuscation of Debray et al., but they come with the disadvantage that they are not at all stealthy. For example, it happens pretty rarely that values are XOR-ed before serving as a branch target, so schemes based on XOR-ing can be targeted with pattern matchers. %

Unlike conditional jump dispatchers, IDA Pro does not add outgoing edges to this type of dispatcher. So while it can be used to prevent the tool from constructing complete function CFGs out of the box, it cannot, by itself, steer IDA Pro towards incorrect CFGs that incorporate basic blocks from multiple, unrelated functions. As we will discuss in Section~\ref{sec:factoring_integration}, we can combine this type of dispatcher with other obfuscation constructs to reach exactly that.

In our prototype obfuscator, we only implemented support for schemes with direct address production in a dead register in the glue code preceding the factored fragments. %

\subsubsection{Static switch table dispatcher}
\label{sec:dispswitch}
Whereas computed jumps occur rarely in compiled C and C++ code, indirect jumps via table look-ups occur regularly, because switch statements are typically compiled into such look-ups. Two variations exist: address tables and branch tables. In the former the address of the case to be executed is loaded from a table and jumped to, in the latter a computed jump is performed into a table of branches, which then forwards control to the case to be executed. Before the look-up, a bounds check is often performed. If it fails, control is transferred to the default case.

Table-based dispatchers mimicking switch dispatchers are therefore more stealthy than branch-function-like dispatchers. With this type of dispatcher, the glue code before factored fragments passes indexes to the dispatcher. These can again be produced directly or in some obfuscated way, and either locally or hoisted. Indexes can also be derived from known constants already in registers in the original code upon entry to the factored fragment.

The tables can be inflated with fake target addresses or jumps to fake targets. Tools like IDA Pro handle many patterns of switch table implementations and implicitly assume that the dispatchers implement intraprocedural transfers, so by implementing this dispatcher in a suitable pattern, IDA Pro can be steered towards creating many fake edges that result in incorrect CFG partitioning of the code. Disassemblers will typically also use the bounds check to determine the size of the table, so by inserting a fake bounds check, they can be fooled also in that regard.

In our prototype tool, we implemented support for both forms of tables. %
The tool inserts (fake) bounds checks if the condition registers are available. If not, there simply is no bounds check inserted. In that case, tools like IDA Pro typically do not analyze the switch statement and the table, and simply do not add outgoing edges at all.

Finally, we need to note that whereas look-up based indirect control flow transfers are more stealthy than computation-based indirect transfers, their use for factoring can still lack stealthyness, in particular for large fragment sets. This is of course the case because in non-obfuscated and hence well-structured code, switch statements typically have a low fan-in. Our factored fragments, however, have a fan-in equal to their (true) fan-out. High fan-ins are suspicious in the eyes of attackers.

The strength of this form of factoring therefore has to come from its improved potency and resilience. The potency can be improved by combining this factoring with other obfuscations, as we will discuss in Section~\ref{sec:factoring_integration}.

\subsubsection{Dynamic switch table dispatcher}
\label{sec:dispdynswitch}
To improve both the potency of look-up-based dispatchers and their resilience against static analyses, we propose to make the look-up tables dynamic rather than static.

In compiled code, there is a static one-to-one mapping of dispatchers to tables. We are not bound by this restriction, however, and can let dispatchers dynamically switch between multiple tables. To that extent, we designed and implemented what we call \textit{dynamic switch tables}. Given a set of global data tables, one such dispatcher may address any of these tables during the execution of the program. The key idea is to separate data table selection from its usage, both spatially and temporally. We do this by introducing so-called \textit{table selection points} in the CFG: locations where we insert a small instruction sequence to select one of the global data tables. We store the base address of the selected data table in a global variable used by the dispatcher. By separating the selection and use of the tables, a single dynamic switch table dispatcher may address different global data tables at different times during a single run.%

\begin{figure}[t]
  \centering
  \includegraphics[width=\linewidth]{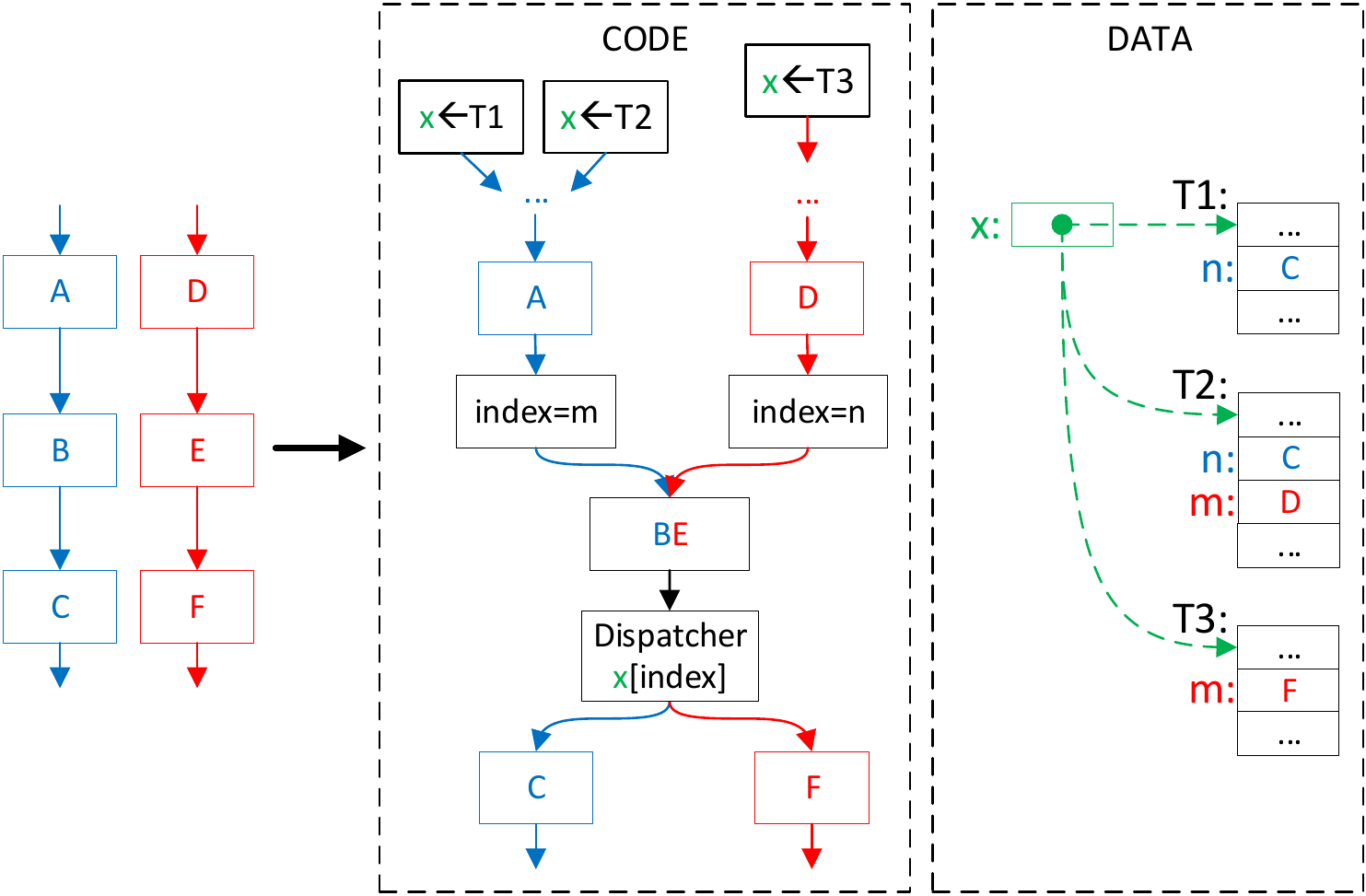}
  \caption{Transformation with dynamic switch tables}
  \label{fig:dynamic_transform}
\end{figure}

Figure~\ref{fig:dynamic_transform} shows the example factoring of two fragments B and E. The end result is shown on the right: three table selection points, the factored block BE, a dynamic switch table dispatcher, its global variable (\texttt{x}), and data tables \texttt{T1}, \texttt{T2} and \texttt{T3}. The glue code with the transfers to the factored block only contains instructions to produce the switch indices for each control flow path (\texttt{m} for fragment A and \texttt{n} for fragment B). The location where \texttt{x} gets assigned a new value does not really matter; the distance between the dispatcher and the table selection points can be arbitrarily large. Using a reachability analysis, the obfuscator determines which table selections reach which assigments of switch indices. In the example, selections of \texttt{T1} and \texttt{T2} reach the point where the index is set to m. This leads to the constraint that \texttt{T1[m]=T2[m]=C}. The tables need to be filled in respecting all such constraints.
Similar to static switch tables, we can also add false entries (e.g., at index \texttt{m} in table \texttt{T2}) to confuse the attacker and his tools.

Compared to static switch table dispatchers, dynamic table dispatchers increase the complexity by introducing an extra layer of indirection, which known static analysis cannot resolve, in particular when multiple obfuscations get combined, as will be discussed in Section~\ref{sec:factoring_integration}. %
We also observed that these dispatchers mislead IDA Pro into constructing incomplete CFGs, because it is incapable of analysing them properly. Consequently, the recursive-descent disassembler does not always disassemble all the instructions in the binary and associations between (sometimes large) portions of code are lost. The potency and resilience of this dispatcher are thus high. By contrast, this dispatcher is not stealthy: An attacker may find it strange that a dispatcher exists with no detected outgoing control flow. Given the high potency and resilience, we believe this lack of stealthiness does not completely void its usefulness.

The preconditions for this dispatcher are identical to the ones for traditional switch-based dispatchers, with the additional requirement that one extra register needs to be available to store a temporary value in. %

\subsection{Integration with other protections}
\label{sec:factoring_integration}
A potential weak point of the factoring is that the computation of the values controlling the dispatchers (such as the index into a table, or a zero constant) is done in a linear control flow path leading up to the transfers to the factored code. We can fall back to all kinds of existing obfuscations to obfuscate this calculation, but the level of obfuscation is limited by the performance budget.

Complementary, e.g., to light-weight obfuscation, we can increase the potency and resilience of the proposed techniques by coupling the factorings. We can couple them with each other as well as with the opaque predicates. In Section~\ref{sec:resilience}, we already discussed how multiple opaque predicates can be coupled by directing fake edge to points in the middle of (other) opaque predicate computations. Likewise, we can also redirect fake opaque predicate edges to the middle of instruction sequences that compute dispatcher control values. And we can choose the targets of fake entries in the tables of switch-based dispatchers in exactly the same way to obfuscate opaque predicate computations as well as dispatcher controller computations. That way, we turn the static analysis and deobfuscation of opaque predicates and factoring into one global hurdle for attackers.

\section{Experimental Evaluation}
\label{sec:evaluation}

\newcommand{\perc}[1]{\color{red}#1\color{black}}
\renewcommand{\arraystretch}{1.2}
\newcommand{\tablecaption}{insert caption here}
\newcommand{\tablelabel}{insert label here}

\definecolor{cDBFNR} {rgb}{0.55,0.00,0.00}
\definecolor{cDBFPR} {rgb}{0.06,0.31,0.55}
\definecolor{cGUIFNR}{rgb}{1.00,0.00,0.00}
\definecolor{cGUIFPR}{rgb}{0.12,0.56,1.00}

\newcommand{\dbfnr}{\textcolor{cDBFNR}{DB FNR}}
\newcommand{\dbfpr}{\textcolor{cDBFPR}{DB FPR}}
\newcommand{\guifnr}{\textcolor{cGUIFNR}{GUI FNR}}
\newcommand{\guifpr}{\textcolor{cGUIFPR}{GUI FPR}}

\newcommand{\linknoop}{\textcolor{cGUIFNR}{no opaque predicates}}
\newcommand{\linkop}{\textcolor{cDBFNR}{20\% opaque predicates}}

\subsection{Prototype Implementation}
We implemented the proposed techniques in the ASPIRE Compiler Tool Chain (ACTC)~\cite{actc}, which can compose multiple protections through source-to-source and binary code rewriting. All proposed techniques are implemented in Diablo~\cite{diablo2005}, the ACTC's link-time binary code rewriter. The code is available as open source at \url{https://github.com/csl-ugent/diablo/tree/oisp}.

Our prototype has limitations. The binary rewriter does not support trees of conditional branch dispatchers, and lacks global register allocation and the option to spill and free status registers. Furthermore, the currently supported opaque predicates are limited to algebraic ones. More complex ones can be supported by combining the ACTC's source-to-source rewriting to inject complex predicates (e.g., graph-based ones~\cite{collbergbook} or predicates resilient to symbolic execution~\cite{banescu2016code}) with binary rewriting to let fake edges cross component boundaries. Finally, the rewriter lacks support for C++ exception handling.

\subsection{Benchmarks}
We have validated correctness on all C and C++ programs from the SPEC CPU2006 benchmark suite~\cite{spec2006} (excluding \texttt{453.povray} and \texttt{471.omnetpp} that depend on exception handling) and on two industrial use cases from the ASPIRE research project~\cite{aspire}. Whereas the SPEC programs are stand-alone Linux binaries, the industrial use cases are dynamically linked Android libraries that are loaded into third-party applications. Nagravision contributed the first use case, a Digital Rights Management (DRM) plug-in that is loaded into the Android DRM and mediaserver daemon processes. SafeNet contributed the second use case, a software license manager (SLM) that is loaded into the Android Dalvik engine. Those daemons and engines are complex third-party multi-threaded processes that load and unload the libraries frequently. They hence stress-test our prototype.

The ASPIRE project deployed and validated the many ACTC-supported protections on those two use cases to mitigate attacks on the assets embedded in them, in line with the assets' security requirements as formulated by the security experts of the companies that contributed them~\cite{aspire_validation}. As part of these protections, numerous archives are linked into the libraries. The protected use cases thus form perfect candidates to evaluate the proposed methods for stealthy, obfuscated integration of components proposed in this \papertechreport{}. Table~\ref{tab:support} lists the deployed protections, and the number of components linked into the libraries thereto. In addition, we consider the SLM use case to consist of three components itself (the manager and linked-in open-source crypto and math libraries) and the DRM case of two components (the manager and some linked-in libgcc.a functionality). From the overall instruction count numbers in Table~\ref{tab:support}, it is clear that our use cases are not micro-benchmarks, but applications and libraries of real-world complexity.

\begin{table*}[]
  \centering
  \scriptsize
  \caption{Number of components in the benchmarks}
  \begin{tabular}{|l|l|c|c|c|c|c|} \cline{3-7}
    \multicolumn{2}{l|}{}                               & \textbf{SLM}  & \textbf{DRM}  & \textbf{436}  & \textbf{445}  & \textbf{454} \\\hline
    \multicolumn{2}{|l|}{\textbf{Number of archives constituting benchmark} ($\bullet$ in Figure~\ref{fig:archives})} & 3             & 2             & 17            & 7             & 5            \\\hline
    \multicolumn{7}{|l|}{\textbf{ACTC protection archives linked into benchmark} ($\circ$ in Figure~\ref{fig:archives})}\\\hline
    \multicolumn{1}{|l|}{Call stack checks} & no support components linked-in & 0 & 0 & \multicolumn{3}{|c|}{\multirow{5}{*}{N/A}} \\\cline{1-4}
    \multicolumn{1}{|l|}{Code mobility} & libwebsockets, libcurl, libssl, libcrypto, implementation & 5 & 5 & \multicolumn{3}{|c|}{} \\\cline{1-4}
    \multicolumn{1}{|l|}{Anti-debugging} & minidebugger & 1 & 1 & \multicolumn{3}{|c|}{} \\\cline{1-4}
    \multicolumn{1}{|l|}{Code guards} & implementation and guards & 1 & 1 & \multicolumn{3}{|c|}{} \\\cline{1-4}
    \multicolumn{1}{|l|}{Custom bytecode interpreter} & application-specific VM implementation & 1 & N/A & \multicolumn{3}{|c|}{} \\\hline
    \multicolumn{2}{r|}{\textbf{Overall component (=archive) count}} & \textbf{11} & \textbf{9} & \textbf{17} & \textbf{7} & \textbf{5} \\\cline{3-7}
    \multicolumn{2}{r|}{\textbf{Overall instruction count without our obfuscations}} & \textbf{276k} & \textbf{255k} & \textbf{99k} & \textbf{152k} & \textbf{366k} \\\cline{3-7}
  \end{tabular}
  \label{tab:support}
\end{table*}

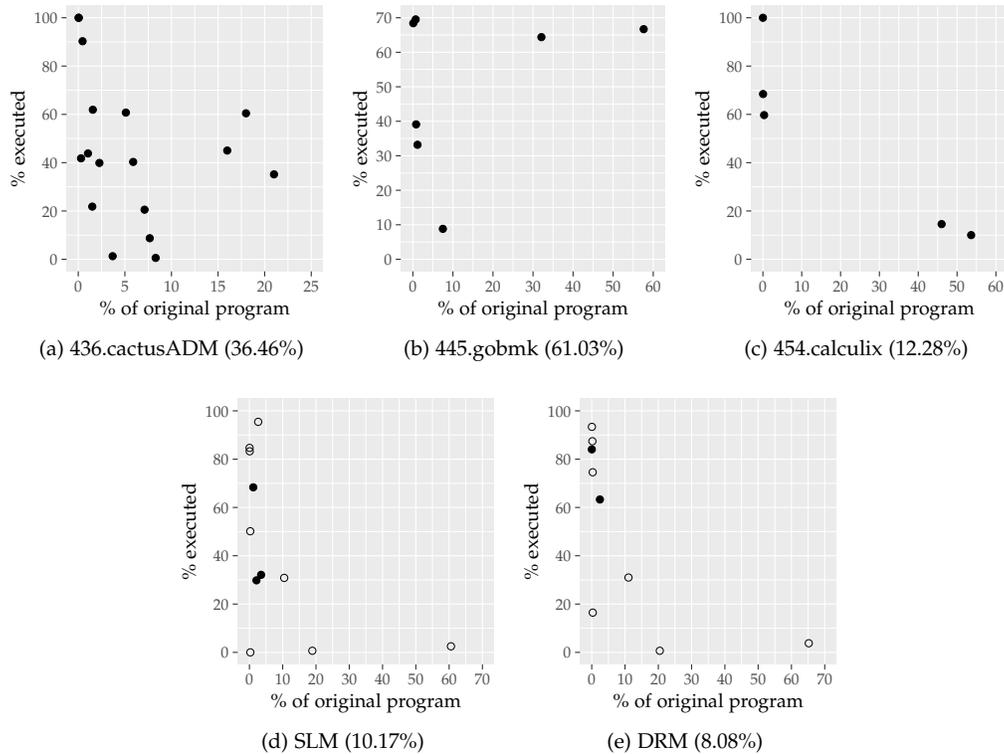
\begin{figure*}[t]
  \centering
  \subfloat[436.cactusADM (\rcaption)]{\resizebox{0.25\linewidth}{!}{\relax
\begin{tikzpicture}[x=1pt,y=1pt]
\definecolor{fillColor}{RGB}{255,255,255}
\path[use as bounding box,fill=fillColor,fill opacity=0.00] (0,0) rectangle (180.67,180.67);
\begin{scope}
\path[clip] (  0.00,  0.00) rectangle (180.67,180.67);
\definecolor{drawColor}{RGB}{255,255,255}
\definecolor{fillColor}{RGB}{255,255,255}

\path[draw=drawColor,line width= 0.6pt,line join=round,line cap=round,fill=fillColor] (  0.00, -0.00) rectangle (180.67,180.67);
\end{scope}
\begin{scope}
\path[clip] ( 35.92, 30.72) rectangle (175.17,175.17);
\definecolor{fillColor}{gray}{0.92}

\path[fill=fillColor] ( 35.92, 30.72) rectangle (175.17,175.17);
\definecolor{drawColor}{RGB}{255,255,255}

\path[draw=drawColor,line width= 0.3pt,line join=round] ( 35.92, 50.42) --
	(175.17, 50.42);

\path[draw=drawColor,line width= 0.3pt,line join=round] ( 35.92, 76.69) --
	(175.17, 76.69);

\path[draw=drawColor,line width= 0.3pt,line join=round] ( 35.92,102.95) --
	(175.17,102.95);

\path[draw=drawColor,line width= 0.3pt,line join=round] ( 35.92,129.21) --
	(175.17,129.21);

\path[draw=drawColor,line width= 0.3pt,line join=round] ( 35.92,155.48) --
	(175.17,155.48);

\path[draw=drawColor,line width= 0.3pt,line join=round] ( 54.91, 30.72) --
	( 54.91,175.17);

\path[draw=drawColor,line width= 0.3pt,line join=round] ( 80.23, 30.72) --
	( 80.23,175.17);

\path[draw=drawColor,line width= 0.3pt,line join=round] (105.55, 30.72) --
	(105.55,175.17);

\path[draw=drawColor,line width= 0.3pt,line join=round] (130.87, 30.72) --
	(130.87,175.17);

\path[draw=drawColor,line width= 0.3pt,line join=round] (156.19, 30.72) --
	(156.19,175.17);

\path[draw=drawColor,line width= 0.6pt,line join=round] ( 35.92, 37.29) --
	(175.17, 37.29);

\path[draw=drawColor,line width= 0.6pt,line join=round] ( 35.92, 63.55) --
	(175.17, 63.55);

\path[draw=drawColor,line width= 0.6pt,line join=round] ( 35.92, 89.82) --
	(175.17, 89.82);

\path[draw=drawColor,line width= 0.6pt,line join=round] ( 35.92,116.08) --
	(175.17,116.08);

\path[draw=drawColor,line width= 0.6pt,line join=round] ( 35.92,142.35) --
	(175.17,142.35);

\path[draw=drawColor,line width= 0.6pt,line join=round] ( 35.92,168.61) --
	(175.17,168.61);

\path[draw=drawColor,line width= 0.6pt,line join=round] ( 42.25, 30.72) --
	( 42.25,175.17);

\path[draw=drawColor,line width= 0.6pt,line join=round] ( 67.57, 30.72) --
	( 67.57,175.17);

\path[draw=drawColor,line width= 0.6pt,line join=round] ( 92.89, 30.72) --
	( 92.89,175.17);

\path[draw=drawColor,line width= 0.6pt,line join=round] (118.21, 30.72) --
	(118.21,175.17);

\path[draw=drawColor,line width= 0.6pt,line join=round] (143.53, 30.72) --
	(143.53,175.17);

\path[draw=drawColor,line width= 0.6pt,line join=round] (168.85, 30.72) --
	(168.85,175.17);
\definecolor{drawColor}{RGB}{0,0,0}
\definecolor{fillColor}{RGB}{0,0,0}

\path[draw=drawColor,line width= 0.4pt,line join=round,line cap=round,fill=fillColor] ( 72.08, 90.22) circle (  1.96);

\path[draw=drawColor,line width= 0.4pt,line join=round,line cap=round,fill=fillColor] (123.20, 96.42) circle (  1.96);

\path[draw=drawColor,line width= 0.4pt,line join=round,line cap=round,fill=fillColor] (133.40,116.67) circle (  1.96);

\path[draw=drawColor,line width= 0.4pt,line join=round,line cap=round,fill=fillColor] ( 50.16,118.60) circle (  1.96);

\path[draw=drawColor,line width= 0.4pt,line join=round,line cap=round,fill=fillColor] ( 78.27, 64.25) circle (  1.96);

\path[draw=drawColor,line width= 0.4pt,line join=round,line cap=round,fill=fillColor] ( 42.48,168.61) circle (  1.96);

\path[draw=drawColor,line width= 0.4pt,line join=round,line cap=round,fill=fillColor] ( 84.30, 38.03) circle (  1.96);

\path[draw=drawColor,line width= 0.4pt,line join=round,line cap=round,fill=fillColor] ( 53.74, 89.67) circle (  1.96);

\path[draw=drawColor,line width= 0.4pt,line join=round,line cap=round,fill=fillColor] ( 49.86, 65.93) circle (  1.96);

\path[draw=drawColor,line width= 0.4pt,line join=round,line cap=round,fill=fillColor] ( 68.11,117.05) circle (  1.96);

\path[draw=drawColor,line width= 0.4pt,line join=round,line cap=round,fill=fillColor] ( 81.09, 48.73) circle (  1.96);

\path[draw=drawColor,line width= 0.4pt,line join=round,line cap=round,fill=fillColor] ( 60.93, 38.96) circle (  1.96);

\path[draw=drawColor,line width= 0.4pt,line join=round,line cap=round,fill=fillColor] (148.61, 83.46) circle (  1.96);

\path[draw=drawColor,line width= 0.4pt,line join=round,line cap=round,fill=fillColor] ( 43.75, 92.18) circle (  1.96);

\path[draw=drawColor,line width= 0.4pt,line join=round,line cap=round,fill=fillColor] ( 44.59,155.86) circle (  1.96);

\path[draw=drawColor,line width= 0.4pt,line join=round,line cap=round,fill=fillColor] ( 42.48,168.61) circle (  1.96);

\path[draw=drawColor,line width= 0.4pt,line join=round,line cap=round,fill=fillColor] ( 47.53, 94.83) circle (  1.96);
\end{scope}
\begin{scope}
\path[clip] (  0.00,  0.00) rectangle (180.67,180.67);
\definecolor{drawColor}{gray}{0.30}

\node[text=drawColor,anchor=base east,inner sep=0pt, outer sep=0pt, scale=  0.88] at ( 30.97, 34.26) {0};

\node[text=drawColor,anchor=base east,inner sep=0pt, outer sep=0pt, scale=  0.88] at ( 30.97, 60.52) {20};

\node[text=drawColor,anchor=base east,inner sep=0pt, outer sep=0pt, scale=  0.88] at ( 30.97, 86.79) {40};

\node[text=drawColor,anchor=base east,inner sep=0pt, outer sep=0pt, scale=  0.88] at ( 30.97,113.05) {60};

\node[text=drawColor,anchor=base east,inner sep=0pt, outer sep=0pt, scale=  0.88] at ( 30.97,139.31) {80};

\node[text=drawColor,anchor=base east,inner sep=0pt, outer sep=0pt, scale=  0.88] at ( 30.97,165.58) {100};
\end{scope}
\begin{scope}
\path[clip] (  0.00,  0.00) rectangle (180.67,180.67);
\definecolor{drawColor}{gray}{0.20}

\path[draw=drawColor,line width= 0.6pt,line join=round] ( 33.17, 37.29) --
	( 35.92, 37.29);

\path[draw=drawColor,line width= 0.6pt,line join=round] ( 33.17, 63.55) --
	( 35.92, 63.55);

\path[draw=drawColor,line width= 0.6pt,line join=round] ( 33.17, 89.82) --
	( 35.92, 89.82);

\path[draw=drawColor,line width= 0.6pt,line join=round] ( 33.17,116.08) --
	( 35.92,116.08);

\path[draw=drawColor,line width= 0.6pt,line join=round] ( 33.17,142.35) --
	( 35.92,142.35);

\path[draw=drawColor,line width= 0.6pt,line join=round] ( 33.17,168.61) --
	( 35.92,168.61);
\end{scope}
\begin{scope}
\path[clip] (  0.00,  0.00) rectangle (180.67,180.67);
\definecolor{drawColor}{gray}{0.20}

\path[draw=drawColor,line width= 0.6pt,line join=round] ( 42.25, 27.97) --
	( 42.25, 30.72);

\path[draw=drawColor,line width= 0.6pt,line join=round] ( 67.57, 27.97) --
	( 67.57, 30.72);

\path[draw=drawColor,line width= 0.6pt,line join=round] ( 92.89, 27.97) --
	( 92.89, 30.72);

\path[draw=drawColor,line width= 0.6pt,line join=round] (118.21, 27.97) --
	(118.21, 30.72);

\path[draw=drawColor,line width= 0.6pt,line join=round] (143.53, 27.97) --
	(143.53, 30.72);

\path[draw=drawColor,line width= 0.6pt,line join=round] (168.85, 27.97) --
	(168.85, 30.72);
\end{scope}
\begin{scope}
\path[clip] (  0.00,  0.00) rectangle (180.67,180.67);
\definecolor{drawColor}{gray}{0.30}

\node[text=drawColor,anchor=base,inner sep=0pt, outer sep=0pt, scale=  0.88] at ( 42.25, 19.71) {0};

\node[text=drawColor,anchor=base,inner sep=0pt, outer sep=0pt, scale=  0.88] at ( 67.57, 19.71) {5};

\node[text=drawColor,anchor=base,inner sep=0pt, outer sep=0pt, scale=  0.88] at ( 92.89, 19.71) {10};

\node[text=drawColor,anchor=base,inner sep=0pt, outer sep=0pt, scale=  0.88] at (118.21, 19.71) {15};

\node[text=drawColor,anchor=base,inner sep=0pt, outer sep=0pt, scale=  0.88] at (143.53, 19.71) {20};

\node[text=drawColor,anchor=base,inner sep=0pt, outer sep=0pt, scale=  0.88] at (168.85, 19.71) {25};
\end{scope}
\begin{scope}
\path[clip] (  0.00,  0.00) rectangle (180.67,180.67);
\definecolor{drawColor}{RGB}{0,0,0}

\node[text=drawColor,anchor=base,inner sep=0pt, outer sep=0pt, scale=  1.10] at (105.55,  7.44) {{\%} of original program};
\end{scope}
\begin{scope}
\path[clip] (  0.00,  0.00) rectangle (180.67,180.67);
\definecolor{drawColor}{RGB}{0,0,0}

\node[text=drawColor,rotate= 90.00,anchor=base,inner sep=0pt, outer sep=0pt, scale=  1.10] at ( 13.08,102.95) {{\%} executed};
\end{scope}
\end{tikzpicture}
\gdef\rcaption{36.46{\%}}}}
  \subfloat[445.gobmk (\rcaption)]{\resizebox{0.25\linewidth}{!}{\relax
\begin{tikzpicture}[x=1pt,y=1pt]
\definecolor{fillColor}{RGB}{255,255,255}
\path[use as bounding box,fill=fillColor,fill opacity=0.00] (0,0) rectangle (180.67,180.67);
\begin{scope}
\path[clip] (  0.00,  0.00) rectangle (180.67,180.67);
\definecolor{drawColor}{RGB}{255,255,255}
\definecolor{fillColor}{RGB}{255,255,255}

\path[draw=drawColor,line width= 0.6pt,line join=round,line cap=round,fill=fillColor] (  0.00, -0.00) rectangle (180.68,180.67);
\end{scope}
\begin{scope}
\path[clip] ( 31.52, 30.72) rectangle (175.17,175.17);
\definecolor{fillColor}{gray}{0.92}

\path[fill=fillColor] ( 31.52, 30.72) rectangle (175.17,175.17);
\definecolor{drawColor}{RGB}{255,255,255}

\path[draw=drawColor,line width= 0.3pt,line join=round] ( 31.52, 46.67) --
	(175.17, 46.67);

\path[draw=drawColor,line width= 0.3pt,line join=round] ( 31.52, 65.43) --
	(175.17, 65.43);

\path[draw=drawColor,line width= 0.3pt,line join=round] ( 31.52, 84.19) --
	(175.17, 84.19);

\path[draw=drawColor,line width= 0.3pt,line join=round] ( 31.52,102.95) --
	(175.17,102.95);

\path[draw=drawColor,line width= 0.3pt,line join=round] ( 31.52,121.71) --
	(175.17,121.71);

\path[draw=drawColor,line width= 0.3pt,line join=round] ( 31.52,140.47) --
	(175.17,140.47);

\path[draw=drawColor,line width= 0.3pt,line join=round] ( 31.52,159.23) --
	(175.17,159.23);

\path[draw=drawColor,line width= 0.3pt,line join=round] ( 48.93, 30.72) --
	( 48.93,175.17);

\path[draw=drawColor,line width= 0.3pt,line join=round] ( 70.70, 30.72) --
	( 70.70,175.17);

\path[draw=drawColor,line width= 0.3pt,line join=round] ( 92.46, 30.72) --
	( 92.46,175.17);

\path[draw=drawColor,line width= 0.3pt,line join=round] (114.23, 30.72) --
	(114.23,175.17);

\path[draw=drawColor,line width= 0.3pt,line join=round] (136.00, 30.72) --
	(136.00,175.17);

\path[draw=drawColor,line width= 0.3pt,line join=round] (157.76, 30.72) --
	(157.76,175.17);

\path[draw=drawColor,line width= 0.6pt,line join=round] ( 31.52, 37.29) --
	(175.17, 37.29);

\path[draw=drawColor,line width= 0.6pt,line join=round] ( 31.52, 56.05) --
	(175.17, 56.05);

\path[draw=drawColor,line width= 0.6pt,line join=round] ( 31.52, 74.81) --
	(175.17, 74.81);

\path[draw=drawColor,line width= 0.6pt,line join=round] ( 31.52, 93.57) --
	(175.17, 93.57);

\path[draw=drawColor,line width= 0.6pt,line join=round] ( 31.52,112.33) --
	(175.17,112.33);

\path[draw=drawColor,line width= 0.6pt,line join=round] ( 31.52,131.09) --
	(175.17,131.09);

\path[draw=drawColor,line width= 0.6pt,line join=round] ( 31.52,149.85) --
	(175.17,149.85);

\path[draw=drawColor,line width= 0.6pt,line join=round] ( 31.52,168.61) --
	(175.17,168.61);

\path[draw=drawColor,line width= 0.6pt,line join=round] ( 38.05, 30.72) --
	( 38.05,175.17);

\path[draw=drawColor,line width= 0.6pt,line join=round] ( 59.81, 30.72) --
	( 59.81,175.17);

\path[draw=drawColor,line width= 0.6pt,line join=round] ( 81.58, 30.72) --
	( 81.58,175.17);

\path[draw=drawColor,line width= 0.6pt,line join=round] (103.35, 30.72) --
	(103.35,175.17);

\path[draw=drawColor,line width= 0.6pt,line join=round] (125.11, 30.72) --
	(125.11,175.17);

\path[draw=drawColor,line width= 0.6pt,line join=round] (146.88, 30.72) --
	(146.88,175.17);

\path[draw=drawColor,line width= 0.6pt,line join=round] (168.65, 30.72) --
	(168.65,175.17);
\definecolor{drawColor}{RGB}{0,0,0}
\definecolor{fillColor}{RGB}{0,0,0}

\path[draw=drawColor,line width= 0.4pt,line join=round,line cap=round,fill=fillColor] ( 38.28,165.65) circle (  1.96);

\path[draw=drawColor,line width= 0.4pt,line join=round,line cap=round,fill=fillColor] ( 40.57, 99.55) circle (  1.96);

\path[draw=drawColor,line width= 0.4pt,line join=round,line cap=round,fill=fillColor] ( 54.35, 53.83) circle (  1.96);

\path[draw=drawColor,line width= 0.4pt,line join=round,line cap=round,fill=fillColor] ( 39.83,110.64) circle (  1.96);

\path[draw=drawColor,line width= 0.4pt,line join=round,line cap=round,fill=fillColor] (163.48,162.42) circle (  1.96);

\path[draw=drawColor,line width= 0.4pt,line join=round,line cap=round,fill=fillColor] (107.90,158.08) circle (  1.96);

\path[draw=drawColor,line width= 0.4pt,line join=round,line cap=round,fill=fillColor] ( 39.59,167.71) circle (  1.96);
\end{scope}
\begin{scope}
\path[clip] (  0.00,  0.00) rectangle (180.67,180.67);
\definecolor{drawColor}{gray}{0.30}

\node[text=drawColor,anchor=base east,inner sep=0pt, outer sep=0pt, scale=  0.88] at ( 26.57, 34.26) {0};

\node[text=drawColor,anchor=base east,inner sep=0pt, outer sep=0pt, scale=  0.88] at ( 26.57, 53.02) {10};

\node[text=drawColor,anchor=base east,inner sep=0pt, outer sep=0pt, scale=  0.88] at ( 26.57, 71.78) {20};

\node[text=drawColor,anchor=base east,inner sep=0pt, outer sep=0pt, scale=  0.88] at ( 26.57, 90.54) {30};

\node[text=drawColor,anchor=base east,inner sep=0pt, outer sep=0pt, scale=  0.88] at ( 26.57,109.30) {40};

\node[text=drawColor,anchor=base east,inner sep=0pt, outer sep=0pt, scale=  0.88] at ( 26.57,128.06) {50};

\node[text=drawColor,anchor=base east,inner sep=0pt, outer sep=0pt, scale=  0.88] at ( 26.57,146.82) {60};

\node[text=drawColor,anchor=base east,inner sep=0pt, outer sep=0pt, scale=  0.88] at ( 26.57,165.58) {70};
\end{scope}
\begin{scope}
\path[clip] (  0.00,  0.00) rectangle (180.67,180.67);
\definecolor{drawColor}{gray}{0.20}

\path[draw=drawColor,line width= 0.6pt,line join=round] ( 28.77, 37.29) --
	( 31.52, 37.29);

\path[draw=drawColor,line width= 0.6pt,line join=round] ( 28.77, 56.05) --
	( 31.52, 56.05);

\path[draw=drawColor,line width= 0.6pt,line join=round] ( 28.77, 74.81) --
	( 31.52, 74.81);

\path[draw=drawColor,line width= 0.6pt,line join=round] ( 28.77, 93.57) --
	( 31.52, 93.57);

\path[draw=drawColor,line width= 0.6pt,line join=round] ( 28.77,112.33) --
	( 31.52,112.33);

\path[draw=drawColor,line width= 0.6pt,line join=round] ( 28.77,131.09) --
	( 31.52,131.09);

\path[draw=drawColor,line width= 0.6pt,line join=round] ( 28.77,149.85) --
	( 31.52,149.85);

\path[draw=drawColor,line width= 0.6pt,line join=round] ( 28.77,168.61) --
	( 31.52,168.61);
\end{scope}
\begin{scope}
\path[clip] (  0.00,  0.00) rectangle (180.67,180.67);
\definecolor{drawColor}{gray}{0.20}

\path[draw=drawColor,line width= 0.6pt,line join=round] ( 38.05, 27.97) --
	( 38.05, 30.72);

\path[draw=drawColor,line width= 0.6pt,line join=round] ( 59.81, 27.97) --
	( 59.81, 30.72);

\path[draw=drawColor,line width= 0.6pt,line join=round] ( 81.58, 27.97) --
	( 81.58, 30.72);

\path[draw=drawColor,line width= 0.6pt,line join=round] (103.35, 27.97) --
	(103.35, 30.72);

\path[draw=drawColor,line width= 0.6pt,line join=round] (125.11, 27.97) --
	(125.11, 30.72);

\path[draw=drawColor,line width= 0.6pt,line join=round] (146.88, 27.97) --
	(146.88, 30.72);

\path[draw=drawColor,line width= 0.6pt,line join=round] (168.65, 27.97) --
	(168.65, 30.72);
\end{scope}
\begin{scope}
\path[clip] (  0.00,  0.00) rectangle (180.67,180.67);
\definecolor{drawColor}{gray}{0.30}

\node[text=drawColor,anchor=base,inner sep=0pt, outer sep=0pt, scale=  0.88] at ( 38.05, 19.71) {0};

\node[text=drawColor,anchor=base,inner sep=0pt, outer sep=0pt, scale=  0.88] at ( 59.81, 19.71) {10};

\node[text=drawColor,anchor=base,inner sep=0pt, outer sep=0pt, scale=  0.88] at ( 81.58, 19.71) {20};

\node[text=drawColor,anchor=base,inner sep=0pt, outer sep=0pt, scale=  0.88] at (103.35, 19.71) {30};

\node[text=drawColor,anchor=base,inner sep=0pt, outer sep=0pt, scale=  0.88] at (125.11, 19.71) {40};

\node[text=drawColor,anchor=base,inner sep=0pt, outer sep=0pt, scale=  0.88] at (146.88, 19.71) {50};

\node[text=drawColor,anchor=base,inner sep=0pt, outer sep=0pt, scale=  0.88] at (168.65, 19.71) {60};
\end{scope}
\begin{scope}
\path[clip] (  0.00,  0.00) rectangle (180.67,180.67);
\definecolor{drawColor}{RGB}{0,0,0}

\node[text=drawColor,anchor=base,inner sep=0pt, outer sep=0pt, scale=  1.10] at (103.35,  7.44) {{\%} of original program};
\end{scope}
\begin{scope}
\path[clip] (  0.00,  0.00) rectangle (180.67,180.67);
\definecolor{drawColor}{RGB}{0,0,0}

\node[text=drawColor,rotate= 90.00,anchor=base,inner sep=0pt, outer sep=0pt, scale=  1.10] at ( 13.08,102.95) {{\%} executed};
\end{scope}
\end{tikzpicture}
\gdef\rcaption{61.03{\%}}}}
  \subfloat[454.calculix (\rcaption)]{\resizebox{0.25\linewidth}{!}{\relax
\begin{tikzpicture}[x=1pt,y=1pt]
\definecolor{fillColor}{RGB}{255,255,255}
\path[use as bounding box,fill=fillColor,fill opacity=0.00] (0,0) rectangle (180.67,180.67);
\begin{scope}
\path[clip] (  0.00,  0.00) rectangle (180.67,180.67);
\definecolor{drawColor}{RGB}{255,255,255}
\definecolor{fillColor}{RGB}{255,255,255}

\path[draw=drawColor,line width= 0.6pt,line join=round,line cap=round,fill=fillColor] (  0.00, -0.00) rectangle (180.67,180.67);
\end{scope}
\begin{scope}
\path[clip] ( 35.92, 30.72) rectangle (175.17,175.17);
\definecolor{fillColor}{gray}{0.92}

\path[fill=fillColor] ( 35.92, 30.72) rectangle (175.17,175.17);
\definecolor{drawColor}{RGB}{255,255,255}

\path[draw=drawColor,line width= 0.3pt,line join=round] ( 35.92, 50.42) --
	(175.17, 50.42);

\path[draw=drawColor,line width= 0.3pt,line join=round] ( 35.92, 76.69) --
	(175.17, 76.69);

\path[draw=drawColor,line width= 0.3pt,line join=round] ( 35.92,102.95) --
	(175.17,102.95);

\path[draw=drawColor,line width= 0.3pt,line join=round] ( 35.92,129.21) --
	(175.17,129.21);

\path[draw=drawColor,line width= 0.3pt,line join=round] ( 35.92,155.48) --
	(175.17,155.48);

\path[draw=drawColor,line width= 0.3pt,line join=round] ( 52.80, 30.72) --
	( 52.80,175.17);

\path[draw=drawColor,line width= 0.3pt,line join=round] ( 73.90, 30.72) --
	( 73.90,175.17);

\path[draw=drawColor,line width= 0.3pt,line join=round] ( 95.00, 30.72) --
	( 95.00,175.17);

\path[draw=drawColor,line width= 0.3pt,line join=round] (116.10, 30.72) --
	(116.10,175.17);

\path[draw=drawColor,line width= 0.3pt,line join=round] (137.20, 30.72) --
	(137.20,175.17);

\path[draw=drawColor,line width= 0.3pt,line join=round] (158.30, 30.72) --
	(158.30,175.17);

\path[draw=drawColor,line width= 0.6pt,line join=round] ( 35.92, 37.29) --
	(175.17, 37.29);

\path[draw=drawColor,line width= 0.6pt,line join=round] ( 35.92, 63.55) --
	(175.17, 63.55);

\path[draw=drawColor,line width= 0.6pt,line join=round] ( 35.92, 89.82) --
	(175.17, 89.82);

\path[draw=drawColor,line width= 0.6pt,line join=round] ( 35.92,116.08) --
	(175.17,116.08);

\path[draw=drawColor,line width= 0.6pt,line join=round] ( 35.92,142.35) --
	(175.17,142.35);

\path[draw=drawColor,line width= 0.6pt,line join=round] ( 35.92,168.61) --
	(175.17,168.61);

\path[draw=drawColor,line width= 0.6pt,line join=round] ( 42.25, 30.72) --
	( 42.25,175.17);

\path[draw=drawColor,line width= 0.6pt,line join=round] ( 63.35, 30.72) --
	( 63.35,175.17);

\path[draw=drawColor,line width= 0.6pt,line join=round] ( 84.45, 30.72) --
	( 84.45,175.17);

\path[draw=drawColor,line width= 0.6pt,line join=round] (105.55, 30.72) --
	(105.55,175.17);

\path[draw=drawColor,line width= 0.6pt,line join=round] (126.65, 30.72) --
	(126.65,175.17);

\path[draw=drawColor,line width= 0.6pt,line join=round] (147.75, 30.72) --
	(147.75,175.17);

\path[draw=drawColor,line width= 0.6pt,line join=round] (168.85, 30.72) --
	(168.85,175.17);
\definecolor{drawColor}{RGB}{0,0,0}
\definecolor{fillColor}{RGB}{0,0,0}

\path[draw=drawColor,line width= 0.4pt,line join=round,line cap=round,fill=fillColor] ( 42.34,127.14) circle (  1.96);

\path[draw=drawColor,line width= 0.4pt,line join=round,line cap=round,fill=fillColor] (139.34, 56.41) circle (  1.96);

\path[draw=drawColor,line width= 0.4pt,line join=round,line cap=round,fill=fillColor] (155.40, 50.43) circle (  1.96);

\path[draw=drawColor,line width= 0.4pt,line join=round,line cap=round,fill=fillColor] ( 42.25,168.61) circle (  1.96);

\path[draw=drawColor,line width= 0.4pt,line join=round,line cap=round,fill=fillColor] ( 42.90,115.69) circle (  1.96);
\end{scope}
\begin{scope}
\path[clip] (  0.00,  0.00) rectangle (180.67,180.67);
\definecolor{drawColor}{gray}{0.30}

\node[text=drawColor,anchor=base east,inner sep=0pt, outer sep=0pt, scale=  0.88] at ( 30.97, 34.26) {0};

\node[text=drawColor,anchor=base east,inner sep=0pt, outer sep=0pt, scale=  0.88] at ( 30.97, 60.52) {20};

\node[text=drawColor,anchor=base east,inner sep=0pt, outer sep=0pt, scale=  0.88] at ( 30.97, 86.79) {40};

\node[text=drawColor,anchor=base east,inner sep=0pt, outer sep=0pt, scale=  0.88] at ( 30.97,113.05) {60};

\node[text=drawColor,anchor=base east,inner sep=0pt, outer sep=0pt, scale=  0.88] at ( 30.97,139.31) {80};

\node[text=drawColor,anchor=base east,inner sep=0pt, outer sep=0pt, scale=  0.88] at ( 30.97,165.58) {100};
\end{scope}
\begin{scope}
\path[clip] (  0.00,  0.00) rectangle (180.67,180.67);
\definecolor{drawColor}{gray}{0.20}

\path[draw=drawColor,line width= 0.6pt,line join=round] ( 33.17, 37.29) --
	( 35.92, 37.29);

\path[draw=drawColor,line width= 0.6pt,line join=round] ( 33.17, 63.55) --
	( 35.92, 63.55);

\path[draw=drawColor,line width= 0.6pt,line join=round] ( 33.17, 89.82) --
	( 35.92, 89.82);

\path[draw=drawColor,line width= 0.6pt,line join=round] ( 33.17,116.08) --
	( 35.92,116.08);

\path[draw=drawColor,line width= 0.6pt,line join=round] ( 33.17,142.35) --
	( 35.92,142.35);

\path[draw=drawColor,line width= 0.6pt,line join=round] ( 33.17,168.61) --
	( 35.92,168.61);
\end{scope}
\begin{scope}
\path[clip] (  0.00,  0.00) rectangle (180.67,180.67);
\definecolor{drawColor}{gray}{0.20}

\path[draw=drawColor,line width= 0.6pt,line join=round] ( 42.25, 27.97) --
	( 42.25, 30.72);

\path[draw=drawColor,line width= 0.6pt,line join=round] ( 63.35, 27.97) --
	( 63.35, 30.72);

\path[draw=drawColor,line width= 0.6pt,line join=round] ( 84.45, 27.97) --
	( 84.45, 30.72);

\path[draw=drawColor,line width= 0.6pt,line join=round] (105.55, 27.97) --
	(105.55, 30.72);

\path[draw=drawColor,line width= 0.6pt,line join=round] (126.65, 27.97) --
	(126.65, 30.72);

\path[draw=drawColor,line width= 0.6pt,line join=round] (147.75, 27.97) --
	(147.75, 30.72);

\path[draw=drawColor,line width= 0.6pt,line join=round] (168.85, 27.97) --
	(168.85, 30.72);
\end{scope}
\begin{scope}
\path[clip] (  0.00,  0.00) rectangle (180.67,180.67);
\definecolor{drawColor}{gray}{0.30}

\node[text=drawColor,anchor=base,inner sep=0pt, outer sep=0pt, scale=  0.88] at ( 42.25, 19.71) {0};

\node[text=drawColor,anchor=base,inner sep=0pt, outer sep=0pt, scale=  0.88] at ( 63.35, 19.71) {10};

\node[text=drawColor,anchor=base,inner sep=0pt, outer sep=0pt, scale=  0.88] at ( 84.45, 19.71) {20};

\node[text=drawColor,anchor=base,inner sep=0pt, outer sep=0pt, scale=  0.88] at (105.55, 19.71) {30};

\node[text=drawColor,anchor=base,inner sep=0pt, outer sep=0pt, scale=  0.88] at (126.65, 19.71) {40};

\node[text=drawColor,anchor=base,inner sep=0pt, outer sep=0pt, scale=  0.88] at (147.75, 19.71) {50};

\node[text=drawColor,anchor=base,inner sep=0pt, outer sep=0pt, scale=  0.88] at (168.85, 19.71) {60};
\end{scope}
\begin{scope}
\path[clip] (  0.00,  0.00) rectangle (180.67,180.67);
\definecolor{drawColor}{RGB}{0,0,0}

\node[text=drawColor,anchor=base,inner sep=0pt, outer sep=0pt, scale=  1.10] at (105.55,  7.44) {{\%} of original program};
\end{scope}
\begin{scope}
\path[clip] (  0.00,  0.00) rectangle (180.67,180.67);
\definecolor{drawColor}{RGB}{0,0,0}

\node[text=drawColor,rotate= 90.00,anchor=base,inner sep=0pt, outer sep=0pt, scale=  1.10] at ( 13.08,102.95) {{\%} executed};
\end{scope}
\end{tikzpicture}
\gdef\rcaption{12.28{\%}}}}
  \\
  \subfloat[SLM (\rcaption)]{\resizebox{0.25\linewidth}{!}{\relax
\begin{tikzpicture}[x=1pt,y=1pt]
\definecolor{fillColor}{RGB}{255,255,255}
\path[use as bounding box,fill=fillColor,fill opacity=0.00] (0,0) rectangle (180.67,180.67);
\begin{scope}
\path[clip] (  0.00,  0.00) rectangle (180.67,180.67);
\definecolor{drawColor}{RGB}{255,255,255}
\definecolor{fillColor}{RGB}{255,255,255}

\path[draw=drawColor,line width= 0.6pt,line join=round,line cap=round,fill=fillColor] (  0.00, -0.00) rectangle (180.67,180.67);
\end{scope}
\begin{scope}
\path[clip] ( 35.92, 30.72) rectangle (175.17,175.17);
\definecolor{fillColor}{gray}{0.92}

\path[fill=fillColor] ( 35.92, 30.72) rectangle (175.17,175.17);
\definecolor{drawColor}{RGB}{255,255,255}

\path[draw=drawColor,line width= 0.3pt,line join=round] ( 35.92, 50.42) --
	(175.17, 50.42);

\path[draw=drawColor,line width= 0.3pt,line join=round] ( 35.92, 76.69) --
	(175.17, 76.69);

\path[draw=drawColor,line width= 0.3pt,line join=round] ( 35.92,102.95) --
	(175.17,102.95);

\path[draw=drawColor,line width= 0.3pt,line join=round] ( 35.92,129.21) --
	(175.17,129.21);

\path[draw=drawColor,line width= 0.3pt,line join=round] ( 35.92,155.48) --
	(175.17,155.48);

\path[draw=drawColor,line width= 0.3pt,line join=round] ( 51.29, 30.72) --
	( 51.29,175.17);

\path[draw=drawColor,line width= 0.3pt,line join=round] ( 69.37, 30.72) --
	( 69.37,175.17);

\path[draw=drawColor,line width= 0.3pt,line join=round] ( 87.46, 30.72) --
	( 87.46,175.17);

\path[draw=drawColor,line width= 0.3pt,line join=round] (105.55, 30.72) --
	(105.55,175.17);

\path[draw=drawColor,line width= 0.3pt,line join=round] (123.63, 30.72) --
	(123.63,175.17);

\path[draw=drawColor,line width= 0.3pt,line join=round] (141.72, 30.72) --
	(141.72,175.17);

\path[draw=drawColor,line width= 0.3pt,line join=round] (159.80, 30.72) --
	(159.80,175.17);

\path[draw=drawColor,line width= 0.6pt,line join=round] ( 35.92, 37.29) --
	(175.17, 37.29);

\path[draw=drawColor,line width= 0.6pt,line join=round] ( 35.92, 63.55) --
	(175.17, 63.55);

\path[draw=drawColor,line width= 0.6pt,line join=round] ( 35.92, 89.82) --
	(175.17, 89.82);

\path[draw=drawColor,line width= 0.6pt,line join=round] ( 35.92,116.08) --
	(175.17,116.08);

\path[draw=drawColor,line width= 0.6pt,line join=round] ( 35.92,142.35) --
	(175.17,142.35);

\path[draw=drawColor,line width= 0.6pt,line join=round] ( 35.92,168.61) --
	(175.17,168.61);

\path[draw=drawColor,line width= 0.6pt,line join=round] ( 42.25, 30.72) --
	( 42.25,175.17);

\path[draw=drawColor,line width= 0.6pt,line join=round] ( 60.33, 30.72) --
	( 60.33,175.17);

\path[draw=drawColor,line width= 0.6pt,line join=round] ( 78.42, 30.72) --
	( 78.42,175.17);

\path[draw=drawColor,line width= 0.6pt,line join=round] ( 96.50, 30.72) --
	( 96.50,175.17);

\path[draw=drawColor,line width= 0.6pt,line join=round] (114.59, 30.72) --
	(114.59,175.17);

\path[draw=drawColor,line width= 0.6pt,line join=round] (132.67, 30.72) --
	(132.67,175.17);

\path[draw=drawColor,line width= 0.6pt,line join=round] (150.76, 30.72) --
	(150.76,175.17);

\path[draw=drawColor,line width= 0.6pt,line join=round] (168.85, 30.72) --
	(168.85,175.17);
\definecolor{drawColor}{RGB}{0,0,0}
\definecolor{fillColor}{RGB}{0,0,0}

\path[draw=drawColor,line width= 0.4pt,line join=round,line cap=round,fill=fillColor] ( 44.34,127.04) circle (  1.96);

\path[draw=drawColor,line width= 0.4pt,line join=round,line cap=round] ( 61.15, 77.78) circle (  1.96);

\path[draw=drawColor,line width= 0.4pt,line join=round,line cap=round] (151.76, 40.58) circle (  1.96);

\path[draw=drawColor,line width= 0.4pt,line join=round,line cap=round] ( 42.40,146.63) circle (  1.96);

\path[draw=drawColor,line width= 0.4pt,line join=round,line cap=round] ( 42.70,103.14) circle (  1.96);

\path[draw=drawColor,line width= 0.4pt,line join=round,line cap=round] ( 42.34,148.47) circle (  1.96);

\path[draw=drawColor,line width= 0.4pt,line join=round,line cap=round] ( 42.75, 37.29) circle (  1.96);

\path[draw=drawColor,line width= 0.4pt,line join=round,line cap=round] ( 76.44, 38.15) circle (  1.96);

\path[draw=drawColor,line width= 0.4pt,line join=round,line cap=round] ( 47.01,162.64) circle (  1.96);

\path[draw=drawColor,line width= 0.4pt,line join=round,line cap=round,fill=fillColor] ( 48.60, 79.45) circle (  1.96);

\path[draw=drawColor,line width= 0.4pt,line join=round,line cap=round,fill=fillColor] ( 46.08, 76.45) circle (  1.96);
\end{scope}
\begin{scope}
\path[clip] (  0.00,  0.00) rectangle (180.67,180.67);
\definecolor{drawColor}{gray}{0.30}

\node[text=drawColor,anchor=base east,inner sep=0pt, outer sep=0pt, scale=  0.88] at ( 30.97, 34.26) {0};

\node[text=drawColor,anchor=base east,inner sep=0pt, outer sep=0pt, scale=  0.88] at ( 30.97, 60.52) {20};

\node[text=drawColor,anchor=base east,inner sep=0pt, outer sep=0pt, scale=  0.88] at ( 30.97, 86.79) {40};

\node[text=drawColor,anchor=base east,inner sep=0pt, outer sep=0pt, scale=  0.88] at ( 30.97,113.05) {60};

\node[text=drawColor,anchor=base east,inner sep=0pt, outer sep=0pt, scale=  0.88] at ( 30.97,139.31) {80};

\node[text=drawColor,anchor=base east,inner sep=0pt, outer sep=0pt, scale=  0.88] at ( 30.97,165.58) {100};
\end{scope}
\begin{scope}
\path[clip] (  0.00,  0.00) rectangle (180.67,180.67);
\definecolor{drawColor}{gray}{0.20}

\path[draw=drawColor,line width= 0.6pt,line join=round] ( 33.17, 37.29) --
	( 35.92, 37.29);

\path[draw=drawColor,line width= 0.6pt,line join=round] ( 33.17, 63.55) --
	( 35.92, 63.55);

\path[draw=drawColor,line width= 0.6pt,line join=round] ( 33.17, 89.82) --
	( 35.92, 89.82);

\path[draw=drawColor,line width= 0.6pt,line join=round] ( 33.17,116.08) --
	( 35.92,116.08);

\path[draw=drawColor,line width= 0.6pt,line join=round] ( 33.17,142.35) --
	( 35.92,142.35);

\path[draw=drawColor,line width= 0.6pt,line join=round] ( 33.17,168.61) --
	( 35.92,168.61);
\end{scope}
\begin{scope}
\path[clip] (  0.00,  0.00) rectangle (180.67,180.67);
\definecolor{drawColor}{gray}{0.20}

\path[draw=drawColor,line width= 0.6pt,line join=round] ( 42.25, 27.97) --
	( 42.25, 30.72);

\path[draw=drawColor,line width= 0.6pt,line join=round] ( 60.33, 27.97) --
	( 60.33, 30.72);

\path[draw=drawColor,line width= 0.6pt,line join=round] ( 78.42, 27.97) --
	( 78.42, 30.72);

\path[draw=drawColor,line width= 0.6pt,line join=round] ( 96.50, 27.97) --
	( 96.50, 30.72);

\path[draw=drawColor,line width= 0.6pt,line join=round] (114.59, 27.97) --
	(114.59, 30.72);

\path[draw=drawColor,line width= 0.6pt,line join=round] (132.67, 27.97) --
	(132.67, 30.72);

\path[draw=drawColor,line width= 0.6pt,line join=round] (150.76, 27.97) --
	(150.76, 30.72);

\path[draw=drawColor,line width= 0.6pt,line join=round] (168.85, 27.97) --
	(168.85, 30.72);
\end{scope}
\begin{scope}
\path[clip] (  0.00,  0.00) rectangle (180.67,180.67);
\definecolor{drawColor}{gray}{0.30}

\node[text=drawColor,anchor=base,inner sep=0pt, outer sep=0pt, scale=  0.88] at ( 42.25, 19.71) {0};

\node[text=drawColor,anchor=base,inner sep=0pt, outer sep=0pt, scale=  0.88] at ( 60.33, 19.71) {10};

\node[text=drawColor,anchor=base,inner sep=0pt, outer sep=0pt, scale=  0.88] at ( 78.42, 19.71) {20};

\node[text=drawColor,anchor=base,inner sep=0pt, outer sep=0pt, scale=  0.88] at ( 96.50, 19.71) {30};

\node[text=drawColor,anchor=base,inner sep=0pt, outer sep=0pt, scale=  0.88] at (114.59, 19.71) {40};

\node[text=drawColor,anchor=base,inner sep=0pt, outer sep=0pt, scale=  0.88] at (132.67, 19.71) {50};

\node[text=drawColor,anchor=base,inner sep=0pt, outer sep=0pt, scale=  0.88] at (150.76, 19.71) {60};

\node[text=drawColor,anchor=base,inner sep=0pt, outer sep=0pt, scale=  0.88] at (168.85, 19.71) {70};
\end{scope}
\begin{scope}
\path[clip] (  0.00,  0.00) rectangle (180.67,180.67);
\definecolor{drawColor}{RGB}{0,0,0}

\node[text=drawColor,anchor=base,inner sep=0pt, outer sep=0pt, scale=  1.10] at (105.55,  7.44) {{\%} of original program};
\end{scope}
\begin{scope}
\path[clip] (  0.00,  0.00) rectangle (180.67,180.67);
\definecolor{drawColor}{RGB}{0,0,0}

\node[text=drawColor,rotate= 90.00,anchor=base,inner sep=0pt, outer sep=0pt, scale=  1.10] at ( 13.08,102.95) {{\%} executed};
\end{scope}
\end{tikzpicture}
\gdef\rcaption{10.17{\%}}}}
  \subfloat[DRM (\rcaption)]{\resizebox{0.25\linewidth}{!}{\relax
\begin{tikzpicture}[x=1pt,y=1pt]
\definecolor{fillColor}{RGB}{255,255,255}
\path[use as bounding box,fill=fillColor,fill opacity=0.00] (0,0) rectangle (180.67,180.67);
\begin{scope}
\path[clip] (  0.00,  0.00) rectangle (180.67,180.67);
\definecolor{drawColor}{RGB}{255,255,255}
\definecolor{fillColor}{RGB}{255,255,255}

\path[draw=drawColor,line width= 0.6pt,line join=round,line cap=round,fill=fillColor] (  0.00, -0.00) rectangle (180.67,180.67);
\end{scope}
\begin{scope}
\path[clip] ( 35.92, 30.72) rectangle (175.17,175.17);
\definecolor{fillColor}{gray}{0.92}

\path[fill=fillColor] ( 35.92, 30.72) rectangle (175.17,175.17);
\definecolor{drawColor}{RGB}{255,255,255}

\path[draw=drawColor,line width= 0.3pt,line join=round] ( 35.92, 50.42) --
	(175.17, 50.42);

\path[draw=drawColor,line width= 0.3pt,line join=round] ( 35.92, 76.69) --
	(175.17, 76.69);

\path[draw=drawColor,line width= 0.3pt,line join=round] ( 35.92,102.95) --
	(175.17,102.95);

\path[draw=drawColor,line width= 0.3pt,line join=round] ( 35.92,129.21) --
	(175.17,129.21);

\path[draw=drawColor,line width= 0.3pt,line join=round] ( 35.92,155.48) --
	(175.17,155.48);

\path[draw=drawColor,line width= 0.3pt,line join=round] ( 51.29, 30.72) --
	( 51.29,175.17);

\path[draw=drawColor,line width= 0.3pt,line join=round] ( 69.37, 30.72) --
	( 69.37,175.17);

\path[draw=drawColor,line width= 0.3pt,line join=round] ( 87.46, 30.72) --
	( 87.46,175.17);

\path[draw=drawColor,line width= 0.3pt,line join=round] (105.55, 30.72) --
	(105.55,175.17);

\path[draw=drawColor,line width= 0.3pt,line join=round] (123.63, 30.72) --
	(123.63,175.17);

\path[draw=drawColor,line width= 0.3pt,line join=round] (141.72, 30.72) --
	(141.72,175.17);

\path[draw=drawColor,line width= 0.3pt,line join=round] (159.80, 30.72) --
	(159.80,175.17);

\path[draw=drawColor,line width= 0.6pt,line join=round] ( 35.92, 37.29) --
	(175.17, 37.29);

\path[draw=drawColor,line width= 0.6pt,line join=round] ( 35.92, 63.55) --
	(175.17, 63.55);

\path[draw=drawColor,line width= 0.6pt,line join=round] ( 35.92, 89.82) --
	(175.17, 89.82);

\path[draw=drawColor,line width= 0.6pt,line join=round] ( 35.92,116.08) --
	(175.17,116.08);

\path[draw=drawColor,line width= 0.6pt,line join=round] ( 35.92,142.35) --
	(175.17,142.35);

\path[draw=drawColor,line width= 0.6pt,line join=round] ( 35.92,168.61) --
	(175.17,168.61);

\path[draw=drawColor,line width= 0.6pt,line join=round] ( 42.25, 30.72) --
	( 42.25,175.17);

\path[draw=drawColor,line width= 0.6pt,line join=round] ( 60.33, 30.72) --
	( 60.33,175.17);

\path[draw=drawColor,line width= 0.6pt,line join=round] ( 78.42, 30.72) --
	( 78.42,175.17);

\path[draw=drawColor,line width= 0.6pt,line join=round] ( 96.50, 30.72) --
	( 96.50,175.17);

\path[draw=drawColor,line width= 0.6pt,line join=round] (114.59, 30.72) --
	(114.59,175.17);

\path[draw=drawColor,line width= 0.6pt,line join=round] (132.67, 30.72) --
	(132.67,175.17);

\path[draw=drawColor,line width= 0.6pt,line join=round] (150.76, 30.72) --
	(150.76,175.17);

\path[draw=drawColor,line width= 0.6pt,line join=round] (168.85, 30.72) --
	(168.85,175.17);
\definecolor{drawColor}{RGB}{0,0,0}

\path[draw=drawColor,line width= 0.4pt,line join=round,line cap=round] ( 62.21, 77.98) circle (  1.96);
\definecolor{fillColor}{RGB}{0,0,0}

\path[draw=drawColor,line width= 0.4pt,line join=round,line cap=round,fill=fillColor] ( 46.68,120.41) circle (  1.96);

\path[draw=drawColor,line width= 0.4pt,line join=round,line cap=round] (160.16, 42.25) circle (  1.96);

\path[draw=drawColor,line width= 0.4pt,line join=round,line cap=round] ( 42.60,152.10) circle (  1.96);

\path[draw=drawColor,line width= 0.4pt,line join=round,line cap=round] ( 42.77,135.22) circle (  1.96);

\path[draw=drawColor,line width= 0.4pt,line join=round,line cap=round] ( 42.34,159.92) circle (  1.96);

\path[draw=drawColor,line width= 0.4pt,line join=round,line cap=round] ( 42.80, 58.90) circle (  1.96);

\path[draw=drawColor,line width= 0.4pt,line join=round,line cap=round] ( 79.22, 38.15) circle (  1.96);

\path[draw=drawColor,line width= 0.4pt,line join=round,line cap=round,fill=fillColor] ( 42.29,147.67) circle (  1.96);
\end{scope}
\begin{scope}
\path[clip] (  0.00,  0.00) rectangle (180.67,180.67);
\definecolor{drawColor}{gray}{0.30}

\node[text=drawColor,anchor=base east,inner sep=0pt, outer sep=0pt, scale=  0.88] at ( 30.97, 34.26) {0};

\node[text=drawColor,anchor=base east,inner sep=0pt, outer sep=0pt, scale=  0.88] at ( 30.97, 60.52) {20};

\node[text=drawColor,anchor=base east,inner sep=0pt, outer sep=0pt, scale=  0.88] at ( 30.97, 86.79) {40};

\node[text=drawColor,anchor=base east,inner sep=0pt, outer sep=0pt, scale=  0.88] at ( 30.97,113.05) {60};

\node[text=drawColor,anchor=base east,inner sep=0pt, outer sep=0pt, scale=  0.88] at ( 30.97,139.31) {80};

\node[text=drawColor,anchor=base east,inner sep=0pt, outer sep=0pt, scale=  0.88] at ( 30.97,165.58) {100};
\end{scope}
\begin{scope}
\path[clip] (  0.00,  0.00) rectangle (180.67,180.67);
\definecolor{drawColor}{gray}{0.20}

\path[draw=drawColor,line width= 0.6pt,line join=round] ( 33.17, 37.29) --
	( 35.92, 37.29);

\path[draw=drawColor,line width= 0.6pt,line join=round] ( 33.17, 63.55) --
	( 35.92, 63.55);

\path[draw=drawColor,line width= 0.6pt,line join=round] ( 33.17, 89.82) --
	( 35.92, 89.82);

\path[draw=drawColor,line width= 0.6pt,line join=round] ( 33.17,116.08) --
	( 35.92,116.08);

\path[draw=drawColor,line width= 0.6pt,line join=round] ( 33.17,142.35) --
	( 35.92,142.35);

\path[draw=drawColor,line width= 0.6pt,line join=round] ( 33.17,168.61) --
	( 35.92,168.61);
\end{scope}
\begin{scope}
\path[clip] (  0.00,  0.00) rectangle (180.67,180.67);
\definecolor{drawColor}{gray}{0.20}

\path[draw=drawColor,line width= 0.6pt,line join=round] ( 42.25, 27.97) --
	( 42.25, 30.72);

\path[draw=drawColor,line width= 0.6pt,line join=round] ( 60.33, 27.97) --
	( 60.33, 30.72);

\path[draw=drawColor,line width= 0.6pt,line join=round] ( 78.42, 27.97) --
	( 78.42, 30.72);

\path[draw=drawColor,line width= 0.6pt,line join=round] ( 96.50, 27.97) --
	( 96.50, 30.72);

\path[draw=drawColor,line width= 0.6pt,line join=round] (114.59, 27.97) --
	(114.59, 30.72);

\path[draw=drawColor,line width= 0.6pt,line join=round] (132.67, 27.97) --
	(132.67, 30.72);

\path[draw=drawColor,line width= 0.6pt,line join=round] (150.76, 27.97) --
	(150.76, 30.72);

\path[draw=drawColor,line width= 0.6pt,line join=round] (168.85, 27.97) --
	(168.85, 30.72);
\end{scope}
\begin{scope}
\path[clip] (  0.00,  0.00) rectangle (180.67,180.67);
\definecolor{drawColor}{gray}{0.30}

\node[text=drawColor,anchor=base,inner sep=0pt, outer sep=0pt, scale=  0.88] at ( 42.25, 19.71) {0};

\node[text=drawColor,anchor=base,inner sep=0pt, outer sep=0pt, scale=  0.88] at ( 60.33, 19.71) {10};

\node[text=drawColor,anchor=base,inner sep=0pt, outer sep=0pt, scale=  0.88] at ( 78.42, 19.71) {20};

\node[text=drawColor,anchor=base,inner sep=0pt, outer sep=0pt, scale=  0.88] at ( 96.50, 19.71) {30};

\node[text=drawColor,anchor=base,inner sep=0pt, outer sep=0pt, scale=  0.88] at (114.59, 19.71) {40};

\node[text=drawColor,anchor=base,inner sep=0pt, outer sep=0pt, scale=  0.88] at (132.67, 19.71) {50};

\node[text=drawColor,anchor=base,inner sep=0pt, outer sep=0pt, scale=  0.88] at (150.76, 19.71) {60};

\node[text=drawColor,anchor=base,inner sep=0pt, outer sep=0pt, scale=  0.88] at (168.85, 19.71) {70};
\end{scope}
\begin{scope}
\path[clip] (  0.00,  0.00) rectangle (180.67,180.67);
\definecolor{drawColor}{RGB}{0,0,0}

\node[text=drawColor,anchor=base,inner sep=0pt, outer sep=0pt, scale=  1.10] at (105.55,  7.44) {{\%} of original program};
\end{scope}
\begin{scope}
\path[clip] (  0.00,  0.00) rectangle (180.67,180.67);
\definecolor{drawColor}{RGB}{0,0,0}

\node[text=drawColor,rotate= 90.00,anchor=base,inner sep=0pt, outer sep=0pt, scale=  1.10] at ( 13.08,102.95) {{\%} executed};
\end{scope}
\end{tikzpicture}
\gdef\rcaption{8.08{\%}}}}

  \caption{Relative archive sizes and their individual coverage in the benchmarks, plus overall coverage per benchmark}
  \label{fig:archives}
\end{figure*}

By contrast, the ACTC does not deploy additional protection on the SPEC benchmarks, as those embed no security-sensitive assets. Still, three of those benchmarks have their source code split over multiple directories: \texttt{436.cactusADM}, \texttt{445.gobmk}, and \texttt{454.calculix}. By treating each directory as a separate archive, we can still evaluate our techniques on them. Figure~\ref{fig:archives} plots the relative sizes of the benchmarks' components on the x-axis; the y-axis is the code coverage in the different components obtained when we profiled the benchmarks on our training inputs. These data enable the interpretation of measurement results below.

\subsection{Applicability}
\label{sec:applicability}

First, we analyse the applicability of the different
transformations. Code layout randomization is applicable everywhere
trivially. Opaque predicates and related conditional branches can also be
inserted almost everywhere easily.
In our prototype obfuscator, the user can specify the probability
with which an opaque predicate is injected into each basic block. A
pseudo-random process then chooses blocks and opaque predicate constructs
accordingly.

\begin{figure*}[t]
  \centering
  \resizebox{\linewidth}{!}{\input{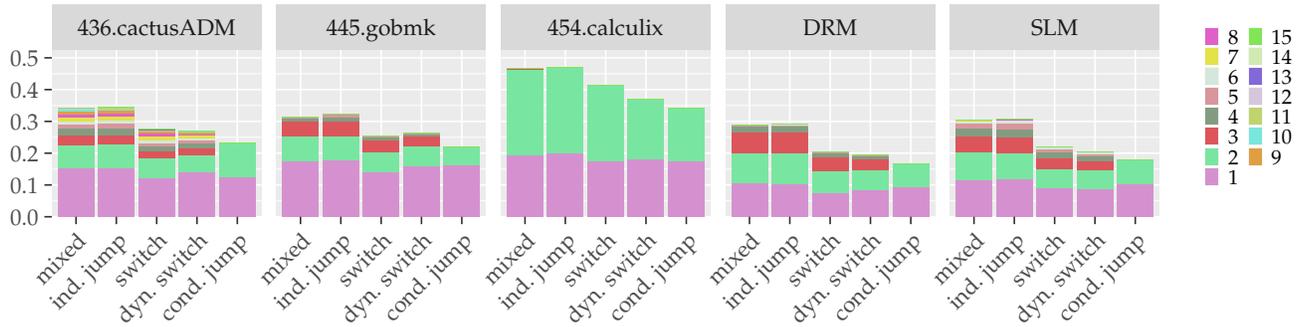}}
  \vspace{-1.2cm}
  \caption{Fraction of all instructions that get factored from within the indicated number of archives}
  \label{fig:factored_ins_archs}
\end{figure*}

\begin{figure*}[t]
  \centering
  \resizebox{\linewidth}{!}{\input{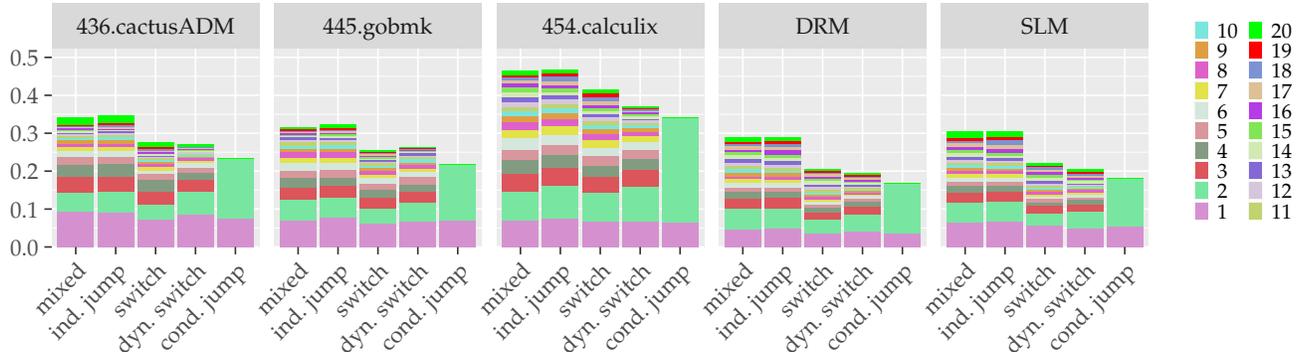}}
  \vspace{-1.0cm}
  \caption{Fraction of all instructions that get factored from within the indicated number of object files}
  \label{fig:factored_ins_objects}
\end{figure*}

By contrast, the proposed factoring techniques are not applicable trivially:
factorable fragments need to be available, preferably over component
boundaries. So first, we measured the applicability of
factoring. Figure~\ref{fig:factored_ins_archs} shows the fraction of the
original instructions that get factored in five cases: when all four types of
dispatchers (indirect branches, switches, switches with dynamic tables, and
conditional jumps) are mixed with some randomization, and when each of those
four is deployed in isolation. In each bar, the colored segments in the stack
mark the number of different \emph{archives} from which the slices/sequences
factored together originate. The lowest segment corresponds to instructions that
are factored from within only one archive. The second to instructions that are
factored from within two archives etc. It is clear that a considerable fraction
of all instructions gets factored. It is also clear that the amounts of
instructions factored from within multiple archives clearly correlates with
the number of available archives and with the uniformity with which the
application is partitioned into archives.
\if \techreport1
Figures~\ref{fig:factored_ins_objects} and~\ref{fig:factored_ins_functions}
similarly show that many instructions are factored from within multiple object
files and multiple functions, at least for dispatchers that support slice sets
with more than two slices.  Also at those levels of granularity, the factoring
approach is hence capable of obfuscating component boundaries.
\else
Figure~\ref{fig:factored_ins_objects}
similarly shows that many instructions are factored from within multiple object
files, at least for dispatchers that support slice sets
with more than two slices.  Also at that level of granularity, and hence also at
the still lower levels of individual functions and code contexts, the factoring
approach is hence capable of obfuscating component boundaries.
\fi

In the context of dynamic attacks such as generic deobfuscation that focus on
covered instructions with quasi-invariant behavior, it is also useful to know
how many of the factored instructions were originally covered (i.e., executed)
in one or more contexts. To that extent,
Figure~\ref{fig:dispatchers_slices_cov_exec} shows the distributions of the
factored instructions in terms of the number of the covered slices/sequences
from which they were factored.
Each segment marks the fraction of all
instructions that got factored in a set of slices/sequences, where the number of slices/sequences covered in the original program is indicated by the color of the segment.
This means that the lowest segment corresponds to the instructions that got factored in a set of which no slice/sequence is covered in the original program. The next segment to instructions that got factored in a set in which one slice/sequence is covered, etc.
The observed distributions are in line with the
data in Figure~\ref{fig:archives}: When few instructions are covered in the
first place, even fewer get factored from within one or more covered contexts.

Figure~\ref{fig:dispatchers_slices_exec_cov_exec} shows similar data, but rather
than considering all instructions, it only considers the covered instructions in
the protected program, i.e., the instructions targeted by dynamic attacks. From
the overall height of the bars, it is obvious that significant parts of the
covered instructions are factored. Moreover, the vast majority of the factored
covered instructions are factored from multiple covered contexts. This implies
that in the protected program, most of the factored fragments are executed on
data from two contexts. This implies that the injected dispatchers for the vast
majority of the covered and factored slices do not display quasi-invariant
behavior.

\if \techreport 0
Figure~\ref{fig:heatmap_SLM} presents a further dissection of the factoring
applicability, for the SLM benchmark. The heatmap displays the relations between the number of factored
fragments in the protected program (color), the sizes of the factored
fragments (first x-axis), the number of archives from which they are factored
(second x-axis), and the number of archives in which the factored fragments were
covered (y-axis).
\else
Figures~\ref{fig:heatmap_SLM}--\ref{fig:heatmap_454} presents a further dissection of the factoring
applicability. The heatmaps display the relations between the number of factored
fragments in the protected program (color), the sizes of the factored
fragments (first x-axis), the number of archives from which they are factored
(second x-axis), and the number of archives in which the factored fragments were
covered (y-axis).
\fi

As to be expected, the number of factored shorter fragments is
significantly larger than that for longer ones. Second, the longer factored
fragments all come from within a single archive. From further examination, we
actually observed that the exceptionally long fragments originate from
loop-unrolled code.

It is also clear that the most interesting factorizations, i.e., those from
multiple contexts executed in multiple archives, are relatively rare, and
involve only rather short sequences. This clearly indicates that there are
practical limitations to the level of protection that our techniques can
provide. Still, the colored cells in the upper right corner shows that even if
attackers completely neglect all uncovered control flow edges and code
fragments, some dispatchers that are executed in more than one direction will
keep hampering their reconstruction of the original program. In future work, we
will investigate techniques to generate more and larger factorizable fragments
by transforming code fragments rather than simply selecting existing ones like
we do now.

\if\techreport1
\begin{figure*}[t]
   \centering
   \resizebox{\linewidth}{!}{\input{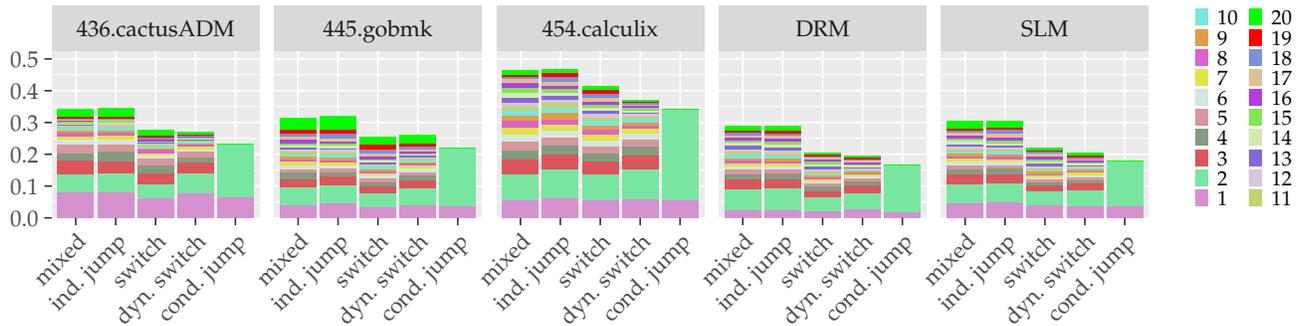}}
   \caption{Fraction of all instructions that get factored from within the indicated number of functions}
   \label{fig:factored_ins_functions}
 \end{figure*}
\fi
\begin{figure*}[t]
  \centering
  \resizebox{\linewidth}{!}{\input{dispatchers_slices_cov_exec5}}
  \vspace{-1.2cm}
  \caption{Fraction of all instructions that get factored from within the indicated number of covered slices/sequences (irrespective of additional uncovered slices/sequences)}
  \label{fig:dispatchers_slices_cov_exec}
\end{figure*}
\begin{figure*}[t]
  \centering
  \resizebox{\linewidth}{!}{\input{dispatchers_slices_exec_cov_exec5}}
  \vspace{-1.2cm}
  \caption{Fraction of covered instructions that get factored from within the indicated number of covered slices/sequences (irrespective of additional uncovered slices/sequences)}
  \label{fig:dispatchers_slices_exec_cov_exec}
\end{figure*}

\begin{figure*}[t]
  \centering
  \subfloat{\input{heatmap_SLM_archives}}\\
  \subfloat{\input{heatmap_SLM_archives_legend}}
  \caption{Heatmap dissecting the applicability of factoring on the SLM benchmark, showing the number of fragments (color) of size (minor, top X-axis) in sets covering (major, bottom X-axis) archives versus the number of covered archives (Y-axis) in that set.}
  \label{fig:heatmap_SLM}
\end{figure*}

\if\techreport1
\begin{figure*}[t]
  \centering
  \subfloat{\input{heatmap_DRM_archives}}\\
  \subfloat{\input{heatmap_DRM_archives_legend}}
  \caption{Heatmap dissecting the applicability of factoring on the DRM benchmark, showing the number of fragments (color) of size (minor, top X-axis) in sets covering (major, bottom X-axis) archives versus the number of covered archives (Y-axis) in that set.}
  \label{fig:heatmap_DRM}
\end{figure*}
\begin{figure*}[t]
  \centering
  \subfloat{\input{heatmap_436_archives}}\\
  \subfloat{\input{heatmap_436_archives_legend}}
  \caption{Heatmap dissecting the applicability of factoring on the 436.cactusADM benchmark, showing the number of fragments (color) of size (minor, top X-axis) in sets covering (major, bottom X-axis) archives versus the number of covered archives (Y-axis) in that set.}
  \label{fig:heatmap_436}
\end{figure*}
\begin{figure*}[t]
  \centering
  \subfloat{\input{heatmap_445_archives}}\\
  \subfloat{\input{heatmap_445_archives_legend}}
  \caption{Heatmap dissecting the applicability of factoring on the 445.gobmk benchmark, showing the number of fragments (color) of size (minor, top X-axis) in sets covering (major, bottom X-axis) archives versus the number of covered archives (Y-axis) in that set.}
  \label{fig:heatmap_445}
\end{figure*}
\begin{figure*}[t]
  \centering
  \subfloat{\input{heatmap_454_archives}}\\
  \subfloat{\input{heatmap_454_archives_legend}}
  \caption{Heatmap dissecting the applicability of factoring on the 454.calculix benchmark, showing the number of fragments (color) of size (minor, top X-axis) in sets covering (major, bottom X-axis) archives versus the number of covered archives (Y-axis) in that set.}
  \label{fig:heatmap_454}
\end{figure*}
\fi

\subsection{Potency}
\label{sec:potency}

To estimate the potency of the presented obfuscations, i.e., the extent to which they confuse human attackers, we performed two measurements on binaries protected with our Diablo-based tool. For these experiments, we configured the tool as follows. For factoring, we enable all dispatchers (with switch tables filled with 30\% fake entries) and only factor fragments of at least 2 instructions but with no other restrictions, e.g., regarding hotness. We insert opaque predicates and corresponding conditional branches into 20\% of randomly selected basic blocks, making the fall-through edge the fake edge whenever possible. After code layout randomization, we redirect fake edges throughout the binary to create cycles of four coupled obfuscations as discussed in Section~\ref{sec:resilience}.

\begin{figure*}[t]
  \centering
  \input{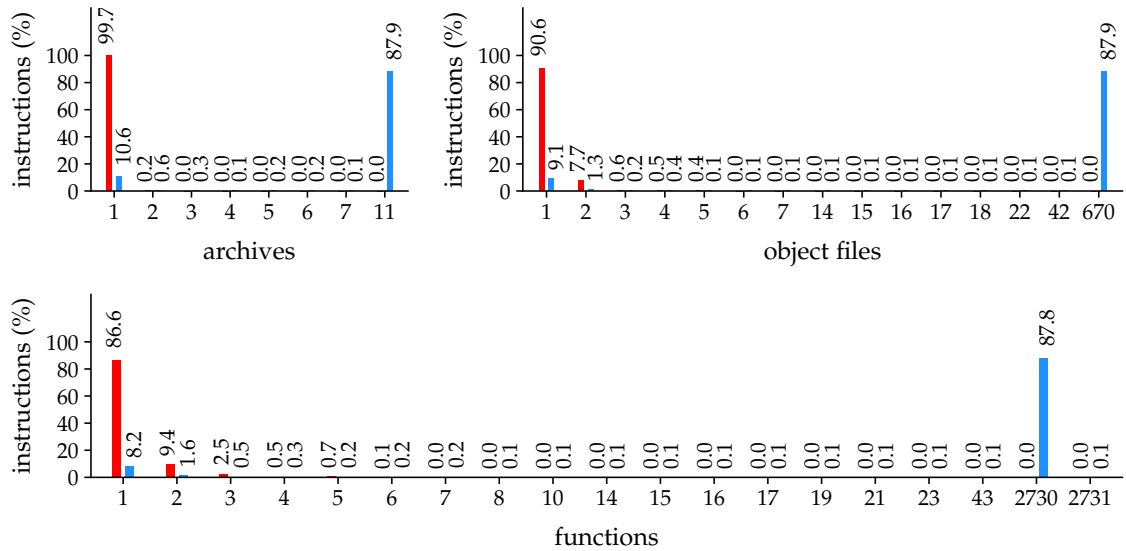}
  \caption{Instructions reachable from function entry points in different numbers of archives/object files/functions, \textcolor{cGUIFNR}{before (left bars)} and \textcolor{cGUIFPR}{after (right bars)} factoring (SLM).}
  \label{fig:reachability_SLM}
\end{figure*}

First, we measure the extent to which the code of different components has become interconnected by in\-tra\-pro\-ce\-du\-ral-looking edges. For each instruction, we count from how many function entry points those instructions are reachable through intraprocedural control flow idioms only (i.e., through direct branches, fall throughs, switches, and from call sites to their corresponding return addresses). We then count from how many archives, objects, and functions those entry points originate. This metric thus measures the number of different components to which an attacker or his tools can potentially assign each instruction, and from which he has to make a choice to reconstruct the CFGs correctly. For the SLM benchmark, Figure~\ref{fig:reachability_SLM} shows the results. For other benchmarks, the results are similar.
Before factoring, most code is reachable from a single function entry point, as one expects for code written in C. The few exceptions mainly originate from manually written and optimized assembly functions in the linked-in crypto library.
After factoring, the vast majority of the code is reachable from within a vast number of function entry points, that originate from a large number of different object files, and from all archives. The reason is that a large part of the code
ends up in one big intraprocedurally-single-connected component in the combined CFGs of the program. So at least in theory, our transformations succeed in obfuscating the boundaries between components at the three levels of granularity.

Secondly, we measure a practically oriented metric in the form of the amount of incorrect information that the popular reverse engineering tool IDA Pro (v.6.8) presents to the user due to the obfuscations. Concretely, we measure the fraction of fake CFG edges that IDA Pro stores in its database and/or shows in its GUI, as well as the fraction of true CFG edges that IDA Pro does not store and/or show. The former are FP rates, the latter are FN rates.

It should be noted that IDA Pro is not designed for reverse engineering obfuscated binaries. In particular, it is not designed to handle basic blocks that are reachable via intraprocedural control flow idioms from multiple function entry points. It simply assigns basic blocks to functions based on the order in which the recursive descent assembler visits them, not based on heuristics that take into account the effects of our transformations. IDA Pro can easily be augmented by an attacker, however, as it exports the constructed CFGs in a database that attacker scripts can manipulate. In other words, a skilled attacker can easily override and extend the disassembler and function reconstruction heuristics of IDA Pro.

To mimic skilled attackers, we experimented with various algorithms to maximize the amount of code in a binary that IDA Pro actually disassembles, as well as with various heuristics that repartition the disassembled code fragments (i.e., basic blocks) into functions such that the reconstructed functions better ressemble the actual functions. We observed that many similar algorithms yielded very similar results, so the exact implementation details do not matter, as long as they incorporate three main ideas. First, one should try to put all identified code in functions, even if that code was not identified as being reachable by the original IDA Pro. This is the case, e.g., for code fragments that are only reachable through switch tables that IDA Pro cannot analyse precisely. Secondly, for such code fragments as well as for code fragments that the original IDA Pro already did put into functions, one should determine the function to which the fragment is most connected through incoming and outgoing intraprocedurally looking CFG edges in the IDA Pro database, and then put the fragment in that function. Finally, for determining the function to which a fragment is most connected and in which it hence belongs, one should assign different weights to different types of edges. Most importantly, edges originating from indirect control flow transfers such as those used to implement switches should have lower weights that other direct control flow edges. In addition, if the attacker knows somehow that the fake edges in opaque predicates are mostly fall-through edges or mostly taken edges, he can assign different weights to those types of conditional branch edges as well. Our code implementing these heuristics is available online at \url{https://github.com/csl-ugent/oisp}.

\if \techreport1
Table~\ref{resultsSLM} presents the measurement data for the SLM benchmark, with Table~\ref{fig:SLMflags} listing the compilation and link flags used to generate the program, and Tables~\ref{fig:potency_unprotected} to~\ref{fig:potency_repartition} listing the results. Similar measurement data are available in Table~\ref{results436} for 436.cactusADM, Table~\ref{results445} for 445.gobmk, Table~\ref{results454} for 454.calculix and Table~\ref{resultsDRM} for DRM. As the conclusions are the same for all benchmarks, we only discuss the results for the SLM benchmark quantitatively here.
\else
Tables~\ref{fig:potency_unprotected} to ~\ref{fig:potency_repartition} present the results for the SLM benchmark. Similar results for the other bechmarks are available in a technical report~\cite{TR}.
\fi
The top part of each table shows the aforementioned FP and FN rates of correctly or incorrectly handled CFG edges. The bottom parts additionally present the total amounts of edges and instructions in the binaries to ease the interpretation of the false rates, where we also mention how many edges are drawn in the GUI. The overall counts and corresponding false rates are further refined into 6 partially overlapping categories $xy$, with $x$ being either I (Inter) or i (intra), and $y$ being A (archive), O (object file), or F (function). The category IA, for example, is that of edges from a block originating from one archive to a fragment originating from another archive, i.e., inter-archive, while category iO is that of edges between two blocks originating from the same object file. Furthermore, we present separate numbers for the edges that IDA Pro stores in its database because it has detected them in the code, and the ones it shows in the GUI because it considers them to be intraprocedural edges, meaning that it has correctly or incorrectly put the source and sink nodes of the edges in the same functions.

Table~\ref{fig:potency_unprotected} shows that for the unprotected program, IDA Pro does a pretty good job in detecting the true edges. There are no fake edges of course, and most code is put into functions. Exceptions are rare, and mostly related to manually written and optimized assembly functions in the linked-in crypto library that feature interprocedural jumps.

Table~\ref{fig:potency_vanilla} shows that IDA Pro out-of-the-box performs poorly on a protected program. In the GUI, it draws about 74\% of the fake edges (75--76\% for other benchmarks), of which more than half connect blocks from different archives. Furthermore, the GUI does not draw 56\% of the true edges (53--56\% for other benchmarks). As a result of the obfuscation, IDA Pro also gave up on about 28\% of the identified instructions (23--31\% for the other benchmarks), and simply did not put that code in any function. Obviously this also contributes to the FN rates.

Notice how these total numbers are comparable for different benchmarks, despite their different constitution. This is of course due to the fact that the totals do not depend on the number of archives or object files making up the programs. For the intra- and interarchive FPs, the rates vary more from one benchmark to another, but they are still comparable. For example, the GUI IA FPR with IDA Pro out-of-the-box ranges from 39\% to 55\%.
\if \techreport0
All numbers are available in the technical report~\cite{TR}.
\fi
This relatively small variation implies that the obtained potency ports rather well from one benchmark to another, which is of course beneficial for users of tools that implement the obfuscations, as it will limit the need to retune the tool configuration for each benchmark.

At first sight, it might seem strange that there are also intra-function GUI FPs, since we never purposely inject fake intra-function edges. Those FPs are a side-effect, however, as they correspond to the never executed fall-through paths of injected switch dispatchers, which are intra-function in our prototype.

Table~\ref{fig:potency_repartition} shows that an attacker-improved IDA Pro puts almost all code into functions. The FP rates go up as a result, and the FN rates drop significantly. Different versions of the repartitioning algorithm never got significantly better results than the ones reported here. Without more advanced data flow analysis or other attacks to identify fake edges, those edges simply confused the disassembler's code partitioning strategies. The proposed protections thus display a significant amount of practically relevant potency.

The above results and in particular the FNs are to some extent inherent to IDA Pro, which can put each basic block in only one function. For the example of Figure~\ref{fig:fragments_factored}, at least one of the incoming edges of block 3a and one of the outgoing edges of block 3b inherently become FNs. So additionally, we measure how many (source, sink) pairs of code fragments that were split apart by factorization (e.g., pairs (1a,1b) and (2a,2b) in Figure~\ref{fig:fragments_split}) are correctly put in the same function by IDA Pro. The results are presented in the bottom left parts of the tables. Most importantly, the results in Table~\ref{fig:potency_repartition} indicate that even with the repartitioning heuristics, the vast majority (85\%, 85--88\% for the other benchmarks) of related block pairs are not put in the same function. There are two reasons: First, when the factoring is applied as frequently as we applied it, many non-factored fragments end up in between two factored fragments, and thus are no longer connected directly to any non-factored fragment. Secondly, even if we drop the frequency of factoring to a low number (such as 1\% of all factorizable cases), the number only drops to about 82\%. It remains that high because of the negative impact of the opaque predicate insertion on IDA Pro's performance. When no opaque predicates are inserted at all, and very little factoring is performed, the number still does not drop below 49\% (51--59\% for the other benchmarks). The reason is that at about half of the points where factorization can be applied, the points before and after the factorized fragments are only connected via one direct control flow path, which then gets interrupted because of the factoring.
\if\techreport1
A detailed analysis on this can be found at the end of Section~\ref{sec:sensitivity}.
\else
More detailed results are available in our technical report~\cite{TR}.
\fi

We can conclude that unless IDA Pro gets the capability of putting blocks in more than one function, which by the design of its APIs seems like a rather fundamental and hence hard to change underlying principle of its implementation, the proposed factoring obfuscation has a strong potency.

\if\techreport1
\begin{table*}
  \begin{minipage}[t]{\textwidth}
    \centering
    \scriptsize
    \subfloat[Compile and link flags for SLM\label{fig:SLMflags}]{\parbox{\linewidth}{
      \centering
      \begin{tabular}{|l|l|}
        \hline
        \textbf{Compiler} & \texttt{-DLTC\_NO\_ASM -DLTC\_SOURCE -DLTM\_DESC -DPATH\_MAX=2048 -DUSE\_LTM -D\_\_32BIT\_\_}\\
        & \texttt{-Os -fPIC -fno-aggressive-loop-optimizations -fno-stack-protector -fno-strict-aliasing}\\
        & \texttt{-fomit-frame-pointer -g -marm -mcpu=cortex-a8 -msoft-float -std=c99}\\
        \hline
        \textbf{Linker} & \texttt{-Wl,--fix-cortex-a8 -Wl,--hash-style=sysv -Wl,--no-demangle -Wl,--no-merge-exidx-entries}\\
        & \texttt{-Wl,--no-undefined -lc -shared}\\
        \hline
      \end{tabular}}}
    \vspace{0.25em}
    \subfloat[Potency metrics for SLM without the protections proposed in this \papertechreport{}.\label{fig:potency_unprotected}]{\parbox{\linewidth}{
      \centering
  \scriptsize
  \begin{tabular}{|l|c|ccc|ccc|c|ccc|ccc|}
    \cline{2-15}
    \multicolumn{1}{l|}{} & \multicolumn{7}{|c|}{\textbf{FP/FN CFG edges drawn in GUI}} & \multicolumn{7}{|c|}{\textbf{FP/FN CFG edges stored in database}}\\
    \cline{2-15}
    \multicolumn{1}{l|}{} & \textbf{Total} & \textbf{IA} & \textbf{IO} & \textbf{IF} & \textbf{iA} & \textbf{iO} & \textbf{iF} & \textbf{Total} & \textbf{IA} & \textbf{IO} & \textbf{IF} & \textbf{iA} & \textbf{iO} & \textbf{iF}\\
    \cline{1-15}
    \textbf{\# FP} & 0 & 0 & 0 & 0 & 0 & 0 & 0 & 0 & 0 & 0 & 0 & 0 & 0 & 0\\
    \textbf{\perc{FPR}} & \perc{{0\%}} & \perc{{0\%}} & \perc{{0\%}} & \perc{{0\%}} & \perc{{0\%}} & \perc{{0\%}} & \perc{{0\%}} & \perc{{0\%}} & \perc{{0\%}} & \perc{{0\%}} & \perc{{0\%}} & \perc{{0\%}} & \perc{{0\%}} & \perc{{0\%}}\\
    \cline{1-15}
    \textbf{\# FN} & 1.8k & 24 & 618 & 753 & 1.8k & 1.2k & 1.1k & 916 & 0 & 45 & 48 & 916 & 871 & 868\\
    \textbf{\perc{FNR}} & \perc{{3\%}} & \perc{{0\%}} & \perc{{1\%}} & \perc{{1\%}} & \perc{{3\%}} & \perc{{2\%}} & \perc{{2\%}} & \perc{{1\%}} & \perc{{0\%}} & \perc{{0\%}} & \perc{{0\%}} & \perc{{1\%}} & \perc{{1\%}} & \perc{{1\%}}\\
    \cline{1-15}
  \end{tabular}

  \vspace{1em}
  \begin{tabular}{|c|cc|}
    \hline
    \multicolumn{3}{|c|}{\textbf{Pairs of fragments split by factorization}}\\
    \hline
    Total & Wrong & Correct\\
    0 & 0 (\perc{{0\%}}) & 0 (\perc{{0\%}})\\
    \hline
  \end{tabular}
  \begin{tabular}{|c|cc|c|}
    \hline
    \multicolumn{4}{|c|}{\textbf{CFG edges}}\\
    \hline
    Total & True & Fake & Drawn in GUI\\
    67.2k & 67.2k (\perc{{100\%}}) & 0 (\perc{{0\%}}) & 65.3k (\perc{{97\%}})\\
    \hline
  \end{tabular}
  \begin{tabular}{|c|c|}
    \hline
    \multicolumn{2}{|c|}{\textbf{Instructions}}\\
    \hline
    Total & Functionless\\
    281.8k & 4.0k (\perc{{1\%}})\\
    \hline
  \end{tabular}}}
    \vspace{0.25em}
    \subfloat[Potency metrics for a fully protected SLM with IDA Pro out-of-the-box.\label{fig:potency_vanilla}]{\parbox{\linewidth}{
      \centering
  \scriptsize
  \begin{tabular}{|l|c|ccc|ccc|c|ccc|ccc|}
    \cline{2-15}
    \multicolumn{1}{l|}{} & \multicolumn{7}{|c|}{\textbf{FP/FN CFG edges drawn in GUI}} & \multicolumn{7}{|c|}{\textbf{FP/FN CFG edges stored in database}}\\
    \cline{2-15}
    \multicolumn{1}{l|}{} & \textbf{Total} & \textbf{IA} & \textbf{IO} & \textbf{IF} & \textbf{iA} & \textbf{iO} & \textbf{iF} & \textbf{Total} & \textbf{IA} & \textbf{IO} & \textbf{IF} & \textbf{iA} & \textbf{iO} & \textbf{iF}\\
    \cline{1-15}
    \textbf{\# FP} & 16.5k & 9.6k & 12.5k & 12.6k & 6.9k & 4.0k & 3.9k & 20.0k & 11.8k & 16.0k & 16.0k & 8.2k & 4.0k & 3.9k\\
    \textbf{\perc{FPR}} & \perc{{74\%}} & \perc{{43\%}} & \perc{{57\%}} & \perc{{57\%}} & \perc{{31\%}} & \perc{{18\%}} & \perc{{18\%}} & \perc{{90\%}} & \perc{{53\%}} & \perc{{72\%}} & \perc{{72\%}} & \perc{{37\%}} & \perc{{18\%}} & \perc{{18\%}}\\
    \cline{1-15}
    \textbf{\# FN} & 101.4k & 24 & 622 & 760 & 101.3k & 100.7k & 100.6k & 63.2k & 9 & 131 & 165 & 63.2k & 63.1k & 63.0k\\
    \textbf{\perc{FNR}} & \perc{{56\%}} & \perc{{0\%}} & \perc{{0\%}} & \perc{{0\%}} & \perc{{56\%}} & \perc{{55\%}} & \perc{{55\%}} & \perc{{35\%}} & \perc{{0\%}} & \perc{{0\%}} & \perc{{0\%}} & \perc{{35\%}} & \perc{{35\%}} & \perc{{35\%}}\\
    \cline{1-15}
  \end{tabular}

  \vspace{1em}
  \begin{tabular}{|c|cc|}
    \hline
    \multicolumn{3}{|c|}{\textbf{Pairs of fragments split by factorization}}\\
    \hline
    Total & Wrong & Correct\\
    28.4k & 26.6k (\perc{{94\%}}) & 1.7k (\perc{{6\%}})\\
    \hline
  \end{tabular}
  \begin{tabular}{|c|cc|c|}
    \hline
    \multicolumn{4}{|c|}{\textbf{CFG edges}}\\
    \hline
    Total & True & Fake & Drawn in GUI\\
    204.4k & 182.2k (\perc{{89\%}}) & 22.2k (\perc{{11\%}}) & 97.4k (\perc{{48\%}})\\
    \hline
  \end{tabular}
  \begin{tabular}{|c|c|}
    \hline
    \multicolumn{2}{|c|}{\textbf{Instructions}}\\
    \hline
    Total & Functionless\\
    772.3k & 213.6k (\perc{{28\%}})\\
    \hline
  \end{tabular}}}
    \vspace{0.25em}
    \subfloat[Potency metrics for a fully protected SLM with attacker-improved IDA Pro.\label{fig:potency_repartition}]{\parbox{\linewidth}{
      \centering
  \scriptsize
  \begin{tabular}{|l|c|ccc|ccc|c|ccc|ccc|}
    \cline{2-15}
    \multicolumn{1}{l|}{} & \multicolumn{7}{|c|}{\textbf{FP/FN CFG edges drawn in GUI}} & \multicolumn{7}{|c|}{\textbf{FP/FN CFG edges stored in database}}\\
    \cline{2-15}
    \multicolumn{1}{l|}{} & \textbf{Total} & \textbf{IA} & \textbf{IO} & \textbf{IF} & \textbf{iA} & \textbf{iO} & \textbf{iF} & \textbf{Total} & \textbf{IA} & \textbf{IO} & \textbf{IF} & \textbf{iA} & \textbf{iO} & \textbf{iF}\\
    \cline{1-15}
    \textbf{\# FP} & 17.1k & 10.0k & 13.0k & 13.0k & 7.1k & 4.1k & 4.0k & 21.2k & 12.6k & 17.0k & 17.1k & 8.6k & 4.2k & 4.1k\\
    \textbf{\perc{FPR}} & \perc{{77\%}} & \perc{{45\%}} & \perc{{59\%}} & \perc{{59\%}} & \perc{{32\%}} & \perc{{18\%}} & \perc{{18\%}} & \perc{{96\%}} & \perc{{57\%}} & \perc{{77\%}} & \perc{{77\%}} & \perc{{39\%}} & \perc{{19\%}} & \perc{{18\%}}\\
    \cline{1-15}
    \textbf{\# FN} & 74.5k & 16 & 492 & 588 & 74.5k & 74.0k & 73.9k & 27.5k & 0 & 17 & 20 & 27.5k & 27.5k & 27.5k\\
    \textbf{\perc{FNR}} & \perc{{41\%}} & \perc{{0\%}} & \perc{{0\%}} & \perc{{0\%}} & \perc{{41\%}} & \perc{{41\%}} & \perc{{41\%}} & \perc{{15\%}} & \perc{{0\%}} & \perc{{0\%}} & \perc{{0\%}} & \perc{{15\%}} & \perc{{15\%}} & \perc{{15\%}}\\
    \cline{1-15}
  \end{tabular}

  \vspace{1em}
  \begin{tabular}{|c|cc|}
    \hline
    \multicolumn{3}{|c|}{\textbf{Pairs of fragments split by factorization}}\\
    \hline
    Total & Wrong & Correct\\
    28.4k & 24.1k (\perc{{85\%}}) & 4.3k (\perc{{15\%}})\\
    \hline
  \end{tabular}
  \begin{tabular}{|c|cc|c|}
    \hline
    \multicolumn{4}{|c|}{\textbf{CFG edges}}\\
    \hline
    Total & True & Fake & Drawn in GUI\\
    204.5k & 182.3k (\perc{{89\%}}) & 22.2k (\perc{{11\%}}) & 124.9k (\perc{{61\%}})\\
    \hline
  \end{tabular}
  \begin{tabular}{|c|c|}
    \hline
    \multicolumn{2}{|c|}{\textbf{Instructions}}\\
    \hline
    Total & Functionless\\
    772.4k & 122 (\perc{{0\%}})\\
    \hline
  \end{tabular}}}
    \vspace{0.25em}
    \subfloat[Metrics for a fully protected SLM, after detection and removal of observable opaque predicates (soundish DB attack).\label{fig:resilienceSound}]{\parbox{\linewidth}{
      \centering
  \scriptsize
  \begin{tabular}{|l|c|ccc|ccc|c|ccc|ccc|}
    \cline{2-15}
    \multicolumn{1}{l|}{} & \multicolumn{7}{|c|}{\textbf{FP/FN CFG edges drawn in GUI}} & \multicolumn{7}{|c|}{\textbf{FP/FN CFG edges stored in database}}\\
    \cline{2-15}
    \multicolumn{1}{l|}{} & \textbf{Total} & \textbf{IA} & \textbf{IO} & \textbf{IF} & \textbf{iA} & \textbf{iO} & \textbf{iF} & \textbf{Total} & \textbf{IA} & \textbf{IO} & \textbf{IF} & \textbf{iA} & \textbf{iO} & \textbf{iF}\\
    \cline{1-15}
    \textbf{\# FP} & 16.4k & 9.5k & 12.3k & 12.3k & 6.9k & 4.1k & 4.0k & 21.1k & 12.6k & 17.0k & 17.0k & 8.5k & 4.2k & 4.1k\\
    \textbf{\perc{FPR}} & \perc{{74\%}} & \perc{{43\%}} & \perc{{55\%}} & \perc{{56\%}} & \perc{{31\%}} & \perc{{18\%}} & \perc{{18\%}} & \perc{{95\%}} & \perc{{57\%}} & \perc{{77\%}} & \perc{{77\%}} & \perc{{39\%}} & \perc{{19\%}} & \perc{{18\%}}\\
    \cline{1-15}
    \textbf{\# FN} & 73.4k & 13 & 459 & 544 & 73.4k & 73.0k & 72.9k & 27.5k & 0 & 17 & 20 & 27.5k & 27.5k & 27.5k\\
    \textbf{\perc{FNR}} & \perc{{40\%}} & \perc{{0\%}} & \perc{{0\%}} & \perc{{0\%}} & \perc{{40\%}} & \perc{{40\%}} & \perc{{40\%}} & \perc{{15\%}} & \perc{{0\%}} & \perc{{0\%}} & \perc{{0\%}} & \perc{{15\%}} & \perc{{15\%}} & \perc{{15\%}}\\
    \cline{1-15}
  \end{tabular}

  \vspace{1em}
  \begin{tabular}{|c|cc|}
    \hline
    \multicolumn{3}{|c|}{\textbf{Pairs of fragments split by factorization}}\\
    \hline
    Total & Wrong & Correct\\
    28.4k & 24.0k (\perc{{85\%}}) & 4.3k (\perc{{15\%}})\\
    \hline
  \end{tabular}
  \begin{tabular}{|c|cc|c|}
    \hline
    \multicolumn{4}{|c|}{\textbf{CFG edges}}\\
    \hline
    Total & True & Fake & Drawn in GUI\\
    204.5k & 182.3k (\perc{{89\%}}) & 22.2k (\perc{{11\%}}) & 125.2k (\perc{{61\%}})\\
    \hline
  \end{tabular}
  \begin{tabular}{|c|c|}
    \hline
    \multicolumn{2}{|c|}{\textbf{Instructions}}\\
    \hline
    Total & Functionless\\
    772.4k & 122 (\perc{{0\%}})\\
    \hline
  \end{tabular}
  \begin{tabular}{|c|c|}
    \hline
    \multicolumn{2}{|c|}{\textbf{Opaque predicates}}\\
    \hline
    Total & Resolved\\
    13.3k & 29 (\perc{{0\%}})\\
    \hline
  \end{tabular}}}
    \vspace{0.25em}
    \subfloat[Metrics for a fully protected SLM, after detection and removal of observable opaque predicates (unsound GUI attack).\label{fig:resilience}]{\parbox{\linewidth}{
      \centering
  \scriptsize
  \begin{tabular}{|l|c|ccc|ccc|c|ccc|ccc|}
    \cline{2-15}
    \multicolumn{1}{l|}{} & \multicolumn{7}{|c|}{\textbf{FP/FN CFG edges drawn in GUI}} & \multicolumn{7}{|c|}{\textbf{FP/FN CFG edges stored in database}}\\
    \cline{2-15}
    \multicolumn{1}{l|}{} & \textbf{Total} & \textbf{IA} & \textbf{IO} & \textbf{IF} & \textbf{iA} & \textbf{iO} & \textbf{iF} & \textbf{Total} & \textbf{IA} & \textbf{IO} & \textbf{IF} & \textbf{iA} & \textbf{iO} & \textbf{iF}\\
    \cline{1-15}
    \textbf{\# FP} & 14.9k & 8.3k & 10.9k & 11.0k & 6.6k & 4.0k & 4.0k & 18.2k & 10.3k & 14.1k & 14.2k & 7.9k & 4.1k & 4.0k\\
    \textbf{\perc{FPR}} & \perc{{67\%}} & \perc{{38\%}} & \perc{{49\%}} & \perc{{49\%}} & \perc{{30\%}} & \perc{{18\%}} & \perc{{18\%}} & \perc{{82\%}} & \perc{{46\%}} & \perc{{64\%}} & \perc{{64\%}} & \perc{{36\%}} & \perc{{18\%}} & \perc{{18\%}}\\
    \cline{1-15}
    \textbf{\# FN} & 73.0k & 13 & 448 & 526 & 73.0k & 72.6k & 72.5k & 27.5k & 0 & 17 & 20 & 27.5k & 27.5k & 27.5k\\
    \textbf{\perc{FNR}} & \perc{{40\%}} & \perc{{0\%}} & \perc{{0\%}} & \perc{{0\%}} & \perc{{40\%}} & \perc{{40\%}} & \perc{{40\%}} & \perc{{15\%}} & \perc{{0\%}} & \perc{{0\%}} & \perc{{0\%}} & \perc{{15\%}} & \perc{{15\%}} & \perc{{15\%}}\\
    \cline{1-15}
  \end{tabular}

  \vspace{1em}
  \begin{tabular}{|c|cc|}
    \hline
    \multicolumn{3}{|c|}{\textbf{Pairs of fragments split by factorization}}\\
    \hline
    Total & Wrong & Correct\\
    28.4k & 24.0k (\perc{{85\%}}) & 4.4k (\perc{{15\%}})\\
    \hline
  \end{tabular}
  \begin{tabular}{|c|cc|c|}
    \hline
    \multicolumn{4}{|c|}{\textbf{CFG edges}}\\
    \hline
    Total & True & Fake & Drawn in GUI\\
    204.5k & 182.3k (\perc{{89\%}}) & 22.2k (\perc{{11\%}}) & 124.2k (\perc{{61\%}})\\
    \hline
  \end{tabular}
  \begin{tabular}{|c|c|}
    \hline
    \multicolumn{2}{|c|}{\textbf{Instructions}}\\
    \hline
    Total & Functionless\\
    772.4k & 122 (\perc{{0\%}})\\
    \hline
  \end{tabular}
  \begin{tabular}{|c|c|}
    \hline
    \multicolumn{2}{|c|}{\textbf{Opaque predicates}}\\
    \hline
    Total & Resolved\\
    13.3k & 3.0k (\perc{{22\%}})\\
    \hline
  \end{tabular}}}
  \end{minipage}
  \caption{Measurement data for SLM}
  \label{resultsSLM}
\end{table*}

\begin{table*}
  \begin{minipage}[t]{\textwidth}
    \centering
    \scriptsize
    \subfloat[Compile and link flags for 436.cactusADM\label{fig:436flags}]{\parbox{\linewidth}{
      \centering
      \begin{tabular}{|l|l|}
        \hline
        \textbf{Compiler} & \texttt{-DCCODE -DNDEBUG -DSPEC\_CPU}\\
        & \texttt{-O2 -fno-aggressive-loop-optimizations -g -marm -mcpu=cortex-a8 -std=c99}\\
        \hline
        \textbf{Linker} & \texttt{-Wl,--no-demangle -Wl,--hash-style=sysv -Wl,--no-merge-exidx-entries -Bdynamic}\\
        \hline
      \end{tabular}}}
    \vspace{0.25em}
    \subfloat[Potency metrics for 436.cactusADM without the protections proposed in this \papertechreport{}.\label{fig:436potency_unprotected}]{\parbox{\linewidth}{
      \centering
  \scriptsize
  \begin{tabular}{|l|c|ccc|ccc|c|ccc|ccc|}
    \cline{2-15}
    \multicolumn{1}{l|}{} & \multicolumn{7}{|c|}{\textbf{FP/FN CFG edges drawn in GUI}} & \multicolumn{7}{|c|}{\textbf{FP/FN CFG edges stored in database}}\\
    \cline{2-15}
    \multicolumn{1}{l|}{} & \textbf{Total} & \textbf{IA} & \textbf{IO} & \textbf{IF} & \textbf{iA} & \textbf{iO} & \textbf{iF} & \textbf{Total} & \textbf{IA} & \textbf{IO} & \textbf{IF} & \textbf{iA} & \textbf{iO} & \textbf{iF}\\
    \cline{1-15}
    \textbf{\# FP} & 0 & 0 & 0 & 0 & 0 & 0 & 0 & 0 & 0 & 0 & 0 & 0 & 0 & 0\\
    \textbf{\perc{FPR}} & \perc{{0\%}} & \perc{{0\%}} & \perc{{0\%}} & \perc{{0\%}} & \perc{{0\%}} & \perc{{0\%}} & \perc{{0\%}} & \perc{{0\%}} & \perc{{0\%}} & \perc{{0\%}} & \perc{{0\%}} & \perc{{0\%}} & \perc{{0\%}} & \perc{{0\%}}\\
    \cline{1-15}
    \textbf{\# FN} & 953 & 11 & 26 & 43 & 942 & 927 & 910 & 504 & 2 & 2 & 2 & 502 & 502 & 502\\
    \textbf{\perc{FNR}} & \perc{{6\%}} & \perc{{0\%}} & \perc{{0\%}} & \perc{{0\%}} & \perc{{6\%}} & \perc{{6\%}} & \perc{{5\%}} & \perc{{3\%}} & \perc{{0\%}} & \perc{{0\%}} & \perc{{0\%}} & \perc{{3\%}} & \perc{{3\%}} & \perc{{3\%}}\\
    \cline{1-15}
  \end{tabular}

  \vspace{1em}
  \begin{tabular}{|c|cc|}
    \hline
    \multicolumn{3}{|c|}{\textbf{Pairs of fragments split by factorization}}\\
    \hline
    Total & Wrong & Correct\\
    0 & 0 (\perc{{0\%}}) & 0 (\perc{{0\%}})\\
    \hline
  \end{tabular}
  \begin{tabular}{|c|cc|c|}
    \hline
    \multicolumn{4}{|c|}{\textbf{CFG edges}}\\
    \hline
    Total & True & Fake & Drawn in GUI\\
    16.7k & 16.7k (\perc{{100\%}}) & 0 (\perc{{0\%}}) & 15.8k (\perc{{94\%}})\\
    \hline
  \end{tabular}
  \begin{tabular}{|c|c|}
    \hline
    \multicolumn{2}{|c|}{\textbf{Instructions}}\\
    \hline
    Total & Functionless\\
    98.2k & 2.5k (\perc{{2\%}})\\
    \hline
  \end{tabular}}}
    \vspace{0.25em}
    \subfloat[Potency metrics for a fully protected 436.cactusADM with IDA Pro out-of-the-box.\label{fig:436potency_vanilla}]{\parbox{\linewidth}{
      \centering
  \scriptsize
  \begin{tabular}{|l|c|ccc|ccc|c|ccc|ccc|}
    \cline{2-15}
    \multicolumn{1}{l|}{} & \multicolumn{7}{|c|}{\textbf{FP/FN CFG edges drawn in GUI}} & \multicolumn{7}{|c|}{\textbf{FP/FN CFG edges stored in database}}\\
    \cline{2-15}
    \multicolumn{1}{l|}{} & \textbf{Total} & \textbf{IA} & \textbf{IO} & \textbf{IF} & \textbf{iA} & \textbf{iO} & \textbf{iF} & \textbf{Total} & \textbf{IA} & \textbf{IO} & \textbf{IF} & \textbf{iA} & \textbf{iO} & \textbf{iF}\\
    \cline{1-15}
    \textbf{\# FP} & 7.6k & 5.5k & 5.8k & 5.8k & 2.1k & 1.9k & 1.8k & 9.2k & 6.9k & 7.3k & 7.3k & 2.3k & 1.9k & 1.9k\\
    \textbf{\perc{FPR}} & \perc{{76\%}} & \perc{{55\%}} & \perc{{58\%}} & \perc{{58\%}} & \perc{{21\%}} & \perc{{19\%}} & \perc{{18\%}} & \perc{{92\%}} & \perc{{69\%}} & \perc{{73\%}} & \perc{{73\%}} & \perc{{23\%}} & \perc{{19\%}} & \perc{{19\%}}\\
    \cline{1-15}
    \textbf{\# FN} & 36.1k & 9 & 22 & 39 & 36.0k & 36.0k & 36.0k & 20.1k & 0 & 4 & 4 & 20.1k & 20.1k & 20.1k\\
    \textbf{\perc{FNR}} & \perc{{53\%}} & \perc{{0\%}} & \perc{{0\%}} & \perc{{0\%}} & \perc{{53\%}} & \perc{{53\%}} & \perc{{53\%}} & \perc{{30\%}} & \perc{{0\%}} & \perc{{0\%}} & \perc{{0\%}} & \perc{{30\%}} & \perc{{30\%}} & \perc{{30\%}}\\
    \cline{1-15}
  \end{tabular}

  \vspace{1em}
  \begin{tabular}{|c|cc|}
    \hline
    \multicolumn{3}{|c|}{\textbf{Pairs of fragments split by factorization}}\\
    \hline
    Total & Wrong & Correct\\
    8.2k & 7.5k (\perc{{91\%}}) & 699 (\perc{{9\%}})\\
    \hline
  \end{tabular}
  \begin{tabular}{|c|cc|c|}
    \hline
    \multicolumn{4}{|c|}{\textbf{CFG edges}}\\
    \hline
    Total & True & Fake & Drawn in GUI\\
    77.6k & 67.6k (\perc{{87\%}}) & 10.0k (\perc{{13\%}}) & 39.2k (\perc{{50\%}})\\
    \hline
  \end{tabular}
  \begin{tabular}{|c|c|}
    \hline
    \multicolumn{2}{|c|}{\textbf{Instructions}}\\
    \hline
    Total & Functionless\\
    281.3k & 63.3k (\perc{{23\%}})\\
    \hline
  \end{tabular}}}
    \vspace{0.25em}
    \subfloat[Potency metrics for a fully protected 436.cactusADM with attacker-improved IDA Pro.\label{fig:436potency_repartition}]{\parbox{\linewidth}{
      \centering
  \scriptsize
  \begin{tabular}{|l|c|ccc|ccc|c|ccc|ccc|}
    \cline{2-15}
    \multicolumn{1}{l|}{} & \multicolumn{7}{|c|}{\textbf{FP/FN CFG edges drawn in GUI}} & \multicolumn{7}{|c|}{\textbf{FP/FN CFG edges stored in database}}\\
    \cline{2-15}
    \multicolumn{1}{l|}{} & \textbf{Total} & \textbf{IA} & \textbf{IO} & \textbf{IF} & \textbf{iA} & \textbf{iO} & \textbf{iF} & \textbf{Total} & \textbf{IA} & \textbf{IO} & \textbf{IF} & \textbf{iA} & \textbf{iO} & \textbf{iF}\\
    \cline{1-15}
    \textbf{\# FP} & 7.8k & 5.6k & 5.9k & 5.9k & 2.1k & 1.9k & 1.9k & 9.5k & 7.2k & 7.6k & 7.6k & 2.3k & 2.0k & 1.9k\\
    \textbf{\perc{FPR}} & \perc{{77\%}} & \perc{{56\%}} & \perc{{59\%}} & \perc{{59\%}} & \perc{{21\%}} & \perc{{19\%}} & \perc{{19\%}} & \perc{{95\%}} & \perc{{72\%}} & \perc{{75\%}} & \perc{{76\%}} & \perc{{23\%}} & \perc{{20\%}} & \perc{{19\%}}\\
    \cline{1-15}
    \textbf{\# FN} & 29.3k & 7 & 14 & 30 & 29.3k & 29.3k & 29.2k & 11.5k & 0 & 0 & 0 & 11.5k & 11.5k & 11.5k\\
    \textbf{\perc{FNR}} & \perc{{43\%}} & \perc{{0\%}} & \perc{{0\%}} & \perc{{0\%}} & \perc{{43\%}} & \perc{{43\%}} & \perc{{43\%}} & \perc{{17\%}} & \perc{{0\%}} & \perc{{0\%}} & \perc{{0\%}} & \perc{{17\%}} & \perc{{17\%}} & \perc{{17\%}}\\
    \cline{1-15}
  \end{tabular}

  \vspace{1em}
  \begin{tabular}{|c|cc|}
    \hline
    \multicolumn{3}{|c|}{\textbf{Pairs of fragments split by factorization}}\\
    \hline
    Total & Wrong & Correct\\
    8.2k & 7.0k (\perc{{85\%}}) & 1.2k (\perc{{15\%}})\\
    \hline
  \end{tabular}
  \begin{tabular}{|c|cc|c|}
    \hline
    \multicolumn{4}{|c|}{\textbf{CFG edges}}\\
    \hline
    Total & True & Fake & Drawn in GUI\\
    77.6k & 67.5k (\perc{{87\%}}) & 10.0k (\perc{{13\%}}) & 46.0k (\perc{{59\%}})\\
    \hline
  \end{tabular}
  \begin{tabular}{|c|c|}
    \hline
    \multicolumn{2}{|c|}{\textbf{Instructions}}\\
    \hline
    Total & Functionless\\
    281.4k & 1.8k (\perc{{1\%}})\\
    \hline
  \end{tabular}}}
    \vspace{0.25em}
    \subfloat[Metrics for a fully protected 436.cactusADM, after detection and removal of observable opaque predicates (soundish DB attack).\label{fig:436resilienceSound}]{\parbox{\linewidth}{
      \centering
  \scriptsize
  \begin{tabular}{|l|c|ccc|ccc|c|ccc|ccc|}
    \cline{2-15}
    \multicolumn{1}{l|}{} & \multicolumn{7}{|c|}{\textbf{FP/FN CFG edges drawn in GUI}} & \multicolumn{7}{|c|}{\textbf{FP/FN CFG edges stored in database}}\\
    \cline{2-15}
    \multicolumn{1}{l|}{} & \textbf{Total} & \textbf{IA} & \textbf{IO} & \textbf{IF} & \textbf{iA} & \textbf{iO} & \textbf{iF} & \textbf{Total} & \textbf{IA} & \textbf{IO} & \textbf{IF} & \textbf{iA} & \textbf{iO} & \textbf{iF}\\
    \cline{1-15}
    \textbf{\# FP} & 7.5k & 5.3k & 5.6k & 5.6k & 2.1k & 1.9k & 1.8k & 9.5k & 7.1k & 7.5k & 7.5k & 2.3k & 2.0k & 1.9k\\
    \textbf{\perc{FPR}} & \perc{{74\%}} & \perc{{53\%}} & \perc{{56\%}} & \perc{{56\%}} & \perc{{21\%}} & \perc{{19\%}} & \perc{{18\%}} & \perc{{94\%}} & \perc{{71\%}} & \perc{{75\%}} & \perc{{75\%}} & \perc{{23\%}} & \perc{{19\%}} & \perc{{19\%}}\\
    \cline{1-15}
    \textbf{\# FN} & 28.4k & 6 & 13 & 29 & 28.4k & 28.4k & 28.4k & 11.5k & 0 & 0 & 0 & 11.5k & 11.5k & 11.5k\\
    \textbf{\perc{FNR}} & \perc{{42\%}} & \perc{{0\%}} & \perc{{0\%}} & \perc{{0\%}} & \perc{{42\%}} & \perc{{42\%}} & \perc{{42\%}} & \perc{{17\%}} & \perc{{0\%}} & \perc{{0\%}} & \perc{{0\%}} & \perc{{17\%}} & \perc{{17\%}} & \perc{{17\%}}\\
    \cline{1-15}
  \end{tabular}

  \vspace{1em}
  \begin{tabular}{|c|cc|}
    \hline
    \multicolumn{3}{|c|}{\textbf{Pairs of fragments split by factorization}}\\
    \hline
    Total & Wrong & Correct\\
    8.2k & 6.9k (\perc{{84\%}}) & 1.3k (\perc{{16\%}})\\
    \hline
  \end{tabular}
  \begin{tabular}{|c|cc|c|}
    \hline
    \multicolumn{4}{|c|}{\textbf{CFG edges}}\\
    \hline
    Total & True & Fake & Drawn in GUI\\
    77.6k & 67.5k (\perc{{87\%}}) & 10.0k (\perc{{13\%}}) & 46.6k (\perc{{60\%}})\\
    \hline
  \end{tabular}
  \begin{tabular}{|c|c|}
    \hline
    \multicolumn{2}{|c|}{\textbf{Instructions}}\\
    \hline
    Total & Functionless\\
    281.4k & 1.8k (\perc{{1\%}})\\
    \hline
  \end{tabular}
  \begin{tabular}{|c|c|}
    \hline
    \multicolumn{2}{|c|}{\textbf{Opaque predicates}}\\
    \hline
    Total & Resolved\\
    6.2k & 48 (\perc{{1\%}})\\
    \hline
  \end{tabular}}}
    \vspace{0.25em}
    \subfloat[Metrics for a fully protected 436.cactusADM, after detection and removal of observable opaque predicates (unsound GUI attack).\label{fig:436resilience}]{\parbox{\linewidth}{
      \centering
  \scriptsize
  \begin{tabular}{|l|c|ccc|ccc|c|ccc|ccc|}
    \cline{2-15}
    \multicolumn{1}{l|}{} & \multicolumn{7}{|c|}{\textbf{FP/FN CFG edges drawn in GUI}} & \multicolumn{7}{|c|}{\textbf{FP/FN CFG edges stored in database}}\\
    \cline{2-15}
    \multicolumn{1}{l|}{} & \textbf{Total} & \textbf{IA} & \textbf{IO} & \textbf{IF} & \textbf{iA} & \textbf{iO} & \textbf{iF} & \textbf{Total} & \textbf{IA} & \textbf{IO} & \textbf{IF} & \textbf{iA} & \textbf{iO} & \textbf{iF}\\
    \cline{1-15}
    \textbf{\# FP} & 6.8k & 4.7k & 5.0k & 5.0k & 2.0k & 1.8k & 1.8k & 8.2k & 5.9k & 6.3k & 6.3k & 2.2k & 1.9k & 1.9k\\
    \textbf{\perc{FPR}} & \perc{{68\%}} & \perc{{47\%}} & \perc{{49\%}} & \perc{{50\%}} & \perc{{20\%}} & \perc{{18\%}} & \perc{{18\%}} & \perc{{82\%}} & \perc{{59\%}} & \perc{{63\%}} & \perc{{63\%}} & \perc{{22\%}} & \perc{{19\%}} & \perc{{19\%}}\\
    \cline{1-15}
    \textbf{\# FN} & 28.5k & 4 & 10 & 26 & 28.5k & 28.5k & 28.5k & 11.5k & 0 & 0 & 0 & 11.5k & 11.5k & 11.5k\\
    \textbf{\perc{FNR}} & \perc{{42\%}} & \perc{{0\%}} & \perc{{0\%}} & \perc{{0\%}} & \perc{{42\%}} & \perc{{42\%}} & \perc{{42\%}} & \perc{{17\%}} & \perc{{0\%}} & \perc{{0\%}} & \perc{{0\%}} & \perc{{17\%}} & \perc{{17\%}} & \perc{{17\%}}\\
    \cline{1-15}
  \end{tabular}

  \vspace{1em}
  \begin{tabular}{|c|cc|}
    \hline
    \multicolumn{3}{|c|}{\textbf{Pairs of fragments split by factorization}}\\
    \hline
    Total & Wrong & Correct\\
    8.2k & 6.9k (\perc{{84\%}}) & 1.3k (\perc{{16\%}})\\
    \hline
  \end{tabular}
  \begin{tabular}{|c|cc|c|}
    \hline
    \multicolumn{4}{|c|}{\textbf{CFG edges}}\\
    \hline
    Total & True & Fake & Drawn in GUI\\
    77.6k & 67.5k (\perc{{87\%}}) & 10.0k (\perc{{13\%}}) & 45.8k (\perc{{59\%}})\\
    \hline
  \end{tabular}
  \begin{tabular}{|c|c|}
    \hline
    \multicolumn{2}{|c|}{\textbf{Instructions}}\\
    \hline
    Total & Functionless\\
    281.4k & 1.8k (\perc{{1\%}})\\
    \hline
  \end{tabular}
  \begin{tabular}{|c|c|}
    \hline
    \multicolumn{2}{|c|}{\textbf{Opaque predicates}}\\
    \hline
    Total & Resolved\\
    6.2k & 1.3k (\perc{{21\%}})\\
    \hline
  \end{tabular}}}
  \end{minipage}
  \caption{Measurement data for 436.cactusADM}
  \label{results436}
\end{table*}

\begin{table*}
  \begin{minipage}[t]{\textwidth}
    \centering
    \scriptsize
    \subfloat[Compile and link flags for 445.gobmk\label{fig:445flags}]{\parbox{\linewidth}{
      \centering
      \begin{tabular}{|l|l|}
        \hline
        \textbf{Compiler} & \texttt{-DHAVE\_CONFIG\_H -DNDEBUG -DSPEC\_CPU}\\
        & \texttt{-O2 -fno-aggressive-loop-optimizations -g -marm -mcpu=cortex-a8 -std=c99}\\
        \hline
        \textbf{Linker} & \texttt{-Wl,--no-demangle -Wl,--hash-style=sysv -Wl,--no-merge-exidx-entries -Bdynamic -lm}\\
        \hline
      \end{tabular}}}
    \vspace{0.25em}
    \subfloat[Potency metrics for 445.gobmk without the protections proposed in this \papertechreport{}.\label{fig:445potency_unprotected}]{\parbox{\linewidth}{
      \centering
  \scriptsize
  \begin{tabular}{|l|c|ccc|ccc|c|ccc|ccc|}
    \cline{2-15}
    \multicolumn{1}{l|}{} & \multicolumn{7}{|c|}{\textbf{FP/FN CFG edges drawn in GUI}} & \multicolumn{7}{|c|}{\textbf{FP/FN CFG edges stored in database}}\\
    \cline{2-15}
    \multicolumn{1}{l|}{} & \textbf{Total} & \textbf{IA} & \textbf{IO} & \textbf{IF} & \textbf{iA} & \textbf{iO} & \textbf{iF} & \textbf{Total} & \textbf{IA} & \textbf{IO} & \textbf{IF} & \textbf{iA} & \textbf{iO} & \textbf{iF}\\
    \cline{1-15}
    \textbf{\# FP} & 0 & 0 & 0 & 0 & 0 & 0 & 0 & 0 & 0 & 0 & 0 & 0 & 0 & 0\\
    \textbf{\perc{FPR}} & \perc{{0\%}} & \perc{{0\%}} & \perc{{0\%}} & \perc{{0\%}} & \perc{{0\%}} & \perc{{0\%}} & \perc{{0\%}} & \perc{{0\%}} & \perc{{0\%}} & \perc{{0\%}} & \perc{{0\%}} & \perc{{0\%}} & \perc{{0\%}} & \perc{{0\%}}\\
    \cline{1-15}
    \textbf{\# FN} & 6.0k & 93 & 176 & 283 & 5.9k & 5.8k & 5.7k & 5.2k & 81 & 113 & 122 & 5.1k & 5.1k & 5.1k\\
    \textbf{\perc{FNR}} & \perc{{20\%}} & \perc{{0\%}} & \perc{{1\%}} & \perc{{1\%}} & \perc{{20\%}} & \perc{{19\%}} & \perc{{19\%}} & \perc{{17\%}} & \perc{{0\%}} & \perc{{0\%}} & \perc{{0\%}} & \perc{{17\%}} & \perc{{17\%}} & \perc{{17\%}}\\
    \cline{1-15}
  \end{tabular}

  \vspace{1em}
  \begin{tabular}{|c|cc|}
    \hline
    \multicolumn{3}{|c|}{\textbf{Pairs of fragments split by factorization}}\\
    \hline
    Total & Wrong & Correct\\
    0 & 0 (\perc{{0\%}}) & 0 (\perc{{0\%}})\\
    \hline
  \end{tabular}
  \begin{tabular}{|c|cc|c|}
    \hline
    \multicolumn{4}{|c|}{\textbf{CFG edges}}\\
    \hline
    Total & True & Fake & Drawn in GUI\\
    30.4k & 30.4k (\perc{{100\%}}) & 0 (\perc{{0\%}}) & 24.4k (\perc{{80\%}})\\
    \hline
  \end{tabular}
  \begin{tabular}{|c|c|}
    \hline
    \multicolumn{2}{|c|}{\textbf{Instructions}}\\
    \hline
    Total & Functionless\\
    151.6k & 51.0k (\perc{{34\%}})\\
    \hline
  \end{tabular}}}
    \vspace{0.25em}
    \subfloat[Potency metrics for a fully protected 445.gobmk with IDA Pro out-of-the-box.\label{fig:445potency_vanilla}]{\parbox{\linewidth}{
      \centering
  \scriptsize
  \begin{tabular}{|l|c|ccc|ccc|c|ccc|ccc|}
    \cline{2-15}
    \multicolumn{1}{l|}{} & \multicolumn{7}{|c|}{\textbf{FP/FN CFG edges drawn in GUI}} & \multicolumn{7}{|c|}{\textbf{FP/FN CFG edges stored in database}}\\
    \cline{2-15}
    \multicolumn{1}{l|}{} & \textbf{Total} & \textbf{IA} & \textbf{IO} & \textbf{IF} & \textbf{iA} & \textbf{iO} & \textbf{iF} & \textbf{Total} & \textbf{IA} & \textbf{IO} & \textbf{IF} & \textbf{iA} & \textbf{iO} & \textbf{iF}\\
    \cline{1-15}
    \textbf{\# FP} & 10.9k & 5.7k & 8.1k & 8.4k & 5.2k & 2.8k & 2.5k & 12.9k & 6.8k & 10.0k & 10.4k & 6.1k & 2.9k & 2.5k\\
    \textbf{\perc{FPR}} & \perc{{75\%}} & \perc{{39\%}} & \perc{{56\%}} & \perc{{58\%}} & \perc{{36\%}} & \perc{{19\%}} & \perc{{17\%}} & \perc{{89\%}} & \perc{{47\%}} & \perc{{68\%}} & \perc{{71\%}} & \perc{{42\%}} & \perc{{20\%}} & \perc{{17\%}}\\
    \cline{1-15}
    \textbf{\# FN} & 58.0k & 93 & 178 & 287 & 57.9k & 57.8k & 57.7k & 36.2k & 59 & 99 & 128 & 36.1k & 36.1k & 36.1k\\
    \textbf{\perc{FNR}} & \perc{{56\%}} & \perc{{0\%}} & \perc{{0\%}} & \perc{{0\%}} & \perc{{56\%}} & \perc{{56\%}} & \perc{{56\%}} & \perc{{35\%}} & \perc{{0\%}} & \perc{{0\%}} & \perc{{0\%}} & \perc{{35\%}} & \perc{{35\%}} & \perc{{35\%}}\\
    \cline{1-15}
  \end{tabular}

  \vspace{1em}
  \begin{tabular}{|c|cc|}
    \hline
    \multicolumn{3}{|c|}{\textbf{Pairs of fragments split by factorization}}\\
    \hline
    Total & Wrong & Correct\\
    15.7k & 14.7k (\perc{{94\%}}) & 919 (\perc{{6\%}})\\
    \hline
  \end{tabular}
  \begin{tabular}{|c|cc|c|}
    \hline
    \multicolumn{4}{|c|}{\textbf{CFG edges}}\\
    \hline
    Total & True & Fake & Drawn in GUI\\
    118.0k & 103.5k (\perc{{88\%}}) & 14.6k (\perc{{12\%}}) & 56.4k (\perc{{48\%}})\\
    \hline
  \end{tabular}
  \begin{tabular}{|c|c|}
    \hline
    \multicolumn{2}{|c|}{\textbf{Instructions}}\\
    \hline
    Total & Functionless\\
    413.9k & 127.6k (\perc{{31\%}})\\
    \hline
  \end{tabular}}}
    \vspace{0.25em}
    \subfloat[Potency metrics for a fully protected 445.gobmk with attacker-improved IDA Pro.\label{fig:445potency_repartition}]{\parbox{\linewidth}{
      \centering
  \scriptsize
  \begin{tabular}{|l|c|ccc|ccc|c|ccc|ccc|}
    \cline{2-15}
    \multicolumn{1}{l|}{} & \multicolumn{7}{|c|}{\textbf{FP/FN CFG edges drawn in GUI}} & \multicolumn{7}{|c|}{\textbf{FP/FN CFG edges stored in database}}\\
    \cline{2-15}
    \multicolumn{1}{l|}{} & \textbf{Total} & \textbf{IA} & \textbf{IO} & \textbf{IF} & \textbf{iA} & \textbf{iO} & \textbf{iF} & \textbf{Total} & \textbf{IA} & \textbf{IO} & \textbf{IF} & \textbf{iA} & \textbf{iO} & \textbf{iF}\\
    \cline{1-15}
    \textbf{\# FP} & 11.5k & 6.2k & 8.7k & 8.9k & 5.4k & 2.9k & 2.6k & 13.8k & 7.5k & 10.8k & 11.2k & 6.4k & 3.0k & 2.6k\\
    \textbf{\perc{FPR}} & \perc{{79\%}} & \perc{{42\%}} & \perc{{59\%}} & \perc{{61\%}} & \perc{{37\%}} & \perc{{20\%}} & \perc{{18\%}} & \perc{{95\%}} & \perc{{51\%}} & \perc{{74\%}} & \perc{{77\%}} & \perc{{44\%}} & \perc{{21\%}} & \perc{{18\%}}\\
    \cline{1-15}
    \textbf{\# FN} & 42.7k & 38 & 90 & 170 & 42.7k & 42.6k & 42.5k & 16.1k & 0 & 3 & 5 & 16.1k & 16.1k & 16.1k\\
    \textbf{\perc{FNR}} & \perc{{41\%}} & \perc{{0\%}} & \perc{{0\%}} & \perc{{0\%}} & \perc{{41\%}} & \perc{{41\%}} & \perc{{41\%}} & \perc{{16\%}} & \perc{{0\%}} & \perc{{0\%}} & \perc{{0\%}} & \perc{{16\%}} & \perc{{16\%}} & \perc{{16\%}}\\
    \cline{1-15}
  \end{tabular}

  \vspace{1em}
  \begin{tabular}{|c|cc|}
    \hline
    \multicolumn{3}{|c|}{\textbf{Pairs of fragments split by factorization}}\\
    \hline
    Total & Wrong & Correct\\
    15.7k & 13.4k (\perc{{86\%}}) & 2.2k (\perc{{14\%}})\\
    \hline
  \end{tabular}
  \begin{tabular}{|c|cc|c|}
    \hline
    \multicolumn{4}{|c|}{\textbf{CFG edges}}\\
    \hline
    Total & True & Fake & Drawn in GUI\\
    118.1k & 103.6k (\perc{{88\%}}) & 14.6k (\perc{{12\%}}) & 72.4k (\perc{{61\%}})\\
    \hline
  \end{tabular}
  \begin{tabular}{|c|c|}
    \hline
    \multicolumn{2}{|c|}{\textbf{Instructions}}\\
    \hline
    Total & Functionless\\
    414.0k & 86 (\perc{{0\%}})\\
    \hline
  \end{tabular}}}
    \vspace{0.25em}
    \subfloat[Metrics for a fully protected 445.gobmk, after detection and removal of observable opaque predicates (soundish DB attack).\label{fig:445resilienceSound}]{\parbox{\linewidth}{
      \centering
  \scriptsize
  \begin{tabular}{|l|c|ccc|ccc|c|ccc|ccc|}
    \cline{2-15}
    \multicolumn{1}{l|}{} & \multicolumn{7}{|c|}{\textbf{FP/FN CFG edges drawn in GUI}} & \multicolumn{7}{|c|}{\textbf{FP/FN CFG edges stored in database}}\\
    \cline{2-15}
    \multicolumn{1}{l|}{} & \textbf{Total} & \textbf{IA} & \textbf{IO} & \textbf{IF} & \textbf{iA} & \textbf{iO} & \textbf{iF} & \textbf{Total} & \textbf{IA} & \textbf{IO} & \textbf{IF} & \textbf{iA} & \textbf{iO} & \textbf{iF}\\
    \cline{1-15}
    \textbf{\# FP} & 11.1k & 5.9k & 8.3k & 8.5k & 5.2k & 2.8k & 2.6k & 13.8k & 7.4k & 10.7k & 11.2k & 6.4k & 3.0k & 2.6k\\
    \textbf{\perc{FPR}} & \perc{{76\%}} & \perc{{40\%}} & \perc{{57\%}} & \perc{{59\%}} & \perc{{36\%}} & \perc{{20\%}} & \perc{{18\%}} & \perc{{95\%}} & \perc{{51\%}} & \perc{{74\%}} & \perc{{77\%}} & \perc{{44\%}} & \perc{{21\%}} & \perc{{18\%}}\\
    \cline{1-15}
    \textbf{\# FN} & 42.0k & 34 & 80 & 152 & 41.9k & 41.9k & 41.8k & 16.1k & 0 & 3 & 5 & 16.1k & 16.1k & 16.1k\\
    \textbf{\perc{FNR}} & \perc{{41\%}} & \perc{{0\%}} & \perc{{0\%}} & \perc{{0\%}} & \perc{{40\%}} & \perc{{40\%}} & \perc{{40\%}} & \perc{{16\%}} & \perc{{0\%}} & \perc{{0\%}} & \perc{{0\%}} & \perc{{16\%}} & \perc{{16\%}} & \perc{{16\%}}\\
    \cline{1-15}
  \end{tabular}

  \vspace{1em}
  \begin{tabular}{|c|cc|}
    \hline
    \multicolumn{3}{|c|}{\textbf{Pairs of fragments split by factorization}}\\
    \hline
    Total & Wrong & Correct\\
    15.7k & 13.4k (\perc{{85\%}}) & 2.3k (\perc{{15\%}})\\
    \hline
  \end{tabular}
  \begin{tabular}{|c|cc|c|}
    \hline
    \multicolumn{4}{|c|}{\textbf{CFG edges}}\\
    \hline
    Total & True & Fake & Drawn in GUI\\
    118.1k & 103.6k (\perc{{88\%}}) & 14.6k (\perc{{12\%}}) & 72.7k (\perc{{62\%}})\\
    \hline
  \end{tabular}
  \begin{tabular}{|c|c|}
    \hline
    \multicolumn{2}{|c|}{\textbf{Instructions}}\\
    \hline
    Total & Functionless\\
    414.0k & 86 (\perc{{0\%}})\\
    \hline
  \end{tabular}
  \begin{tabular}{|c|c|}
    \hline
    \multicolumn{2}{|c|}{\textbf{Opaque predicates}}\\
    \hline
    Total & Resolved\\
    9.0k & 65 (\perc{{1\%}})\\
    \hline
  \end{tabular}}}
    \vspace{0.25em}
    \subfloat[Metrics for a fully protected 445.gobmk, after detection and removal of observable opaque predicates (unsound GUI attack).\label{fig:445resilience}]{\parbox{\linewidth}{
      \centering
  \scriptsize
  \begin{tabular}{|l|c|ccc|ccc|c|ccc|ccc|}
    \cline{2-15}
    \multicolumn{1}{l|}{} & \multicolumn{7}{|c|}{\textbf{FP/FN CFG edges drawn in GUI}} & \multicolumn{7}{|c|}{\textbf{FP/FN CFG edges stored in database}}\\
    \cline{2-15}
    \multicolumn{1}{l|}{} & \textbf{Total} & \textbf{IA} & \textbf{IO} & \textbf{IF} & \textbf{iA} & \textbf{iO} & \textbf{iF} & \textbf{Total} & \textbf{IA} & \textbf{IO} & \textbf{IF} & \textbf{iA} & \textbf{iO} & \textbf{iF}\\
    \cline{1-15}
    \textbf{\# FP} & 10.1k & 5.2k & 7.3k & 7.6k & 5.0k & 2.8k & 2.5k & 12.0k & 6.2k & 9.1k & 9.4k & 5.8k & 3.0k & 2.6k\\
    \textbf{\perc{FPR}} & \perc{{70\%}} & \perc{{35\%}} & \perc{{50\%}} & \perc{{52\%}} & \perc{{34\%}} & \perc{{19\%}} & \perc{{17\%}} & \perc{{83\%}} & \perc{{42\%}} & \perc{{62\%}} & \perc{{65\%}} & \perc{{40\%}} & \perc{{20\%}} & \perc{{18\%}}\\
    \cline{1-15}
    \textbf{\# FN} & 41.7k & 31 & 74 & 139 & 41.7k & 41.6k & 41.6k & 16.1k & 0 & 3 & 5 & 16.1k & 16.1k & 16.1k\\
    \textbf{\perc{FNR}} & \perc{{40\%}} & \perc{{0\%}} & \perc{{0\%}} & \perc{{0\%}} & \perc{{40\%}} & \perc{{40\%}} & \perc{{40\%}} & \perc{{16\%}} & \perc{{0\%}} & \perc{{0\%}} & \perc{{0\%}} & \perc{{16\%}} & \perc{{16\%}} & \perc{{16\%}}\\
    \cline{1-15}
  \end{tabular}

  \vspace{1em}
  \begin{tabular}{|c|cc|}
    \hline
    \multicolumn{3}{|c|}{\textbf{Pairs of fragments split by factorization}}\\
    \hline
    Total & Wrong & Correct\\
    15.7k & 13.3k (\perc{{85\%}}) & 2.3k (\perc{{15\%}})\\
    \hline
  \end{tabular}
  \begin{tabular}{|c|cc|c|}
    \hline
    \multicolumn{4}{|c|}{\textbf{CFG edges}}\\
    \hline
    Total & True & Fake & Drawn in GUI\\
    118.1k & 103.6k (\perc{{88\%}}) & 14.6k (\perc{{12\%}}) & 72.0k (\perc{{61\%}})\\
    \hline
  \end{tabular}
  \begin{tabular}{|c|c|}
    \hline
    \multicolumn{2}{|c|}{\textbf{Instructions}}\\
    \hline
    Total & Functionless\\
    414.0k & 86 (\perc{{0\%}})\\
    \hline
  \end{tabular}
  \begin{tabular}{|c|c|}
    \hline
    \multicolumn{2}{|c|}{\textbf{Opaque predicates}}\\
    \hline
    Total & Resolved\\
    9.0k & 1.8k (\perc{{20\%}})\\
    \hline
  \end{tabular}}}
  \end{minipage}
  \caption{Measurement data for 445.gobmk}
  \label{results445}
\end{table*}

\begin{table*}
  \begin{minipage}[t]{\textwidth}
    \centering
    \scriptsize
    \subfloat[Compile and link flags for 454.calculix\label{fig:454flags}]{\parbox{\linewidth}{
      \centering
      \begin{tabular}{|l|l|}
        \hline
        \textbf{Compiler} & \texttt{-DNDEBUG -DSPEC\_CPU}\\
        & \texttt{-O2 -fno-aggressive-loop-optimizations -g -marm -mcpu=cortex-a8 -std=c99}\\
        \hline
        \textbf{Linker} & \texttt{-Wl,--no-demangle -Wl,--hash-style=sysv -Wl,--no-merge-exidx-entries -Bdynamic -lm}\\
        \hline
      \end{tabular}}}
    \vspace{0.25em}
    \subfloat[Potency metrics for 454.calculix without the protections proposed in this \papertechreport{}.\label{fig:454potency_unprotected}]{\parbox{\linewidth}{
      \centering
  \scriptsize
  \begin{tabular}{|l|c|ccc|ccc|c|ccc|ccc|}
    \cline{2-15}
    \multicolumn{1}{l|}{} & \multicolumn{7}{|c|}{\textbf{FP/FN CFG edges drawn in GUI}} & \multicolumn{7}{|c|}{\textbf{FP/FN CFG edges stored in database}}\\
    \cline{2-15}
    \multicolumn{1}{l|}{} & \textbf{Total} & \textbf{IA} & \textbf{IO} & \textbf{IF} & \textbf{iA} & \textbf{iO} & \textbf{iF} & \textbf{Total} & \textbf{IA} & \textbf{IO} & \textbf{IF} & \textbf{iA} & \textbf{iO} & \textbf{iF}\\
    \cline{1-15}
    \textbf{\# FP} & 0 & 0 & 0 & 0 & 0 & 0 & 0 & 0 & 0 & 0 & 0 & 0 & 0 & 0\\
    \textbf{\perc{FPR}} & \perc{{0\%}} & \perc{{0\%}} & \perc{{0\%}} & \perc{{0\%}} & \perc{{0\%}} & \perc{{0\%}} & \perc{{0\%}} & \perc{{0\%}} & \perc{{0\%}} & \perc{{0\%}} & \perc{{0\%}} & \perc{{0\%}} & \perc{{0\%}} & \perc{{0\%}}\\
    \cline{1-15}
    \textbf{\# FN} & 2.9k & 0 & 100 & 174 & 2.9k & 2.8k & 2.8k & 2.6k & 0 & 13 & 35 & 2.6k & 2.6k & 2.6k\\
    \textbf{\perc{FNR}} & \perc{{6\%}} & \perc{{0\%}} & \perc{{0\%}} & \perc{{0\%}} & \perc{{6\%}} & \perc{{6\%}} & \perc{{6\%}} & \perc{{6\%}} & \perc{{0\%}} & \perc{{0\%}} & \perc{{0\%}} & \perc{{6\%}} & \perc{{6\%}} & \perc{{6\%}}\\
    \cline{1-15}
  \end{tabular}

  \vspace{1em}
  \begin{tabular}{|c|cc|}
    \hline
    \multicolumn{3}{|c|}{\textbf{Pairs of fragments split by factorization}}\\
    \hline
    Total & Wrong & Correct\\
    0 & 0 (\perc{{0\%}}) & 0 (\perc{{0\%}})\\
    \hline
  \end{tabular}
  \begin{tabular}{|c|cc|c|}
    \hline
    \multicolumn{4}{|c|}{\textbf{CFG edges}}\\
    \hline
    Total & True & Fake & Drawn in GUI\\
    45.5k & 45.5k (\perc{{100\%}}) & 0 (\perc{{0\%}}) & 42.5k (\perc{{94\%}})\\
    \hline
  \end{tabular}
  \begin{tabular}{|c|c|}
    \hline
    \multicolumn{2}{|c|}{\textbf{Instructions}}\\
    \hline
    Total & Functionless\\
    364.0k & 12.1k (\perc{{3\%}})\\
    \hline
  \end{tabular}}}
    \vspace{0.25em}
    \subfloat[Potency metrics for a fully protected 454.calculix with IDA Pro out-of-the-box.\label{fig:454potency_vanilla}]{\parbox{\linewidth}{
      \centering
  \scriptsize
  \begin{tabular}{|l|c|ccc|ccc|c|ccc|ccc|}
    \cline{2-15}
    \multicolumn{1}{l|}{} & \multicolumn{7}{|c|}{\textbf{FP/FN CFG edges drawn in GUI}} & \multicolumn{7}{|c|}{\textbf{FP/FN CFG edges stored in database}}\\
    \cline{2-15}
    \multicolumn{1}{l|}{} & \textbf{Total} & \textbf{IA} & \textbf{IO} & \textbf{IF} & \textbf{iA} & \textbf{iO} & \textbf{iF} & \textbf{Total} & \textbf{IA} & \textbf{IO} & \textbf{IF} & \textbf{iA} & \textbf{iO} & \textbf{iF}\\
    \cline{1-15}
    \textbf{\# FP} & 37.4k & 22.2k & 27.5k & 27.7k & 15.2k & 9.9k & 9.8k & 45.1k & 26.8k & 34.9k & 35.1k & 18.3k & 10.2k & 10.0k\\
    \textbf{\perc{FPR}} & \perc{{75\%}} & \perc{{44\%}} & \perc{{55\%}} & \perc{{55\%}} & \perc{{30\%}} & \perc{{20\%}} & \perc{{19\%}} & \perc{{90\%}} & \perc{{54\%}} & \perc{{70\%}} & \perc{{70\%}} & \perc{{36\%}} & \perc{{20\%}} & \perc{{20\%}}\\
    \cline{1-15}
    \textbf{\# FN} & 150.3k & 1 & 231 & 675 & 150.3k & 150.1k & 149.6k & 88.5k & 1 & 107 & 381 & 88.5k & 88.4k & 88.1k\\
    \textbf{\perc{FNR}} & \perc{{54\%}} & \perc{{0\%}} & \perc{{0\%}} & \perc{{0\%}} & \perc{{54\%}} & \perc{{54\%}} & \perc{{54\%}} & \perc{{32\%}} & \perc{{0\%}} & \perc{{0\%}} & \perc{{0\%}} & \perc{{32\%}} & \perc{{32\%}} & \perc{{32\%}}\\
    \cline{1-15}
  \end{tabular}

  \vspace{1em}
  \begin{tabular}{|c|cc|}
    \hline
    \multicolumn{3}{|c|}{\textbf{Pairs of fragments split by factorization}}\\
    \hline
    Total & Wrong & Correct\\
    31.2k & 29.1k (\perc{{93\%}}) & 2.1k (\perc{{7\%}})\\
    \hline
  \end{tabular}
  \begin{tabular}{|c|cc|c|}
    \hline
    \multicolumn{4}{|c|}{\textbf{CFG edges}}\\
    \hline
    Total & True & Fake & Drawn in GUI\\
    329.1k & 279.0k (\perc{{85\%}}) & 50.1k (\perc{{15\%}}) & 166.1k (\perc{{50\%}})\\
    \hline
  \end{tabular}
  \begin{tabular}{|c|c|}
    \hline
    \multicolumn{2}{|c|}{\textbf{Instructions}}\\
    \hline
    Total & Functionless\\
    1224.4k & 311.7k (\perc{{25\%}})\\
    \hline
  \end{tabular}}}
    \vspace{0.25em}
    \subfloat[Potency metrics for a fully protected 454.calculix with attacker-improved IDA Pro.\label{fig:454potency_repartition}]{\parbox{\linewidth}{
      \centering
  \scriptsize
  \begin{tabular}{|l|c|ccc|ccc|c|ccc|ccc|}
    \cline{2-15}
    \multicolumn{1}{l|}{} & \multicolumn{7}{|c|}{\textbf{FP/FN CFG edges drawn in GUI}} & \multicolumn{7}{|c|}{\textbf{FP/FN CFG edges stored in database}}\\
    \cline{2-15}
    \multicolumn{1}{l|}{} & \textbf{Total} & \textbf{IA} & \textbf{IO} & \textbf{IF} & \textbf{iA} & \textbf{iO} & \textbf{iF} & \textbf{Total} & \textbf{IA} & \textbf{IO} & \textbf{IF} & \textbf{iA} & \textbf{iO} & \textbf{iF}\\
    \cline{1-15}
    \textbf{\# FP} & 38.9k & 23.4k & 28.8k & 29.0k & 15.5k & 10.1k & 10.0k & 47.6k & 28.6k & 37.0k & 37.3k & 19.0k & 10.5k & 10.3k\\
    \textbf{\perc{FPR}} & \perc{{78\%}} & \perc{{47\%}} & \perc{{58\%}} & \perc{{58\%}} & \perc{{31\%}} & \perc{{20\%}} & \perc{{20\%}} & \perc{{95\%}} & \perc{{57\%}} & \perc{{74\%}} & \perc{{74\%}} & \perc{{38\%}} & \perc{{21\%}} & \perc{{21\%}}\\
    \cline{1-15}
    \textbf{\# FN} & 117.6k & 0 & 116 & 287 & 117.6k & 117.5k & 117.3k & 46.9k & 0 & 3 & 10 & 46.9k & 46.9k & 46.9k\\
    \textbf{\perc{FNR}} & \perc{{42\%}} & \perc{{0\%}} & \perc{{0\%}} & \perc{{0\%}} & \perc{{42\%}} & \perc{{42\%}} & \perc{{42\%}} & \perc{{17\%}} & \perc{{0\%}} & \perc{{0\%}} & \perc{{0\%}} & \perc{{17\%}} & \perc{{17\%}} & \perc{{17\%}}\\
    \cline{1-15}
  \end{tabular}

  \vspace{1em}
  \begin{tabular}{|c|cc|}
    \hline
    \multicolumn{3}{|c|}{\textbf{Pairs of fragments split by factorization}}\\
    \hline
    Total & Wrong & Correct\\
    31.2k & 27.1k (\perc{{87\%}}) & 4.1k (\perc{{13\%}})\\
    \hline
  \end{tabular}
  \begin{tabular}{|c|cc|c|}
    \hline
    \multicolumn{4}{|c|}{\textbf{CFG edges}}\\
    \hline
    Total & True & Fake & Drawn in GUI\\
    329.1k & 279.0k (\perc{{85\%}}) & 50.1k (\perc{{15\%}}) & 200.3k (\perc{{61\%}})\\
    \hline
  \end{tabular}
  \begin{tabular}{|c|c|}
    \hline
    \multicolumn{2}{|c|}{\textbf{Instructions}}\\
    \hline
    Total & Functionless\\
    1224.5k & 241 (\perc{{0\%}})\\
    \hline
  \end{tabular}}}
    \vspace{0.25em}
    \subfloat[Metrics for a fully protected 454.calculix, after detection and removal of observable opaque predicates (soundish DB attack).\label{fig:454resilienceSound}]{\parbox{\linewidth}{
      \centering
  \scriptsize
  \begin{tabular}{|l|c|ccc|ccc|c|ccc|ccc|}
    \cline{2-15}
    \multicolumn{1}{l|}{} & \multicolumn{7}{|c|}{\textbf{FP/FN CFG edges drawn in GUI}} & \multicolumn{7}{|c|}{\textbf{FP/FN CFG edges stored in database}}\\
    \cline{2-15}
    \multicolumn{1}{l|}{} & \textbf{Total} & \textbf{IA} & \textbf{IO} & \textbf{IF} & \textbf{iA} & \textbf{iO} & \textbf{iF} & \textbf{Total} & \textbf{IA} & \textbf{IO} & \textbf{IF} & \textbf{iA} & \textbf{iO} & \textbf{iF}\\
    \cline{1-15}
    \textbf{\# FP} & 37.2k & 22.1k & 27.2k & 27.3k & 15.1k & 10.1k & 9.9k & 47.5k & 28.5k & 36.9k & 37.2k & 18.9k & 10.5k & 10.3k\\
    \textbf{\perc{FPR}} & \perc{{74\%}} & \perc{{44\%}} & \perc{{54\%}} & \perc{{54\%}} & \perc{{30\%}} & \perc{{20\%}} & \perc{{20\%}} & \perc{{95\%}} & \perc{{57\%}} & \perc{{74\%}} & \perc{{74\%}} & \perc{{38\%}} & \perc{{21\%}} & \perc{{21\%}}\\
    \cline{1-15}
    \textbf{\# FN} & 115.0k & 0 & 76 & 167 & 115.0k & 114.9k & 114.8k & 46.9k & 0 & 3 & 10 & 46.9k & 46.9k & 46.9k\\
    \textbf{\perc{FNR}} & \perc{{41\%}} & \perc{{0\%}} & \perc{{0\%}} & \perc{{0\%}} & \perc{{41\%}} & \perc{{41\%}} & \perc{{41\%}} & \perc{{17\%}} & \perc{{0\%}} & \perc{{0\%}} & \perc{{0\%}} & \perc{{17\%}} & \perc{{17\%}} & \perc{{17\%}}\\
    \cline{1-15}
  \end{tabular}

  \vspace{1em}
  \begin{tabular}{|c|cc|}
    \hline
    \multicolumn{3}{|c|}{\textbf{Pairs of fragments split by factorization}}\\
    \hline
    Total & Wrong & Correct\\
    31.2k & 27.0k (\perc{{86\%}}) & 4.2k (\perc{{14\%}})\\
    \hline
  \end{tabular}
  \begin{tabular}{|c|cc|c|}
    \hline
    \multicolumn{4}{|c|}{\textbf{CFG edges}}\\
    \hline
    Total & True & Fake & Drawn in GUI\\
    329.1k & 279.0k (\perc{{85\%}}) & 50.1k (\perc{{15\%}}) & 201.2k (\perc{{61\%}})\\
    \hline
  \end{tabular}
  \begin{tabular}{|c|c|}
    \hline
    \multicolumn{2}{|c|}{\textbf{Instructions}}\\
    \hline
    Total & Functionless\\
    1224.5k & 241 (\perc{{0\%}})\\
    \hline
  \end{tabular}
  \begin{tabular}{|c|c|}
    \hline
    \multicolumn{2}{|c|}{\textbf{Opaque predicates}}\\
    \hline
    Total & Resolved\\
    29.0k & 126 (\perc{{0\%}})\\
    \hline
  \end{tabular}}}
    \vspace{0.25em}
    \subfloat[Metrics for a fully protected 454.calculix, after detection and removal of observable opaque predicates (unsound GUI attack).\label{fig:454resilience}]{\parbox{\linewidth}{
      \centering
  \scriptsize
  \begin{tabular}{|l|c|ccc|ccc|c|ccc|ccc|}
    \cline{2-15}
    \multicolumn{1}{l|}{} & \multicolumn{7}{|c|}{\textbf{FP/FN CFG edges drawn in GUI}} & \multicolumn{7}{|c|}{\textbf{FP/FN CFG edges stored in database}}\\
    \cline{2-15}
    \multicolumn{1}{l|}{} & \textbf{Total} & \textbf{IA} & \textbf{IO} & \textbf{IF} & \textbf{iA} & \textbf{iO} & \textbf{iF} & \textbf{Total} & \textbf{IA} & \textbf{IO} & \textbf{IF} & \textbf{iA} & \textbf{iO} & \textbf{iF}\\
    \cline{1-15}
    \textbf{\# FP} & 33.9k & 19.2k & 24.0k & 24.2k & 14.7k & 9.9k & 9.7k & 41.2k & 23.0k & 30.9k & 31.1k & 18.2k & 10.4k & 10.1k\\
    \textbf{\perc{FPR}} & \perc{{68\%}} & \perc{{38\%}} & \perc{{48\%}} & \perc{{48\%}} & \perc{{29\%}} & \perc{{20\%}} & \perc{{19\%}} & \perc{{82\%}} & \perc{{46\%}} & \perc{{62\%}} & \perc{{62\%}} & \perc{{36\%}} & \perc{{21\%}} & \perc{{20\%}}\\
    \cline{1-15}
    \textbf{\# FN} & 114.4k & 0 & 64 & 141 & 114.4k & 114.3k & 114.2k & 46.9k & 0 & 3 & 10 & 46.9k & 46.9k & 46.9k\\
    \textbf{\perc{FNR}} & \perc{{41\%}} & \perc{{0\%}} & \perc{{0\%}} & \perc{{0\%}} & \perc{{41\%}} & \perc{{41\%}} & \perc{{41\%}} & \perc{{17\%}} & \perc{{0\%}} & \perc{{0\%}} & \perc{{0\%}} & \perc{{17\%}} & \perc{{17\%}} & \perc{{17\%}}\\
    \cline{1-15}
  \end{tabular}

  \vspace{1em}
  \begin{tabular}{|c|cc|}
    \hline
    \multicolumn{3}{|c|}{\textbf{Pairs of fragments split by factorization}}\\
    \hline
    Total & Wrong & Correct\\
    31.2k & 26.9k (\perc{{86\%}}) & 4.3k (\perc{{14\%}})\\
    \hline
  \end{tabular}
  \begin{tabular}{|c|cc|c|}
    \hline
    \multicolumn{4}{|c|}{\textbf{CFG edges}}\\
    \hline
    Total & True & Fake & Drawn in GUI\\
    329.1k & 279.0k (\perc{{85\%}}) & 50.1k (\perc{{15\%}}) & 198.5k (\perc{{60\%}})\\
    \hline
  \end{tabular}
  \begin{tabular}{|c|c|}
    \hline
    \multicolumn{2}{|c|}{\textbf{Instructions}}\\
    \hline
    Total & Functionless\\
    1224.5k & 241 (\perc{{0\%}})\\
    \hline
  \end{tabular}
  \begin{tabular}{|c|c|}
    \hline
    \multicolumn{2}{|c|}{\textbf{Opaque predicates}}\\
    \hline
    Total & Resolved\\
    29.0k & 6.4k (\perc{{22\%}})\\
    \hline
  \end{tabular}}}
  \end{minipage}
  \caption{Measurement data for 454.calculix}
  \label{results454}
\end{table*}

\begin{table*}
  \begin{minipage}[t]{\textwidth}
    \centering
    \scriptsize
    \subfloat[Compile and link flags for DRM\label{fig:DRMflags}]{\parbox{\linewidth}{
      \centering
      \begin{tabular}{|l|l|}
        \hline
        \textbf{Compiler (C)} & \texttt{-Os -Wa,--noexecstack -fPIE -fdata-sections -ffunction-sections -fgcse-after-reload}\\
        & \texttt{-fno-builtin-sin -fno-exceptions -fno-short-enums -fno-strict-aliasing}\\
        & \texttt{-fno-strict-volatile-bitfields -fomit-frame-pointer -fpic -frename-registers}\\
        & \texttt{-frerun-cse-after-loop -fstack-protector -funwind-tables -march=armv7-a -marm}\\
        & \texttt{-mfloat-abi=softfp -mfpu=neon -msoft-float -mthumb-interwork -std=c99}\\
        \hline
        \textbf{Compiler (C++)} & \texttt{-fno-rtti -fvisibility-inlines-hidden}\\
        \hline
        \textbf{Linker} & \texttt{-Wl,--fix-cortex-a8 -Wl,--hash-style=sysv -Wl,--no-demangle -Wl,--no-merge-exidx-entries}\\
        & \texttt{-Wl,--no-undefined -Wl,-fuse-ld=bfd -Wl,-shared,-Bsymbolic -Wl,-z,noexecstack -Wl,-z,now}\\
        & \texttt{-Wl,-z,relro -lbinder -lc -lcutils -ldl -ldrmframework -ldrmframeworkcommon -lgcc -licuuc}\\
        & \texttt{-llog -lm -lsqlite -lstdc++ -lstlport -lutils -shared}\\
        \hline
      \end{tabular}}}
    \vspace{0.25em}
    \subfloat[Potency metrics for DRM without the protections proposed in this \papertechreport{}.\label{fig:DRMpotency_unprotected}]{\parbox{\linewidth}{
      \centering
  \scriptsize
  \begin{tabular}{|l|c|ccc|ccc|c|ccc|ccc|}
    \cline{2-15}
    \multicolumn{1}{l|}{} & \multicolumn{7}{|c|}{\textbf{FP/FN CFG edges drawn in GUI}} & \multicolumn{7}{|c|}{\textbf{FP/FN CFG edges stored in database}}\\
    \cline{2-15}
    \multicolumn{1}{l|}{} & \textbf{Total} & \textbf{IA} & \textbf{IO} & \textbf{IF} & \textbf{iA} & \textbf{iO} & \textbf{iF} & \textbf{Total} & \textbf{IA} & \textbf{IO} & \textbf{IF} & \textbf{iA} & \textbf{iO} & \textbf{iF}\\
    \cline{1-15}
    \textbf{\# FP} & 0 & 0 & 0 & 0 & 0 & 0 & 0 & 0 & 0 & 0 & 0 & 0 & 0 & 0\\
    \textbf{\perc{FPR}} & \perc{{0\%}} & \perc{{0\%}} & \perc{{0\%}} & \perc{{0\%}} & \perc{{0\%}} & \perc{{0\%}} & \perc{{0\%}} & \perc{{0\%}} & \perc{{0\%}} & \perc{{0\%}} & \perc{{0\%}} & \perc{{0\%}} & \perc{{0\%}} & \perc{{0\%}}\\
    \cline{1-15}
    \textbf{\# FN} & 2.0k & 22 & 603 & 710 & 1.9k & 1.4k & 1.3k & 1.0k & 1 & 30 & 34 & 1.0k & 1.0k & 1.0k\\
    \textbf{\perc{FNR}} & \perc{{3\%}} & \perc{{0\%}} & \perc{{1\%}} & \perc{{1\%}} & \perc{{3\%}} & \perc{{2\%}} & \perc{{2\%}} & \perc{{2\%}} & \perc{{0\%}} & \perc{{0\%}} & \perc{{0\%}} & \perc{{2\%}} & \perc{{2\%}} & \perc{{2\%}}\\
    \cline{1-15}
  \end{tabular}

  \vspace{1em}
  \begin{tabular}{|c|cc|}
    \hline
    \multicolumn{3}{|c|}{\textbf{Pairs of fragments split by factorization}}\\
    \hline
    Total & Wrong & Correct\\
    0 & 0 (\perc{{0\%}}) & 0 (\perc{{0\%}})\\
    \hline
  \end{tabular}
  \begin{tabular}{|c|cc|c|}
    \hline
    \multicolumn{4}{|c|}{\textbf{CFG edges}}\\
    \hline
    Total & True & Fake & Drawn in GUI\\
    61.4k & 61.4k (\perc{{100\%}}) & 0 (\perc{{0\%}}) & 59.5k (\perc{{97\%}})\\
    \hline
  \end{tabular}
  \begin{tabular}{|c|c|}
    \hline
    \multicolumn{2}{|c|}{\textbf{Instructions}}\\
    \hline
    Total & Functionless\\
    260.8k & 4.6k (\perc{{2\%}})\\
    \hline
  \end{tabular}}}
    \vspace{0.25em}
    \subfloat[Potency metrics for a fully protected DRM with IDA Pro out-of-the-box.\label{fig:DRMpotency_vanilla}]{\parbox{\linewidth}{
      \centering
  \scriptsize
  \begin{tabular}{|l|c|ccc|ccc|c|ccc|ccc|}
    \cline{2-15}
    \multicolumn{1}{l|}{} & \multicolumn{7}{|c|}{\textbf{FP/FN CFG edges drawn in GUI}} & \multicolumn{7}{|c|}{\textbf{FP/FN CFG edges stored in database}}\\
    \cline{2-15}
    \multicolumn{1}{l|}{} & \textbf{Total} & \textbf{IA} & \textbf{IO} & \textbf{IF} & \textbf{iA} & \textbf{iO} & \textbf{iF} & \textbf{Total} & \textbf{IA} & \textbf{IO} & \textbf{IF} & \textbf{iA} & \textbf{iO} & \textbf{iF}\\
    \cline{1-15}
    \textbf{\# FP} & 17.0k & 9.0k & 13.1k & 13.2k & 8.0k & 3.9k & 3.8k & 20.5k & 10.9k & 16.6k & 16.7k & 9.6k & 3.9k & 3.8k\\
    \textbf{\perc{FPR}} & \perc{{76\%}} & \perc{{40\%}} & \perc{{58\%}} & \perc{{59\%}} & \perc{{36\%}} & \perc{{17\%}} & \perc{{17\%}} & \perc{{91\%}} & \perc{{49\%}} & \perc{{74\%}} & \perc{{74\%}} & \perc{{43\%}} & \perc{{17\%}} & \perc{{17\%}}\\
    \cline{1-15}
    \textbf{\# FN} & 92.4k & 23 & 591 & 700 & 92.4k & 91.8k & 91.7k & 56.1k & 5 & 113 & 135 & 56.1k & 56.0k & 56.0k\\
    \textbf{\perc{FNR}} & \perc{{54\%}} & \perc{{0\%}} & \perc{{0\%}} & \perc{{0\%}} & \perc{{54\%}} & \perc{{54\%}} & \perc{{54\%}} & \perc{{33\%}} & \perc{{0\%}} & \perc{{0\%}} & \perc{{0\%}} & \perc{{33\%}} & \perc{{33\%}} & \perc{{33\%}}\\
    \cline{1-15}
  \end{tabular}

  \vspace{1em}
  \begin{tabular}{|c|cc|}
    \hline
    \multicolumn{3}{|c|}{\textbf{Pairs of fragments split by factorization}}\\
    \hline
    Total & Wrong & Correct\\
    26.0k & 24.3k (\perc{{93\%}}) & 1.7k (\perc{{7\%}})\\
    \hline
  \end{tabular}
  \begin{tabular}{|c|cc|c|}
    \hline
    \multicolumn{4}{|c|}{\textbf{CFG edges}}\\
    \hline
    Total & True & Fake & Drawn in GUI\\
    193.8k & 171.4k (\perc{{88\%}}) & 22.5k (\perc{{12\%}}) & 95.9k (\perc{{49\%}})\\
    \hline
  \end{tabular}
  \begin{tabular}{|c|c|}
    \hline
    \multicolumn{2}{|c|}{\textbf{Instructions}}\\
    \hline
    Total & Functionless\\
    731.6k & 189.7k (\perc{{26\%}})\\
    \hline
  \end{tabular}}}
    \vspace{0.25em}
    \subfloat[Potency metrics for a fully protected DRM with attacker-improved IDA Pro.\label{fig:DRMpotency_repartition}]{\parbox{\linewidth}{
      \centering
  \scriptsize
  \begin{tabular}{|l|c|ccc|ccc|c|ccc|ccc|}
    \cline{2-15}
    \multicolumn{1}{l|}{} & \multicolumn{7}{|c|}{\textbf{FP/FN CFG edges drawn in GUI}} & \multicolumn{7}{|c|}{\textbf{FP/FN CFG edges stored in database}}\\
    \cline{2-15}
    \multicolumn{1}{l|}{} & \textbf{Total} & \textbf{IA} & \textbf{IO} & \textbf{IF} & \textbf{iA} & \textbf{iO} & \textbf{iF} & \textbf{Total} & \textbf{IA} & \textbf{IO} & \textbf{IF} & \textbf{iA} & \textbf{iO} & \textbf{iF}\\
    \cline{1-15}
    \textbf{\# FP} & 17.4k & 9.3k & 13.5k & 13.6k & 8.2k & 3.9k & 3.8k & 21.5k & 11.6k & 17.5k & 17.7k & 9.9k & 4.0k & 3.9k\\
    \textbf{\perc{FPR}} & \perc{{78\%}} & \perc{{41\%}} & \perc{{60\%}} & \perc{{60\%}} & \perc{{36\%}} & \perc{{17\%}} & \perc{{17\%}} & \perc{{96\%}} & \perc{{52\%}} & \perc{{78\%}} & \perc{{78\%}} & \perc{{44\%}} & \perc{{18\%}} & \perc{{17\%}}\\
    \cline{1-15}
    \textbf{\# FN} & 69.7k & 16 & 469 & 555 & 69.7k & 69.3k & 69.2k & 25.1k & 0 & 5 & 6 & 25.1k & 25.1k & 25.1k\\
    \textbf{\perc{FNR}} & \perc{{41\%}} & \perc{{0\%}} & \perc{{0\%}} & \perc{{0\%}} & \perc{{41\%}} & \perc{{40\%}} & \perc{{40\%}} & \perc{{15\%}} & \perc{{0\%}} & \perc{{0\%}} & \perc{{0\%}} & \perc{{15\%}} & \perc{{15\%}} & \perc{{15\%}}\\
    \cline{1-15}
  \end{tabular}

  \vspace{1em}
  \begin{tabular}{|c|cc|}
    \hline
    \multicolumn{3}{|c|}{\textbf{Pairs of fragments split by factorization}}\\
    \hline
    Total & Wrong & Correct\\
    26.0k & 22.2k (\perc{{85\%}}) & 3.8k (\perc{{15\%}})\\
    \hline
  \end{tabular}
  \begin{tabular}{|c|cc|c|}
    \hline
    \multicolumn{4}{|c|}{\textbf{CFG edges}}\\
    \hline
    Total & True & Fake & Drawn in GUI\\
    194.3k & 171.8k (\perc{{88\%}}) & 22.5k (\perc{{12\%}}) & 119.5k (\perc{{62\%}})\\
    \hline
  \end{tabular}
  \begin{tabular}{|c|c|}
    \hline
    \multicolumn{2}{|c|}{\textbf{Instructions}}\\
    \hline
    Total & Functionless\\
    731.7k & 388 (\perc{{0\%}})\\
    \hline
  \end{tabular}}}
    \vspace{0.25em}
    \subfloat[Metrics for a fully protected DRM, after detection and removal of observable opaque predicates (soundish DB attack).\label{fig:DRMresilienceSound}]{\parbox{\linewidth}{
      \centering
  \scriptsize
  \begin{tabular}{|l|c|ccc|ccc|c|ccc|ccc|}
    \cline{2-15}
    \multicolumn{1}{l|}{} & \multicolumn{7}{|c|}{\textbf{FP/FN CFG edges drawn in GUI}} & \multicolumn{7}{|c|}{\textbf{FP/FN CFG edges stored in database}}\\
    \cline{2-15}
    \multicolumn{1}{l|}{} & \textbf{Total} & \textbf{IA} & \textbf{IO} & \textbf{IF} & \textbf{iA} & \textbf{iO} & \textbf{iF} & \textbf{Total} & \textbf{IA} & \textbf{IO} & \textbf{IF} & \textbf{iA} & \textbf{iO} & \textbf{iF}\\
    \cline{1-15}
    \textbf{\# FP} & 16.6k & 8.7k & 12.7k & 12.8k & 7.9k & 3.9k & 3.8k & 21.5k & 11.6k & 17.5k & 17.6k & 9.9k & 4.0k & 3.9k\\
    \textbf{\perc{FPR}} & \perc{{74\%}} & \perc{{39\%}} & \perc{{57\%}} & \perc{{57\%}} & \perc{{35\%}} & \perc{{17\%}} & \perc{{17\%}} & \perc{{95\%}} & \perc{{51\%}} & \perc{{78\%}} & \perc{{78\%}} & \perc{{44\%}} & \perc{{18\%}} & \perc{{17\%}}\\
    \cline{1-15}
    \textbf{\# FN} & 69.1k & 15 & 450 & 525 & 69.0k & 68.6k & 68.5k & 25.1k & 0 & 5 & 6 & 25.1k & 25.1k & 25.1k\\
    \textbf{\perc{FNR}} & \perc{{40\%}} & \perc{{0\%}} & \perc{{0\%}} & \perc{{0\%}} & \perc{{40\%}} & \perc{{40\%}} & \perc{{40\%}} & \perc{{15\%}} & \perc{{0\%}} & \perc{{0\%}} & \perc{{0\%}} & \perc{{15\%}} & \perc{{15\%}} & \perc{{15\%}}\\
    \cline{1-15}
  \end{tabular}

  \vspace{1em}
  \begin{tabular}{|c|cc|}
    \hline
    \multicolumn{3}{|c|}{\textbf{Pairs of fragments split by factorization}}\\
    \hline
    Total & Wrong & Correct\\
    26.0k & 22.2k (\perc{{85\%}}) & 3.8k (\perc{{15\%}})\\
    \hline
  \end{tabular}
  \begin{tabular}{|c|cc|c|}
    \hline
    \multicolumn{4}{|c|}{\textbf{CFG edges}}\\
    \hline
    Total & True & Fake & Drawn in GUI\\
    194.3k & 171.8k (\perc{{88\%}}) & 22.5k (\perc{{12\%}}) & 119.4k (\perc{{61\%}})\\
    \hline
  \end{tabular}
  \begin{tabular}{|c|c|}
    \hline
    \multicolumn{2}{|c|}{\textbf{Instructions}}\\
    \hline
    Total & Functionless\\
    731.7k & 388 (\perc{{0\%}})\\
    \hline
  \end{tabular}
  \begin{tabular}{|c|c|}
    \hline
    \multicolumn{2}{|c|}{\textbf{Opaque predicates}}\\
    \hline
    Total & Resolved\\
    14.2k & 62 (\perc{{0\%}})\\
    \hline
  \end{tabular}}}
    \vspace{0.25em}
    \subfloat[Metrics for a fully protected DRM, after detection and removal of observable opaque predicates (unsound GUI attack).\label{fig:DRMresilience}]{\parbox{\linewidth}{
    \input{table_DRM_IDAsmart_ATKsmart}}}
  \end{minipage}
  \caption{Measurement data for DRM}
  \label{resultsDRM}
\end{table*}
\else
\begin{table*}
\caption{Potency metrics for SLM without the protections proposed in this \papertechreport{}.}
\input{table_SLM_unprotected}
\label{fig:potency_unprotected}
\end{table*}

\begin{table*}
\caption{Potency metrics for a fully protected SLM with IDA Pro out-of-the-box.}
\input{table_SLM_IDAdumb_ATKdumb}
\label{fig:potency_vanilla}
\end{table*}

\begin{table*}
\caption{Potency metrics for a fully protected SLM with attacker-improved IDA Pro.}
\input{table_SLM_IDAsmart_ATKdumb}
\label{fig:potency_repartition}
\end{table*}
\fi

\subsection{Resilience}
\label{sec:eval:resilience}
To evaluate the resilience of the presented obfuscation, we analyze to what extent some attack techniques observed in empirical research~\cite{emse2019} and described in literature~\cite{yadegari} can bypass or undo the protections. Obviously, we cannot claim that the protections provide complete protection against attackers with unlimited resources and time. But we can demonstrate that at least some common attack strategies do not overcome the protection trivially.

First, we consider an attacker that can resolve opaque predicate computations when he observes their complete pattern in the code, either because he is good at recognizing them manually, or because he has a pattern matcher. We consider the attacker strong enough to identify opaque predicate computations even if they are mixed with other instructions, including (direct) control flow transfers. He is hence knowledgable, but he is also prudent: If he only observes part of an opaque predicate computation or observes that only part of the computation is guaranteed to be executed leading up to the conditional branch, he does not guess that it will be an opaque predicate with a certain outcome. In the empirical experiments reported by Ceccato et al.~\cite{emse2019}, attackers described how they manually eliminated the identified fake edges and how they could implement simple pattern matchers to automate that attack task.

To assess how far such an attacker might get in the worst case, we implemented a script that iteratively removes all fake edges of opaque predicates that such an attacker can resolve. The script does not need to detect the patterns of the opaque predicate computations itself, instead it gets the necessary information from the ground-truth logs produced by our obfuscator.

We developed two versions of the script. A first one mimicks an automated attack that considers the information in IDA Pro's database. So it observes all edges and all code identified by IDA Pro. We refer to this attack as the ``soundish'' attack, because it considers all available code and control flow. As IDA Pro might have missed some code and edges, it is not completely sound, but it is the closest to sound an automated tool based on IDA Pro disassembler results can get.

The second version of the script mimicks a manual, human attack that considers only the information displayed in the IDA Pro GUI. This attack is on the one hand weaker because it does not resolve opaque predicates of which IDA Pro put parts of the computations in two or more different functions, as those parts are then not shown to the attacker together. On the other hand, this attack is stronger in cases in which IDA Pro has put all the predicate computations in the same function, but in which it does not draw a fake edge that arrives into the middle of the computations, i.e., in which such a fake edge is a GUI TN. So this attacker will miss some opportunities, but he will also remove fake edges because other (fake) edges remain invisible to him. We refer to this attack as the ``unsound'' attack, because the attacker chooses to neglect information readily available in the IDA Pro database that an attacker trying to be sound would not have neglected. As each deleted fake edge can result in opportunities to improve the partitioning of the code into functions, the scripts also execute the repartitioning algorithm discussed in the previous paragraph to potentially improve IDA Pro's performance after every deletion of a fake edge.

The results for the soundish attack are shown in Table~\ref{fig:resilienceSound}, those for the unsound attack are shown in Table~\ref{fig:resilience}. To indicate to which extent the modeled attacker was able to resolve the opaque predicates, we report the number of inserted and resolved opaque predicates in the bottom right parts of the tables.

With the soundish attack, almost no (0--1\% for the other benchmarks) opaque predicates can be resolved. This demonstrates the effectiveness of the strategy to couple opaque predicates. %

With the unsound attack, about 22\% (20--22\% for the other benchmarks) of the opaque predicates can be resolved. In this scenario, a relatively large drop of about 10\% (9--11\% for the other benchmarks) for the number of drawn fake edges in the GUI is observed. Still, about 67\% of the fake edges remain. This is due to the coupling of opaque predicates in cycles, as discussed in Section~\ref{sec:resilience} and because of the addition of fake entries in the switch tables of the dispatchers. Here, too, the number of true edges that do not get drawn remains high.
While the attack has therefore weakened the confusion created by our obfuscations in the eyes of the attacker, he has not been able to remove it completely.
\if\techreport1
Similar results for the other benchmarks are available in Tables~\ref{fig:436resilienceSound}-\ref{fig:436resilience},~\ref{fig:445resilienceSound}-\ref{fig:445resilience},~\ref{fig:454resilienceSound}-\ref{fig:454resilience} and~\ref{fig:DRMresilienceSound}-\ref{fig:DRMresilience}.
\fi

\begin{figure}[t]
  \centering
  \resizebox{\linewidth}{!}{\input{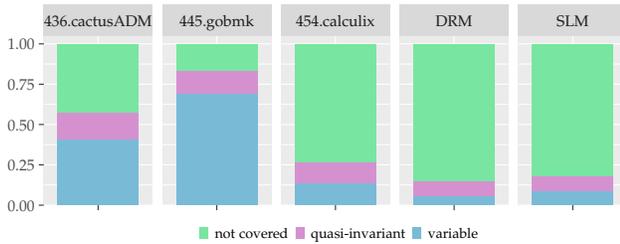}}
  \caption{Variability of dispatcher execution paths}
  \label{fig:dispatchervar}
\end{figure}

Regarding the resilience against the automated, ge\-ne\-ric deobfuscation technique of Yadegari et al.~\cite{yadegari}, we already noted in Section~\ref{sec:applicability} that the majority of covered dispatchers does not display quasi-invariant behavior. Figure~\ref{fig:dispatchervar} shows the fractions of the dispatchers that are not covered (i.e., not executed for our training inputs), feature quasi-invariant behavior (i.e., ``return'' to only one ``return site''), and show variable behavior (i.e., ``return'' to multiple ``return sites''). Note the correlation with the overall coverage numbers in Figure~\ref{fig:archives}. Obviously, if only a small percentage of the code is covered, and factoring is done on both covered and uncovered slices, only a small percentage of the dispatchers will be covered, let alone display variable behavior. Of those covered, between 39\% (DRM) and 83\% (445.gobmk) have variable behavior, and will hence not be simplified by the quasi-invariance based generic deobfuscation.

\if\techreport0
\begin{table*}
\caption{Metrics for a fully protected SLM, after detection and removal of observable opaque predicates (soundish DB attack).}
\input{table_SLM_IDAsmart_ATKsmart_DB}
\label{fig:resilienceSound}
\end{table*}

\begin{table*}
\caption{Metrics for a fully protected SLM, after detection and removal of observable opaque predicates (unsound GUI attack).}
\input{table_SLM_IDAsmart_ATKsmart}
\label{fig:resilience}
\end{table*}
\fi

\subsection{Overhead}
\label{sec:overhead}

Obfuscating transformations always come with performance and code size overhead. The performance penalty can be limited by using profile information to stay clear from the hottest code.
As we only proposed a new way to redirect fake edges of opaque predicates, rather than introduce new ones which require new code sequences to be injected, we do not evaluate the performance penalty of opaque predicate insertion.
Instead we focus on the proposed factoring technique, which can involve the insertion of considerable glue code, and which is hence expected to have a major impact on performance and code size.
\if \techreport1
To measure those impacts, we steered the selection process of our obfuscator tool to exclude a certain fraction of the hottest code, which we express in permille. A value of 0 includes even the hottest code, a value of 1000 excludes all the executed code. Figure~\ref{fig:overheadtime} shows the run time overhead which, as expected, decreases when more executed code is excluded from the selection process. Figure~\ref{fig:overheadsize} shows the overhead in code size which, as expected, drops when more code is excluded. A summary of the data in both figures can be found in Figure~\ref{fig:overheadscatter}.
\else
Those impacts are summarized in Figure~\ref{fig:overheadscatter}. Solid lines represent run time overhead, dashed lines code size overhead. More detailed results and descriptions of the experiments are available in a technical report~\cite{TR}.
\fi
The measured run times are averages of 5 runs. For the SPEC benchmarks, we used slightly altered reference inputs to reduce run times on the (relatively slow) developer boards; for the SLM benchmark we used a custom input; for the DRM benchmark we have no run time measurement as this is an interactive application. Each pair of dashed/solid lines on the chart corresponds to one benchmark. The different points denote different amounts of factoring, guided by profile information. To collect profile information, (standard) training inputs were used that in each case differ from the measurement inputs. The measured versions range from no covered code being factored (lower left points) to all code being factored (upper right points). In between, gradually more, hotter code (i.e., more frequently executed code) gets factored. It is clear that the overheads can become very large if the transformation is deployed blindly, but also that the overheads, in particularly the performance overhead can be easily reduced by excluding the hottest fragments from the factorization. To what extent a certain reduction limits the practical effectiveness of the protection of course depends on the software at hand. In any case, excluding all covered code cannot result in factored code dispatchers with variable behavior. So clearly one should be willing to accept some performance overhead. We do not consider this a big problem: All MATE protections inherently come with some overhead. Note that for the code size, the smallest overheads are still rather large because we only excluded the executed code. If program size is more important than performance, a better strategy would be to exclude non-executed fragments. Then much smaller size overheads can still be obtained.

\if \techreport1
\begin{figure}[t]
  \centering
  \resizebox{\linewidth}{!}{\input{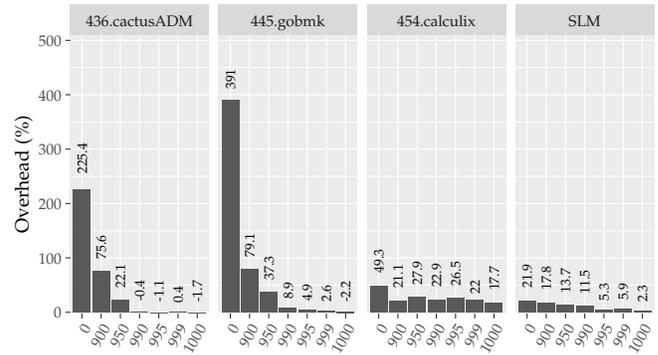}}
  \caption{Steering the factoring candidate selection process based on code hotness (run time overhead)}
  \label{fig:overheadtime}
\end{figure}
\begin{figure}[t]
  \centering
  \resizebox{\linewidth}{!}{\input{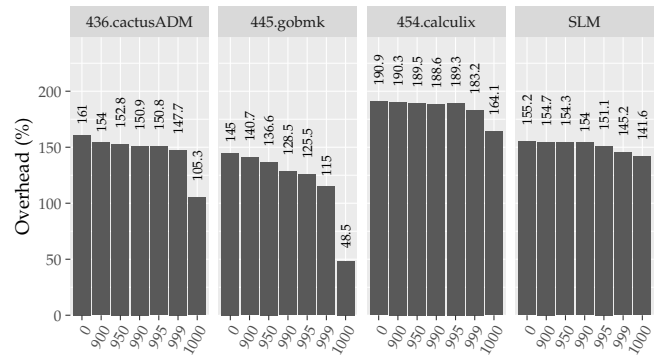}}
  \caption{Steering the factoring candidate selection process based on code hotness (code size overhead)}
  \label{fig:overheadsize}
\end{figure}
\fi

\begin{figure}[t]
  \centering
  \resizebox{\linewidth}{!}{\input{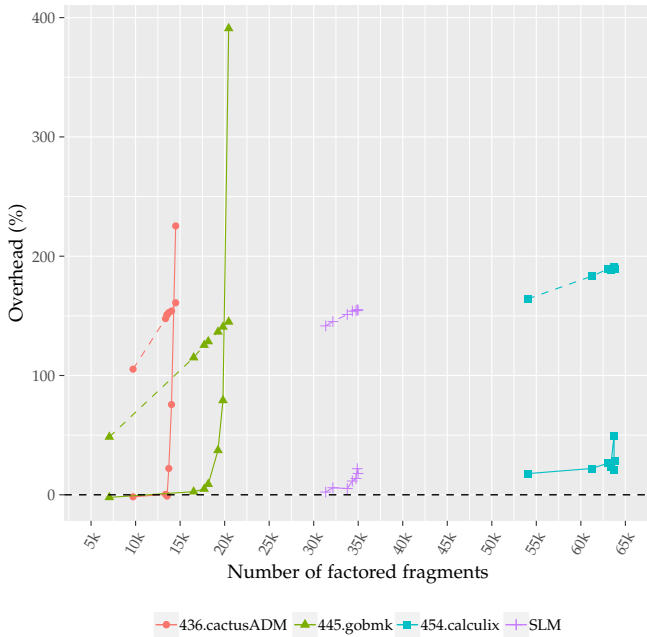}}
  \caption{Overhead versus factored code fragments}
  \label{fig:overheadscatter}
\end{figure}

\subsection{Sensitivity analysis}
\label{sec:sensitivity}
\if \techreport0
The opaque predicate insertion and factoring can be configured in many ways: the mixture of fake fall-through and fake branch-taken edges, amounts of fake edges in switch tables, use of different dispatchers, frequency of deployment, execution frequency threshold, priority function, cycle size of coupled protections and dispatchers, etc. A quantitative sensitivity analysis can be found in our technical report~\cite{TR}. Some major qualitative results are that:
\begin{itemize}
\item the false rates rise with more fake fall-through edges;
\item the FN rates increase with increasing cycle size until cycles of size 4. After that, the false negative rates stabilize;
\item the FP rates decrease with increasing cycle size.
\end{itemize}
\else
Our Diablo-based obfuscator can be configured with many parameters to steer its algorithms and to select which protections to apply. Table~\ref{tab:obfparams} lists the parameters that are most relevant for this \papertechreport{}, along with the default values we used for our measurements, i.e., unless explicitly stated otherwise.
Parameter 1 represents the seed for the main RNG from which random numbers are generated and other RNGs are seeded. Changing its value will steer the pseudo-random processes into making other decisions. For example, other code fragments will be selected for factorisation or opaque predicates will be inserted at other locations.
Parameter 2 is the random seed used to randomize the code layout. Changing this value only affects the code layout ordering of the protected program.
Parameter 3 denotes the probability for a basic block to have an opaque predicate with conditional branch injected into it. The parameter hence controls how many opaque predicates will be present in the protected program.  Parameters 4 and 5 control the coupled opaque predicate cycle size and creation probability, respectively. Predicates are grouped according to the cycle size, and then for each group, a probabilistic decision is taken to put them in an actual cycle or not.
Parameter 6 is the probability of fake edges being the fall-though edges out of conditional branches, rather than the branch-taken edges.
Parameters 7 to 8 control our code factoring technique. Parameter 7 steers the number of fake entries added to switch-based dispatchers. Parameter 8 controls how many of all actual factoring opportunities get factored eventually, by controlling the probability with which each considered actual candidate set is factored or skipped.
One or more dispatcher types can be (de)activated by tuning parameter 9.
Parameter 10 allows to trade-off efficiency and effectiveness, by controlling how much of the hottest code is excluded from factoring.
Finally, parameter 11 enables or disables Diablo's the branch flipping transformation, which randomly flips the condition code of conditional branches, together with the corresponding fall-through and branch-taken edges.

\begin{table}[t]
  \centering
  \begin{tabular}{|l|c|c|}
    \hline
    \textbf{Parameter} & \textbf{Value}\\
    \hline
    1.~
    main RNG seed & \texttt{0xDEADDEADDEADDEAD}\\\hline
    2.~
    code layout random seed & 1\\\hline
    3.~
    predicate insertion probability & 20\%\\\hline
    4.~
    fake edge cycle size & 4\\\hline
    5.~
    fake edge cycle insertion probability & 100\%\\\hline
    6.~
    fake edge fall-through probability & 50\%\\\hline
    7.~
    fake table entry probability & 30\%\\\hline
    8.~
    factoring transformation probability & 100\%\\\hline
    9.~
    enabled dispatchers & all types\\\hline
    10.~
    fraction of hottest code to skip & 0\%\\\hline
    11.~
    branch flipping & on\\\hline
  \end{tabular}
  \caption{Default values for the most important obfuscator parameters}
  \label{tab:obfparams}
\end{table}

We evaluated the effect of each parameter by plotting the false rates (FPR and FNR) for each benchmark, where we assume the default values in Table~\ref{tab:obfparams} for all but the evaluated parameter. Unless mentioned otherwise, only one parameter has its value swept in each evaluation. In each plot, the primary axis (left) is the axis for the false rates. As the false rates are relative to the absolute number of fake and true edges, and as these numbers change with most parameters, we also plot those on a secondary vertical axis (right). To put things into perspective, we discuss the false rates for a protected program that has been analysed with an attacker-improved IDA Pro, where we consider the false rates before and after the program has been attacked with our scripted attack discussed in Section~\ref{sec:eval:resilience}. Hence, the plotted results for the SLM correspond to the data presented in Table~\ref{fig:potency_repartition} (before the attack), Table~\ref{fig:resilienceSound} (after the soundish attack) and Table~\ref{fig:resilience} (after the unsound attack). To keep the plots simple, we only draw the database-based false rates for the soundish attack. Similarly, we only draw the GUI-based false rates for the unsound attack. As the results and trends are very similar for each benchmark, we discuss only the numerical values for the 436.cactusADM benchmark in the text (upper-left plot).

In general, the following conclusions can be made for all parameters: Because the GUI does not draw all edges (especially interprocedural ones, in the context of the partitioning of IDA Pro), the FPR for the database will always be higher than the one for the GUI, at least for the unattacked programs. Similarly, the FNR for the database will always be lower than that for the GUI.

To analyse the effects of parameter 1, we set its value to 10 arbitrarily chosen values. We observed that the false rates hardly vary: the standard deviations measured for the different false rates and benchmarks over the 10 PRNG seeds never exceed 0.7\%. For parameter 2, similar results were obtained (0.6\%). This independence of the obtained results from the used random seeds is important in practice, because it makes the outcome of using the tool, and in particular of the tuning process of the other parameters on a specific case at hand, more predictable.

\begin{figure*}[t]
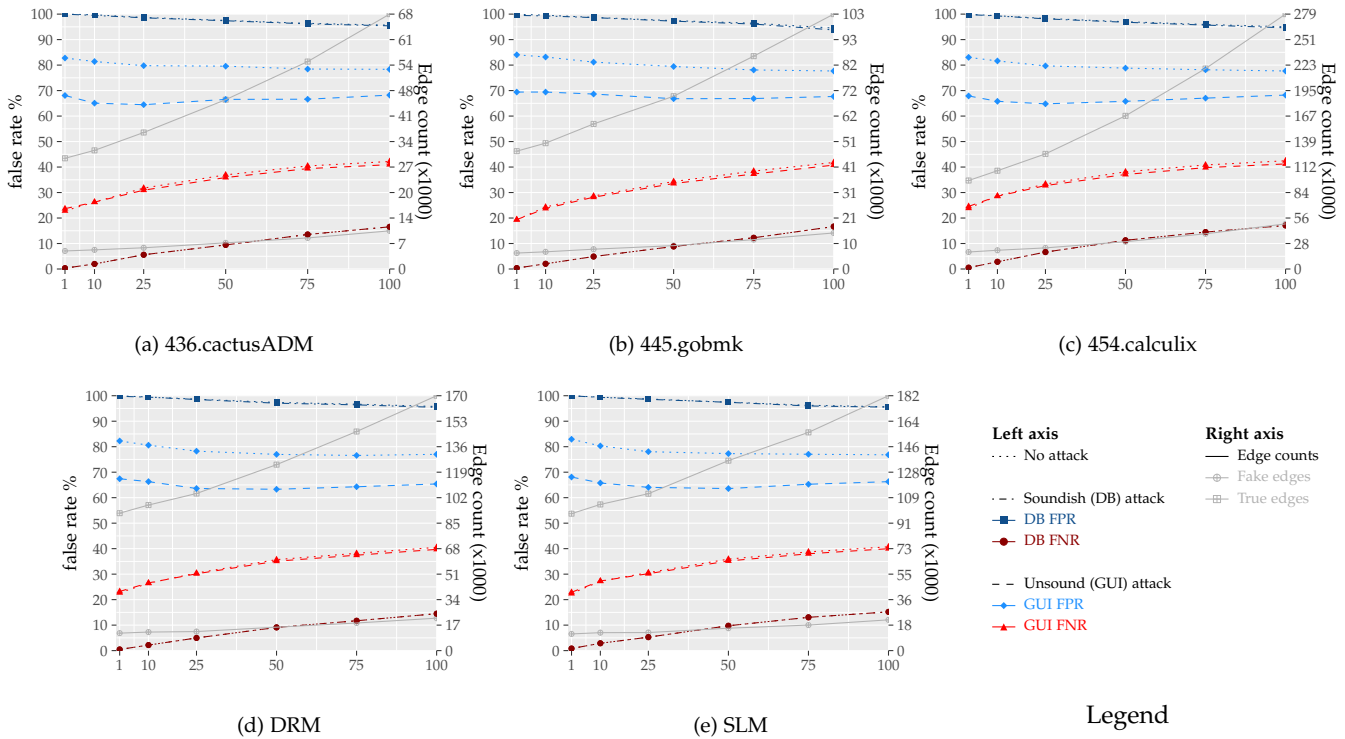

  \centering
  \subfloat[436.cactusADM]
    {\resizebox{0.33\linewidth}{!}{\input{sensitivity_436_afchance}}}
  \subfloat[445.gobmk]
    {\resizebox{0.33\linewidth}{!}{\input{sensitivity_445_afchance}}}
  \subfloat[454.calculix]
    {\resizebox{0.33\linewidth}{!}{\input{sensitivity_454_afchance}}}
  \\
  \subfloat[DRM]
    {\resizebox{0.33\linewidth}{!}{\input{sensitivity_DRM_afchance}}}
  \subfloat[SLM]
    {\resizebox{0.33\linewidth}{!}{\input{sensitivity_SLM_afchance}}}
  \subfloat{\stackunder[0em]{\resizebox{0.25\linewidth}{!}{\relax
\begin{tikzpicture}[x=1pt,y=1pt]

\newcommand{\Loffset}{20}
\newcommand{\Roffset}{140}
\newcommand{\length}{13}
\newcommand{\hlength}{6.5}
\newcommand{\hmarker}{2}
\newcommand{\plus}[2]{\the\numexpr #1 + #2}
\newcommand{\plusT}[3]{\the\numexpr #1 + #2 + #3}
\newcommand{\LToffset}{\plus{\plus{\Loffset}{\length}}{5}}
\newcommand{\RToffset}{\plus{\plus{\Roffset}{\length}}{5}}
\newcommand{\VToffset}[1]{\the\numexpr 150 - #1 * 12}
\newcommand{\VMoffset}[1]{\VToffset{#1} + 2}

\definecolor{fillColor}{RGB}{255,0,0}
\path[use as bounding box,fill=fillColor,fill opacity=0.00] (0,0) rectangle (190,180.67);

\begin{scope}
	\begin{scope}
		\definecolor{drawColor}{RGB}{0,0,0}
		\node[text=drawColor,anchor=base west,inner sep=0pt, outer sep=0pt, scale=  1.00] at (\Loffset,\VToffset{0}) {\textbf{Left axis}};
	\end{scope}

	\begin{scope}
		\definecolor{drawColor}{RGB}{0,0,0}
		\node[text=drawColor,anchor=base west,inner sep=0pt, outer sep=0pt, scale=  1.00] at (\Roffset,\VToffset{0}) {\textbf{Right axis}};
	\end{scope}
\end{scope}

\begin{scope}
	\begin{scope}
		\definecolor{drawColor}{RGB}{0,0,0}
		\path[draw=drawColor,line width= 0.6pt,dash pattern=on 1pt off 3pt ,line join=round] (\Loffset, \VMoffset{1}) -- (\plus{\Loffset}{\length}, \VMoffset{1});
	\end{scope}

	\begin{scope}
		\definecolor{drawColor}{RGB}{0,0,0}
		\node[text=drawColor,anchor=base west,inner sep=0pt, outer sep=0pt, scale=  0.88] at (\LToffset, \VToffset{1}) {No attack};
	\end{scope}
\end{scope}

\begin{scope}
	\begin{scope}
		\definecolor{drawColor}{RGB}{0,0,0}
		\path[draw=drawColor,line width= 0.6pt,dash pattern=on 1pt off 3pt on 4pt off 3pt ,line join=round] (\Loffset, \VMoffset{3}) -- (\plus{\Loffset}{\length}, \VMoffset{3});
	\end{scope}

	\begin{scope}
		\definecolor{drawColor}{RGB}{0,0,0}
			\node[text=drawColor,anchor=base west,inner sep=0pt, outer sep=0pt, scale=  0.88] at (\LToffset, \VToffset{3}) {Soundish (DB) attack};
	\end{scope}
\end{scope}

\begin{scope}
	\begin{scope}
		\definecolor{fillColor}{RGB}{16,78,139}
		\path[fill=fillColor] (\plusT{\Loffset}{\hlength}{-\hmarker},\plus{\VMoffset{4}}{-\hmarker}) --
			(\plusT{\Loffset}{\hlength}{\hmarker},\plus{\VMoffset{4}}{-\hmarker}) --
			(\plusT{\Loffset}{\hlength}{\hmarker},\plus{\VMoffset{4}}{\hmarker}) --
			(\plusT{\Loffset}{\hlength}{-\hmarker},\plus{\VMoffset{4}}{\hmarker}) --
			cycle;
	\end{scope}

	\begin{scope}
		\definecolor{drawColor}{RGB}{16,78,139}
		\path[draw=drawColor,line width= 0.6pt,line join=round] (\Loffset, \VMoffset{4}) -- (\plus{\Loffset}{\length}, \VMoffset{4});
	\end{scope}

	\begin{scope}
		\definecolor{drawColor}{RGB}{16,78,139}
		\node[text=drawColor,anchor=base west,inner sep=0pt, outer sep=0pt, scale=  0.88] at (\LToffset,\VToffset{4}) {DB FPR};
	\end{scope}
\end{scope}

\begin{scope}
	\begin{scope}
		\definecolor{drawColor}{RGB}{139,0,0}
		\definecolor{fillColor}{RGB}{139,0,0}
		\path[draw=drawColor,line width= 0.4pt,line join=round,line cap=round,fill=fillColor] (\plus{\Loffset}{\hlength},\VMoffset{5}) circle (\hmarker);
	\end{scope}

	\begin{scope}
		\definecolor{drawColor}{RGB}{139,0,0}
		\path[draw=drawColor,line width= 0.6pt,line join=round] (\Loffset,\VMoffset{5}) -- (\plus{\Loffset}{\length},\VMoffset{5});
	\end{scope}

	\begin{scope}
		\definecolor{drawColor}{RGB}{139,0,0}
		\node[text=drawColor,anchor=base west,inner sep=0pt, outer sep=0pt, scale=  0.88] at (\LToffset,\VToffset{5}) {DB FNR};
	\end{scope}
\end{scope}

\begin{scope}
	\begin{scope}
		\definecolor{drawColor}{RGB}{0,0,0}
		\path[draw=drawColor,line width= 0.6pt,dash pattern=on 4pt off 4pt ,line join=round] (\Loffset, \VMoffset{7}) -- (\plus{\Loffset}{\length}, \VMoffset{7});
	\end{scope}

	\begin{scope}
		\definecolor{drawColor}{RGB}{0,0,0}
			\node[text=drawColor,anchor=base west,inner sep=0pt, outer sep=0pt, scale=  0.88] at (\LToffset, \VToffset{7}) {Unsound (GUI) attack};
	\end{scope}
\end{scope}

\begin{scope}
	\begin{scope}
		\definecolor{fillColor}{RGB}{30,144,255}
		\path[fill=fillColor] (\plusT{\Loffset}{\hlength}{-\hmarker}, \VMoffset{8}) --
			(\plus{\Loffset}{\hlength},\plus{\VMoffset{8}}{\hmarker}) --
			(\plusT{\Loffset}{\hlength}{\hmarker}, \VMoffset{8}) --
			(\plus{\Loffset}{\hlength}, \plus{\VMoffset{8}}{-\hmarker}) --
			cycle;
	\end{scope}

	\begin{scope}
		\definecolor{drawColor}{RGB}{30,144,255}
		\path[draw=drawColor,line width= 0.6pt,line join=round] (\Loffset, \VMoffset{8}) -- (\plus{\Loffset}{\length}, \VMoffset{8});
	\end{scope}

	\begin{scope}
		\definecolor{drawColor}{RGB}{30,144,255}
		\node[text=drawColor,anchor=base west,inner sep=0pt, outer sep=0pt, scale=  0.88] at (\LToffset, \VToffset{8}) {GUI FPR};
	\end{scope}
\end{scope}

\begin{scope}
	\begin{scope}
		\definecolor{fillColor}{RGB}{255,0,0}
		\path[fill=fillColor] (\plus{\Loffset}{\hlength},\plus{\VMoffset{9}}{\hmarker}) --
			(\plusT{\Loffset}{\hlength}{\hmarker},\plus{\VMoffset{9}}{-\hmarker}) --
			(\plusT{\Loffset}{\hlength}{-\hmarker},\plus{\VMoffset{9}}{-\hmarker}) --
			cycle;
	\end{scope}

	\begin{scope}
		\definecolor{drawColor}{RGB}{255,0,0}
		\path[draw=drawColor,line width= 0.6pt,line join=round] (\Loffset,\VMoffset{9}) -- (\plus{\Loffset}{\length},\VMoffset{9});
	\end{scope}

	\begin{scope}
		\definecolor{drawColor}{RGB}{255,0,0}
		\node[text=drawColor,anchor=base west,inner sep=0pt, outer sep=0pt, scale=  0.88] at (\LToffset,\VToffset{9}) {GUI FNR};
	\end{scope}
\end{scope}

\begin{scope}
	\begin{scope}
		\definecolor{drawColor}{gray}{0.70}
		\path[draw=drawColor,line width= 0.4pt,line join=round,line cap=round] (\plus{\Roffset}{\hlength},\VMoffset{2}) circle (\hmarker);
		\path[draw=drawColor,line width= 0.4pt,line join=round,line cap=round] (\plusT{\Roffset}{\hlength}{-\hmarker},\VMoffset{2}) -- (\plusT{\Roffset}{\hlength}{\hmarker},\VMoffset{2});
		\path[draw=drawColor,line width= 0.4pt,line join=round,line cap=round] (\plus{\Roffset}{\hlength},\plus{\VMoffset{2}}{-\hmarker}) -- (\plus{\Roffset}{\hlength},\plus{\VMoffset{2}}{\hmarker});
	\end{scope}

	\begin{scope}
		\definecolor{drawColor}{gray}{0.70}
		\path[draw=drawColor,line width= 0.6pt,line join=round] (\Roffset,\VMoffset{2}) -- (\plus{\Roffset}{\length},\VMoffset{2});
	\end{scope}

	\begin{scope}
		\definecolor{drawColor}{gray}{0.70}
		\node[text=drawColor,anchor=base west,inner sep=0pt, outer sep=0pt, scale=  0.88] at (\RToffset,\VToffset{2}) {Fake edges};
	\end{scope}
\end{scope}

\begin{scope}
	\begin{scope}
		\definecolor{drawColor}{gray}{0.70}
		\path[draw=drawColor,line width= 0.4pt,line join=round,line cap=round] (\plusT{\Roffset}{\hlength}{-\hmarker},\VMoffset{3}) -- (\plusT{\Roffset}{\hlength}{\hmarker},\VMoffset{3});
		\path[draw=drawColor,line width= 0.4pt,line join=round,line cap=round] (\plus{\Roffset}{\hlength},\plus{\VMoffset{3}}{-\hmarker}) -- (\plus{\Roffset}{\hlength},\plus{\VMoffset{3}}{\hmarker});
		\path[draw=drawColor,line width= 0.4pt,line join=round,line cap=round] (\plusT{\Roffset}{\hlength}{-\hmarker},\plus{\VMoffset{3}}{-\hmarker}) rectangle (\plusT{\Roffset}{\hlength}{\hmarker},\plus{\VMoffset{3}}{\hmarker});
	\end{scope}

	\begin{scope}
		\definecolor{drawColor}{gray}{0.70}
		\path[draw=drawColor,line width= 0.6pt,line join=round] (\Roffset,\VMoffset{3}) -- (\plus{\Roffset}{\length},\VMoffset{3});
	\end{scope}

	\begin{scope}
		\definecolor{drawColor}{gray}{0.70}
		\node[text=drawColor,anchor=base west,inner sep=0pt, outer sep=0pt, scale=  0.88] at (\RToffset,\VToffset{3}) {True edges};
	\end{scope}
\end{scope}

\begin{scope}
	\begin{scope}
		\definecolor{drawColor}{RGB}{0,0,0}
		\path[draw=drawColor,line width= 0.6pt,line join=round] (\Roffset, \VMoffset{1}) -- (\plus{\Roffset}{\length}, \VMoffset{1});
	\end{scope}

	\begin{scope}
		\definecolor{drawColor}{RGB}{0,0,0}
		\node[text=drawColor,anchor=base west,inner sep=0pt, outer sep=0pt, scale=  0.88] at (\RToffset, \VToffset{1}) {Edge counts};
	\end{scope}
\end{scope}

\end{tikzpicture}}}{Legend}}

  \caption{Sensitivity analysis on the factoring transformation probability (X-axis)}
  \label{fig:sense_afchance}
\end{figure*}
Figure~\ref{fig:sense_afchance} shows the results when we evaluate the sensitivity of the probability to transform an actual factoring opportunity. For this parameter, we assumed values of 1\%, 10\%, 25\%, 50\%, 75\%, and 100\%.
\begin{itemize}
  \item \textbf{Before attacking} the program, the \dbfpr{} decreases from 100\% to about 95\%, while the \dbfnr{} increases from 0\% to about 17\%. When little factoring is applied, the 20\% opaque insertion is their main cause. The numbers evolve with more factoring because IDA Pro incorrectly handles some dispatcher constructs. The \guifpr{} slightly drops from about 85\% to about 80\%.  The \guifnr{} increases from about 22\% to about 40\% for reasons similar to those for the \dbfnr{}. This false rate is higher, however, as IDA Pro's GUI does not draw interprocedural edges.
  \item Performing the \textbf{soundish attack} does not change the false rates much.
  \item After the \textbf{unsound attack}, the \guifpr{} remains more or less constant at about 65-70\%. The reason is that the attack eliminates approximately the same number of fake opaque predicate edges, independently of the fake switch table edges, because the latter are most often not drawn in the GUI. The \guifnr{} does not change much compared to the unattacked program.
\end{itemize}

Figure~\ref{fig:sense_cyclechance} shows the results for the sensitivity analysis of the probability to insert fake edge cycles, for which we assumed values of 0\%, 25\%, 50\%, 75\%, and 100\%. To prevent the created cycles from being interrupted due to incorrectly analysed switch dispatchers, we only enabled the conditional jump dispatchers here.
\begin{itemize}
  \item \textbf{Before attacking} the program, the \dbfpr{} remains constant at 100\%, the \dbfnr{} at 0\%. IDA Pro identifies almost all instructions in this case, because we injected no switch-based dispatchers as stated above. The \guifpr{} increases linearly from about 60\% to about 75\% because of the interplay between IDA PRo's heuristics to combine code into functions, and the fact that creating more cycles in the opaque predicate computations lowers the fraction of fake edges that can be implemented as fall-through paths. By contrast, the \guifnr{} remains flat at about 30\% because the total number of opaque predicates stays the same in this experiment.
  \item After the \textbf{soundish attack}, the \dbfpr{} increases linearly from about 5\% to 100\%, which indicates that the coupling of opaque predicates works as expected. As this attack only focuses at resolving opaque predicates, the \dbfnr{} does not change.
  \item After performing the \textbf{unsound attack}, the \guifpr{} increases from about 15\% to about 60\%. The \guifnr{} does not change much, dropping by about 2\%.
\end{itemize}
Notice that the \dbfpr{} and the \guifpr{} cross at a probability of about 25\%. For low cycle creation probabilities, the unsound attack performs worse than the soundish attack due to the splitting of opaque predicate calculations over functions, as explained in Section~\ref{sec:resilience}. For higher probabilities, the number of not drawn fake edges dominates, so then the soundish attack performs worse.
\begin{figure*}[t]
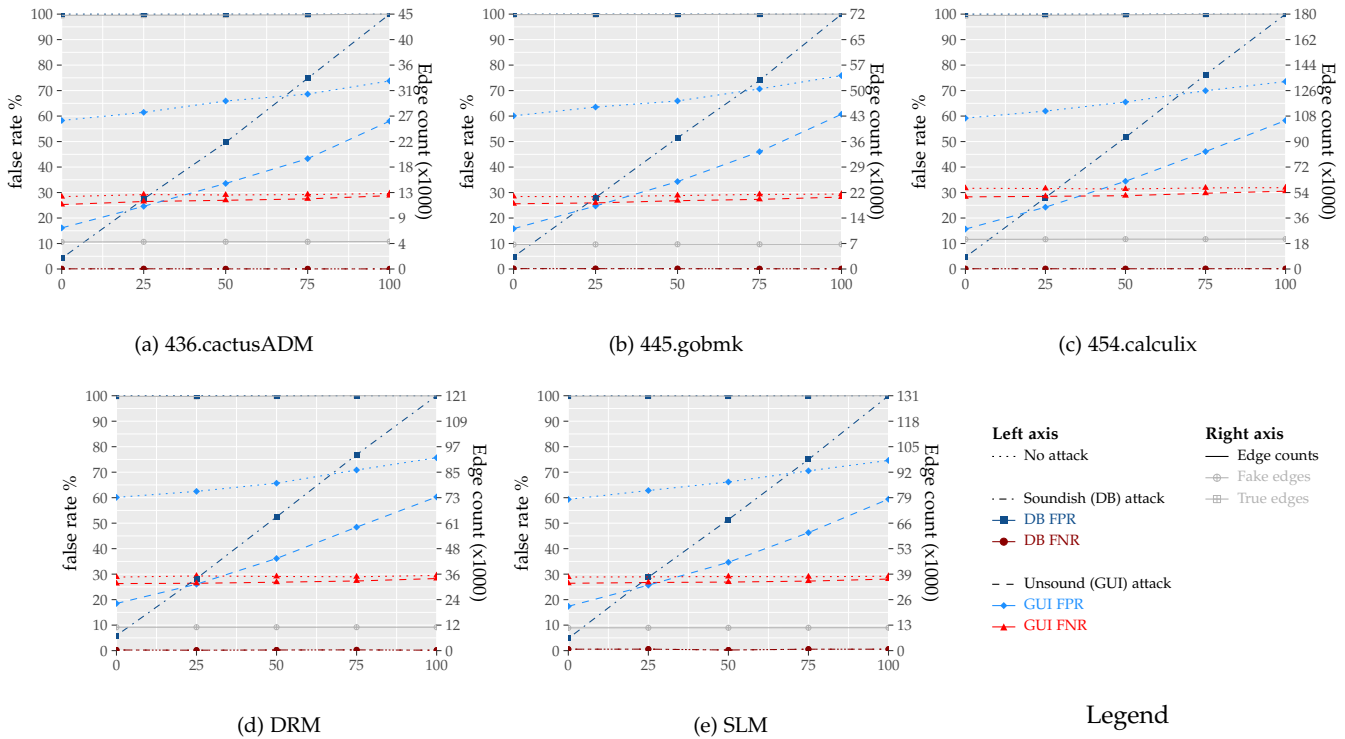

  \centering
  \subfloat[436.cactusADM]
    {\resizebox{0.33\linewidth}{!}{\input{sensitivity_436_clusterchance}}}
  \subfloat[445.gobmk]
    {\resizebox{0.33\linewidth}{!}{\input{sensitivity_445_clusterchance}}}
  \subfloat[454.calculix]
    {\resizebox{0.33\linewidth}{!}{\input{sensitivity_454_clusterchance}}}
  \\
  \subfloat[DRM]
    {\resizebox{0.33\linewidth}{!}{\input{sensitivity_DRM_clusterchance}}}
  \subfloat[SLM]
    {\resizebox{0.33\linewidth}{!}{\input{sensitivity_SLM_clusterchance}}}
  \subfloat{\stackunder[0em]{\resizebox{0.25\linewidth}{!}{\relax
\begin{tikzpicture}[x=1pt,y=1pt]

\newcommand{\Loffset}{20}
\newcommand{\Roffset}{140}
\newcommand{\length}{13}
\newcommand{\hlength}{6.5}
\newcommand{\hmarker}{2}
\newcommand{\plus}[2]{\the\numexpr #1 + #2}
\newcommand{\plusT}[3]{\the\numexpr #1 + #2 + #3}
\newcommand{\LToffset}{\plus{\plus{\Loffset}{\length}}{5}}
\newcommand{\RToffset}{\plus{\plus{\Roffset}{\length}}{5}}
\newcommand{\VToffset}[1]{\the\numexpr 150 - #1 * 12}
\newcommand{\VMoffset}[1]{\VToffset{#1} + 2}

\definecolor{fillColor}{RGB}{255,0,0}
\path[use as bounding box,fill=fillColor,fill opacity=0.00] (0,0) rectangle (190,180.67);

\begin{scope}
	\begin{scope}
		\definecolor{drawColor}{RGB}{0,0,0}
		\node[text=drawColor,anchor=base west,inner sep=0pt, outer sep=0pt, scale=  1.00] at (\Loffset,\VToffset{0}) {\textbf{Left axis}};
	\end{scope}

	\begin{scope}
		\definecolor{drawColor}{RGB}{0,0,0}
		\node[text=drawColor,anchor=base west,inner sep=0pt, outer sep=0pt, scale=  1.00] at (\Roffset,\VToffset{0}) {\textbf{Right axis}};
	\end{scope}
\end{scope}

\begin{scope}
	\begin{scope}
		\definecolor{drawColor}{RGB}{0,0,0}
		\path[draw=drawColor,line width= 0.6pt,dash pattern=on 1pt off 3pt ,line join=round] (\Loffset, \VMoffset{1}) -- (\plus{\Loffset}{\length}, \VMoffset{1});
	\end{scope}

	\begin{scope}
		\definecolor{drawColor}{RGB}{0,0,0}
		\node[text=drawColor,anchor=base west,inner sep=0pt, outer sep=0pt, scale=  0.88] at (\LToffset, \VToffset{1}) {No attack};
	\end{scope}
\end{scope}

\begin{scope}
	\begin{scope}
		\definecolor{drawColor}{RGB}{0,0,0}
		\path[draw=drawColor,line width= 0.6pt,dash pattern=on 1pt off 3pt on 4pt off 3pt ,line join=round] (\Loffset, \VMoffset{3}) -- (\plus{\Loffset}{\length}, \VMoffset{3});
	\end{scope}

	\begin{scope}
		\definecolor{drawColor}{RGB}{0,0,0}
			\node[text=drawColor,anchor=base west,inner sep=0pt, outer sep=0pt, scale=  0.88] at (\LToffset, \VToffset{3}) {Soundish (DB) attack};
	\end{scope}
\end{scope}

\begin{scope}
	\begin{scope}
		\definecolor{fillColor}{RGB}{16,78,139}
		\path[fill=fillColor] (\plusT{\Loffset}{\hlength}{-\hmarker},\plus{\VMoffset{4}}{-\hmarker}) --
			(\plusT{\Loffset}{\hlength}{\hmarker},\plus{\VMoffset{4}}{-\hmarker}) --
			(\plusT{\Loffset}{\hlength}{\hmarker},\plus{\VMoffset{4}}{\hmarker}) --
			(\plusT{\Loffset}{\hlength}{-\hmarker},\plus{\VMoffset{4}}{\hmarker}) --
			cycle;
	\end{scope}

	\begin{scope}
		\definecolor{drawColor}{RGB}{16,78,139}
		\path[draw=drawColor,line width= 0.6pt,line join=round] (\Loffset, \VMoffset{4}) -- (\plus{\Loffset}{\length}, \VMoffset{4});
	\end{scope}

	\begin{scope}
		\definecolor{drawColor}{RGB}{16,78,139}
		\node[text=drawColor,anchor=base west,inner sep=0pt, outer sep=0pt, scale=  0.88] at (\LToffset,\VToffset{4}) {DB FPR};
	\end{scope}
\end{scope}

\begin{scope}
	\begin{scope}
		\definecolor{drawColor}{RGB}{139,0,0}
		\definecolor{fillColor}{RGB}{139,0,0}
		\path[draw=drawColor,line width= 0.4pt,line join=round,line cap=round,fill=fillColor] (\plus{\Loffset}{\hlength},\VMoffset{5}) circle (\hmarker);
	\end{scope}

	\begin{scope}
		\definecolor{drawColor}{RGB}{139,0,0}
		\path[draw=drawColor,line width= 0.6pt,line join=round] (\Loffset,\VMoffset{5}) -- (\plus{\Loffset}{\length},\VMoffset{5});
	\end{scope}

	\begin{scope}
		\definecolor{drawColor}{RGB}{139,0,0}
		\node[text=drawColor,anchor=base west,inner sep=0pt, outer sep=0pt, scale=  0.88] at (\LToffset,\VToffset{5}) {DB FNR};
	\end{scope}
\end{scope}

\begin{scope}
	\begin{scope}
		\definecolor{drawColor}{RGB}{0,0,0}
		\path[draw=drawColor,line width= 0.6pt,dash pattern=on 4pt off 4pt ,line join=round] (\Loffset, \VMoffset{7}) -- (\plus{\Loffset}{\length}, \VMoffset{7});
	\end{scope}

	\begin{scope}
		\definecolor{drawColor}{RGB}{0,0,0}
			\node[text=drawColor,anchor=base west,inner sep=0pt, outer sep=0pt, scale=  0.88] at (\LToffset, \VToffset{7}) {Unsound (GUI) attack};
	\end{scope}
\end{scope}

\begin{scope}
	\begin{scope}
		\definecolor{fillColor}{RGB}{30,144,255}
		\path[fill=fillColor] (\plusT{\Loffset}{\hlength}{-\hmarker}, \VMoffset{8}) --
			(\plus{\Loffset}{\hlength},\plus{\VMoffset{8}}{\hmarker}) --
			(\plusT{\Loffset}{\hlength}{\hmarker}, \VMoffset{8}) --
			(\plus{\Loffset}{\hlength}, \plus{\VMoffset{8}}{-\hmarker}) --
			cycle;
	\end{scope}

	\begin{scope}
		\definecolor{drawColor}{RGB}{30,144,255}
		\path[draw=drawColor,line width= 0.6pt,line join=round] (\Loffset, \VMoffset{8}) -- (\plus{\Loffset}{\length}, \VMoffset{8});
	\end{scope}

	\begin{scope}
		\definecolor{drawColor}{RGB}{30,144,255}
		\node[text=drawColor,anchor=base west,inner sep=0pt, outer sep=0pt, scale=  0.88] at (\LToffset, \VToffset{8}) {GUI FPR};
	\end{scope}
\end{scope}

\begin{scope}
	\begin{scope}
		\definecolor{fillColor}{RGB}{255,0,0}
		\path[fill=fillColor] (\plus{\Loffset}{\hlength},\plus{\VMoffset{9}}{\hmarker}) --
			(\plusT{\Loffset}{\hlength}{\hmarker},\plus{\VMoffset{9}}{-\hmarker}) --
			(\plusT{\Loffset}{\hlength}{-\hmarker},\plus{\VMoffset{9}}{-\hmarker}) --
			cycle;
	\end{scope}

	\begin{scope}
		\definecolor{drawColor}{RGB}{255,0,0}
		\path[draw=drawColor,line width= 0.6pt,line join=round] (\Loffset,\VMoffset{9}) -- (\plus{\Loffset}{\length},\VMoffset{9});
	\end{scope}

	\begin{scope}
		\definecolor{drawColor}{RGB}{255,0,0}
		\node[text=drawColor,anchor=base west,inner sep=0pt, outer sep=0pt, scale=  0.88] at (\LToffset,\VToffset{9}) {GUI FNR};
	\end{scope}
\end{scope}

\begin{scope}
	\begin{scope}
		\definecolor{drawColor}{gray}{0.70}
		\path[draw=drawColor,line width= 0.4pt,line join=round,line cap=round] (\plus{\Roffset}{\hlength},\VMoffset{2}) circle (\hmarker);
		\path[draw=drawColor,line width= 0.4pt,line join=round,line cap=round] (\plusT{\Roffset}{\hlength}{-\hmarker},\VMoffset{2}) -- (\plusT{\Roffset}{\hlength}{\hmarker},\VMoffset{2});
		\path[draw=drawColor,line width= 0.4pt,line join=round,line cap=round] (\plus{\Roffset}{\hlength},\plus{\VMoffset{2}}{-\hmarker}) -- (\plus{\Roffset}{\hlength},\plus{\VMoffset{2}}{\hmarker});
	\end{scope}

	\begin{scope}
		\definecolor{drawColor}{gray}{0.70}
		\path[draw=drawColor,line width= 0.6pt,line join=round] (\Roffset,\VMoffset{2}) -- (\plus{\Roffset}{\length},\VMoffset{2});
	\end{scope}

	\begin{scope}
		\definecolor{drawColor}{gray}{0.70}
		\node[text=drawColor,anchor=base west,inner sep=0pt, outer sep=0pt, scale=  0.88] at (\RToffset,\VToffset{2}) {Fake edges};
	\end{scope}
\end{scope}

\begin{scope}
	\begin{scope}
		\definecolor{drawColor}{gray}{0.70}
		\path[draw=drawColor,line width= 0.4pt,line join=round,line cap=round] (\plusT{\Roffset}{\hlength}{-\hmarker},\VMoffset{3}) -- (\plusT{\Roffset}{\hlength}{\hmarker},\VMoffset{3});
		\path[draw=drawColor,line width= 0.4pt,line join=round,line cap=round] (\plus{\Roffset}{\hlength},\plus{\VMoffset{3}}{-\hmarker}) -- (\plus{\Roffset}{\hlength},\plus{\VMoffset{3}}{\hmarker});
		\path[draw=drawColor,line width= 0.4pt,line join=round,line cap=round] (\plusT{\Roffset}{\hlength}{-\hmarker},\plus{\VMoffset{3}}{-\hmarker}) rectangle (\plusT{\Roffset}{\hlength}{\hmarker},\plus{\VMoffset{3}}{\hmarker});
	\end{scope}

	\begin{scope}
		\definecolor{drawColor}{gray}{0.70}
		\path[draw=drawColor,line width= 0.6pt,line join=round] (\Roffset,\VMoffset{3}) -- (\plus{\Roffset}{\length},\VMoffset{3});
	\end{scope}

	\begin{scope}
		\definecolor{drawColor}{gray}{0.70}
		\node[text=drawColor,anchor=base west,inner sep=0pt, outer sep=0pt, scale=  0.88] at (\RToffset,\VToffset{3}) {True edges};
	\end{scope}
\end{scope}

\begin{scope}
	\begin{scope}
		\definecolor{drawColor}{RGB}{0,0,0}
		\path[draw=drawColor,line width= 0.6pt,line join=round] (\Roffset, \VMoffset{1}) -- (\plus{\Roffset}{\length}, \VMoffset{1});
	\end{scope}

	\begin{scope}
		\definecolor{drawColor}{RGB}{0,0,0}
		\node[text=drawColor,anchor=base west,inner sep=0pt, outer sep=0pt, scale=  0.88] at (\RToffset, \VToffset{1}) {Edge counts};
	\end{scope}
\end{scope}

\end{tikzpicture}}}{Legend}}

  \caption{Sensitivity analysis on the probability to create cycles of coupled obfuscations (X-axis)}
  \label{fig:sense_cyclechance}
\end{figure*}

In Figure~\ref{fig:sense_cyclesize}, we show how the size of fake edge cycles affects the false rates for sizes 1, 2, 4, 6, and 10. For similar reasons as in the previous experiment, we only enabled conditional jump dispatchers here.
\begin{itemize}
\item \textbf{Before attacking} the program, the \dbfpr{} remains constant at 100\%, the \dbfnr{} at 0\%. IDA Pro identifies almost all instructions in this case, because we injected no switch-based dispatchers as stated above. The \guifpr{} decreases from 100\% to about 70\%, stabilizing from cycles of size 4. The \guifnr{} increases from about 20\% to about 30\% in a similar way. The dependence between the cycle size and the GUI FRs results from the interplay between IDA Pro heuristics to partition the code into functions, and the fact that the cycle size impacts the fraction of fake edges that are fall-through edges. This fraction evolves non-linearly (1, 1/2, 1/4, 1/6, 1/10).
  \item The \textbf{soundish attack} does not change the \dbfpr{} and \dbfnr{} due to the strong coupling between the opaque predicates, as discussed previously.
  \item After the \textbf{unsound attack}, the \guifpr{} drops from 100\% to about 50\%. Fake edges coming into an opaque predicate computation are more likely to be drawn in smaller cycles. The \guifnr{} does not change much compared to the unattacked program.
\end{itemize}
Notice that, for cycles of size 1, no opaque predicates can be resolved (i.e., the \dbfpr{} and \guifpr{} remain the same). Obviously, this is because the calculation of each opaque predicate is interrupted by its own fake edge. Hence, no opaque predicate can be resolved by our attack. This is a contrived configuration, as it is easy to set up a similar attack that only requires local analysis on each predicate computation in isolation to identify them. We included it for the sake of completeness anyway.
\begin{figure*}[t]
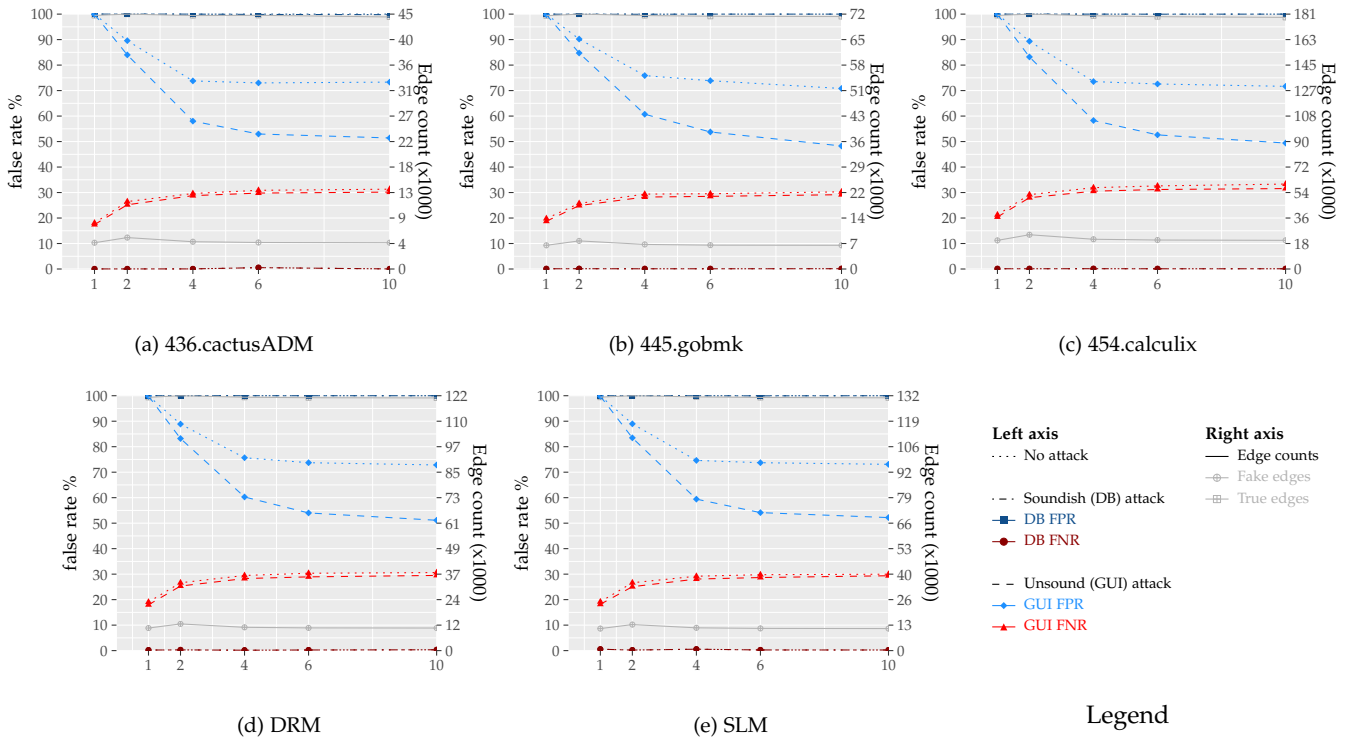

  \centering
  \subfloat[436.cactusADM]
    {\resizebox{0.33\linewidth}{!}{\input{sensitivity_436_clustersize}}}
  \subfloat[445.gobmk]
    {\resizebox{0.33\linewidth}{!}{\input{sensitivity_445_clustersize}}}
  \subfloat[454.calculix]
    {\resizebox{0.33\linewidth}{!}{\input{sensitivity_454_clustersize}}}
  \\
  \subfloat[DRM]
    {\resizebox{0.33\linewidth}{!}{\input{sensitivity_DRM_clustersize}}}
  \subfloat[SLM]
    {\resizebox{0.33\linewidth}{!}{\input{sensitivity_SLM_clustersize}}}
  \subfloat{\stackunder[0em]{\resizebox{0.25\linewidth}{!}{\relax
\begin{tikzpicture}[x=1pt,y=1pt]

\newcommand{\Loffset}{20}
\newcommand{\Roffset}{140}
\newcommand{\length}{13}
\newcommand{\hlength}{6.5}
\newcommand{\hmarker}{2}
\newcommand{\plus}[2]{\the\numexpr #1 + #2}
\newcommand{\plusT}[3]{\the\numexpr #1 + #2 + #3}
\newcommand{\LToffset}{\plus{\plus{\Loffset}{\length}}{5}}
\newcommand{\RToffset}{\plus{\plus{\Roffset}{\length}}{5}}
\newcommand{\VToffset}[1]{\the\numexpr 150 - #1 * 12}
\newcommand{\VMoffset}[1]{\VToffset{#1} + 2}

\definecolor{fillColor}{RGB}{255,0,0}
\path[use as bounding box,fill=fillColor,fill opacity=0.00] (0,0) rectangle (190,180.67);

\begin{scope}
	\begin{scope}
		\definecolor{drawColor}{RGB}{0,0,0}
		\node[text=drawColor,anchor=base west,inner sep=0pt, outer sep=0pt, scale=  1.00] at (\Loffset,\VToffset{0}) {\textbf{Left axis}};
	\end{scope}

	\begin{scope}
		\definecolor{drawColor}{RGB}{0,0,0}
		\node[text=drawColor,anchor=base west,inner sep=0pt, outer sep=0pt, scale=  1.00] at (\Roffset,\VToffset{0}) {\textbf{Right axis}};
	\end{scope}
\end{scope}

\begin{scope}
	\begin{scope}
		\definecolor{drawColor}{RGB}{0,0,0}
		\path[draw=drawColor,line width= 0.6pt,dash pattern=on 1pt off 3pt ,line join=round] (\Loffset, \VMoffset{1}) -- (\plus{\Loffset}{\length}, \VMoffset{1});
	\end{scope}

	\begin{scope}
		\definecolor{drawColor}{RGB}{0,0,0}
		\node[text=drawColor,anchor=base west,inner sep=0pt, outer sep=0pt, scale=  0.88] at (\LToffset, \VToffset{1}) {No attack};
	\end{scope}
\end{scope}

\begin{scope}
	\begin{scope}
		\definecolor{drawColor}{RGB}{0,0,0}
		\path[draw=drawColor,line width= 0.6pt,dash pattern=on 1pt off 3pt on 4pt off 3pt ,line join=round] (\Loffset, \VMoffset{3}) -- (\plus{\Loffset}{\length}, \VMoffset{3});
	\end{scope}

	\begin{scope}
		\definecolor{drawColor}{RGB}{0,0,0}
			\node[text=drawColor,anchor=base west,inner sep=0pt, outer sep=0pt, scale=  0.88] at (\LToffset, \VToffset{3}) {Soundish (DB) attack};
	\end{scope}
\end{scope}

\begin{scope}
	\begin{scope}
		\definecolor{fillColor}{RGB}{16,78,139}
		\path[fill=fillColor] (\plusT{\Loffset}{\hlength}{-\hmarker},\plus{\VMoffset{4}}{-\hmarker}) --
			(\plusT{\Loffset}{\hlength}{\hmarker},\plus{\VMoffset{4}}{-\hmarker}) --
			(\plusT{\Loffset}{\hlength}{\hmarker},\plus{\VMoffset{4}}{\hmarker}) --
			(\plusT{\Loffset}{\hlength}{-\hmarker},\plus{\VMoffset{4}}{\hmarker}) --
			cycle;
	\end{scope}

	\begin{scope}
		\definecolor{drawColor}{RGB}{16,78,139}
		\path[draw=drawColor,line width= 0.6pt,line join=round] (\Loffset, \VMoffset{4}) -- (\plus{\Loffset}{\length}, \VMoffset{4});
	\end{scope}

	\begin{scope}
		\definecolor{drawColor}{RGB}{16,78,139}
		\node[text=drawColor,anchor=base west,inner sep=0pt, outer sep=0pt, scale=  0.88] at (\LToffset,\VToffset{4}) {DB FPR};
	\end{scope}
\end{scope}

\begin{scope}
	\begin{scope}
		\definecolor{drawColor}{RGB}{139,0,0}
		\definecolor{fillColor}{RGB}{139,0,0}
		\path[draw=drawColor,line width= 0.4pt,line join=round,line cap=round,fill=fillColor] (\plus{\Loffset}{\hlength},\VMoffset{5}) circle (\hmarker);
	\end{scope}

	\begin{scope}
		\definecolor{drawColor}{RGB}{139,0,0}
		\path[draw=drawColor,line width= 0.6pt,line join=round] (\Loffset,\VMoffset{5}) -- (\plus{\Loffset}{\length},\VMoffset{5});
	\end{scope}

	\begin{scope}
		\definecolor{drawColor}{RGB}{139,0,0}
		\node[text=drawColor,anchor=base west,inner sep=0pt, outer sep=0pt, scale=  0.88] at (\LToffset,\VToffset{5}) {DB FNR};
	\end{scope}
\end{scope}

\begin{scope}
	\begin{scope}
		\definecolor{drawColor}{RGB}{0,0,0}
		\path[draw=drawColor,line width= 0.6pt,dash pattern=on 4pt off 4pt ,line join=round] (\Loffset, \VMoffset{7}) -- (\plus{\Loffset}{\length}, \VMoffset{7});
	\end{scope}

	\begin{scope}
		\definecolor{drawColor}{RGB}{0,0,0}
			\node[text=drawColor,anchor=base west,inner sep=0pt, outer sep=0pt, scale=  0.88] at (\LToffset, \VToffset{7}) {Unsound (GUI) attack};
	\end{scope}
\end{scope}

\begin{scope}
	\begin{scope}
		\definecolor{fillColor}{RGB}{30,144,255}
		\path[fill=fillColor] (\plusT{\Loffset}{\hlength}{-\hmarker}, \VMoffset{8}) --
			(\plus{\Loffset}{\hlength},\plus{\VMoffset{8}}{\hmarker}) --
			(\plusT{\Loffset}{\hlength}{\hmarker}, \VMoffset{8}) --
			(\plus{\Loffset}{\hlength}, \plus{\VMoffset{8}}{-\hmarker}) --
			cycle;
	\end{scope}

	\begin{scope}
		\definecolor{drawColor}{RGB}{30,144,255}
		\path[draw=drawColor,line width= 0.6pt,line join=round] (\Loffset, \VMoffset{8}) -- (\plus{\Loffset}{\length}, \VMoffset{8});
	\end{scope}

	\begin{scope}
		\definecolor{drawColor}{RGB}{30,144,255}
		\node[text=drawColor,anchor=base west,inner sep=0pt, outer sep=0pt, scale=  0.88] at (\LToffset, \VToffset{8}) {GUI FPR};
	\end{scope}
\end{scope}

\begin{scope}
	\begin{scope}
		\definecolor{fillColor}{RGB}{255,0,0}
		\path[fill=fillColor] (\plus{\Loffset}{\hlength},\plus{\VMoffset{9}}{\hmarker}) --
			(\plusT{\Loffset}{\hlength}{\hmarker},\plus{\VMoffset{9}}{-\hmarker}) --
			(\plusT{\Loffset}{\hlength}{-\hmarker},\plus{\VMoffset{9}}{-\hmarker}) --
			cycle;
	\end{scope}

	\begin{scope}
		\definecolor{drawColor}{RGB}{255,0,0}
		\path[draw=drawColor,line width= 0.6pt,line join=round] (\Loffset,\VMoffset{9}) -- (\plus{\Loffset}{\length},\VMoffset{9});
	\end{scope}

	\begin{scope}
		\definecolor{drawColor}{RGB}{255,0,0}
		\node[text=drawColor,anchor=base west,inner sep=0pt, outer sep=0pt, scale=  0.88] at (\LToffset,\VToffset{9}) {GUI FNR};
	\end{scope}
\end{scope}

\begin{scope}
	\begin{scope}
		\definecolor{drawColor}{gray}{0.70}
		\path[draw=drawColor,line width= 0.4pt,line join=round,line cap=round] (\plus{\Roffset}{\hlength},\VMoffset{2}) circle (\hmarker);
		\path[draw=drawColor,line width= 0.4pt,line join=round,line cap=round] (\plusT{\Roffset}{\hlength}{-\hmarker},\VMoffset{2}) -- (\plusT{\Roffset}{\hlength}{\hmarker},\VMoffset{2});
		\path[draw=drawColor,line width= 0.4pt,line join=round,line cap=round] (\plus{\Roffset}{\hlength},\plus{\VMoffset{2}}{-\hmarker}) -- (\plus{\Roffset}{\hlength},\plus{\VMoffset{2}}{\hmarker});
	\end{scope}

	\begin{scope}
		\definecolor{drawColor}{gray}{0.70}
		\path[draw=drawColor,line width= 0.6pt,line join=round] (\Roffset,\VMoffset{2}) -- (\plus{\Roffset}{\length},\VMoffset{2});
	\end{scope}

	\begin{scope}
		\definecolor{drawColor}{gray}{0.70}
		\node[text=drawColor,anchor=base west,inner sep=0pt, outer sep=0pt, scale=  0.88] at (\RToffset,\VToffset{2}) {Fake edges};
	\end{scope}
\end{scope}

\begin{scope}
	\begin{scope}
		\definecolor{drawColor}{gray}{0.70}
		\path[draw=drawColor,line width= 0.4pt,line join=round,line cap=round] (\plusT{\Roffset}{\hlength}{-\hmarker},\VMoffset{3}) -- (\plusT{\Roffset}{\hlength}{\hmarker},\VMoffset{3});
		\path[draw=drawColor,line width= 0.4pt,line join=round,line cap=round] (\plus{\Roffset}{\hlength},\plus{\VMoffset{3}}{-\hmarker}) -- (\plus{\Roffset}{\hlength},\plus{\VMoffset{3}}{\hmarker});
		\path[draw=drawColor,line width= 0.4pt,line join=round,line cap=round] (\plusT{\Roffset}{\hlength}{-\hmarker},\plus{\VMoffset{3}}{-\hmarker}) rectangle (\plusT{\Roffset}{\hlength}{\hmarker},\plus{\VMoffset{3}}{\hmarker});
	\end{scope}

	\begin{scope}
		\definecolor{drawColor}{gray}{0.70}
		\path[draw=drawColor,line width= 0.6pt,line join=round] (\Roffset,\VMoffset{3}) -- (\plus{\Roffset}{\length},\VMoffset{3});
	\end{scope}

	\begin{scope}
		\definecolor{drawColor}{gray}{0.70}
		\node[text=drawColor,anchor=base west,inner sep=0pt, outer sep=0pt, scale=  0.88] at (\RToffset,\VToffset{3}) {True edges};
	\end{scope}
\end{scope}

\begin{scope}
	\begin{scope}
		\definecolor{drawColor}{RGB}{0,0,0}
		\path[draw=drawColor,line width= 0.6pt,line join=round] (\Roffset, \VMoffset{1}) -- (\plus{\Roffset}{\length}, \VMoffset{1});
	\end{scope}

	\begin{scope}
		\definecolor{drawColor}{RGB}{0,0,0}
		\node[text=drawColor,anchor=base west,inner sep=0pt, outer sep=0pt, scale=  0.88] at (\RToffset, \VToffset{1}) {Edge counts};
	\end{scope}
\end{scope}

\end{tikzpicture}}}{Legend}}

  \caption{Sensitivity analysis on the size of coupled obfuscation cycles (X-axis)}
  \label{fig:sense_cyclesize}
\end{figure*}

In Figure~\ref{fig:sense_fakeentry}, we show how the sparseness of the dispatcher's switch tables affects the false rates for values of 0\%, 10\%, 20\%, and 30\%.
\begin{itemize}
  \item \textbf{Before attacking} the program, the \dbfpr{} decreases from 100\% to about 95\% with increasing sparseness. The \guifpr{} follows a similar trend, but starts at about 82\%, decreasing to about 80\% due to the IDA Pro GUI not drawing all control flow. The \dbfnr{} and \guifnr{} do not change much, which of course is due to IDA Pro either supporting our dispatchers or not, independent of the sparseness of the switch tables. The \dbfnr{} is not zero (about 17\%), however, providing evidence that IDA Pro does not analyse the control flow from all dispatcher types.
  \item The \textbf{soundish attack} does not impact the false rates.
  \item After the \textbf{unsound attack}, the \guifpr{} decreases from about 72\% to about 70\%. Compared to the numbers before the attack, about 10\% of the fake edges are eliminated. The \guifnr{} does not change much compared to the unattacked scenario.
\end{itemize}
From the results discussed above, we conclude that this parameter does not influence the false rates much.
\begin{figure*}[t]
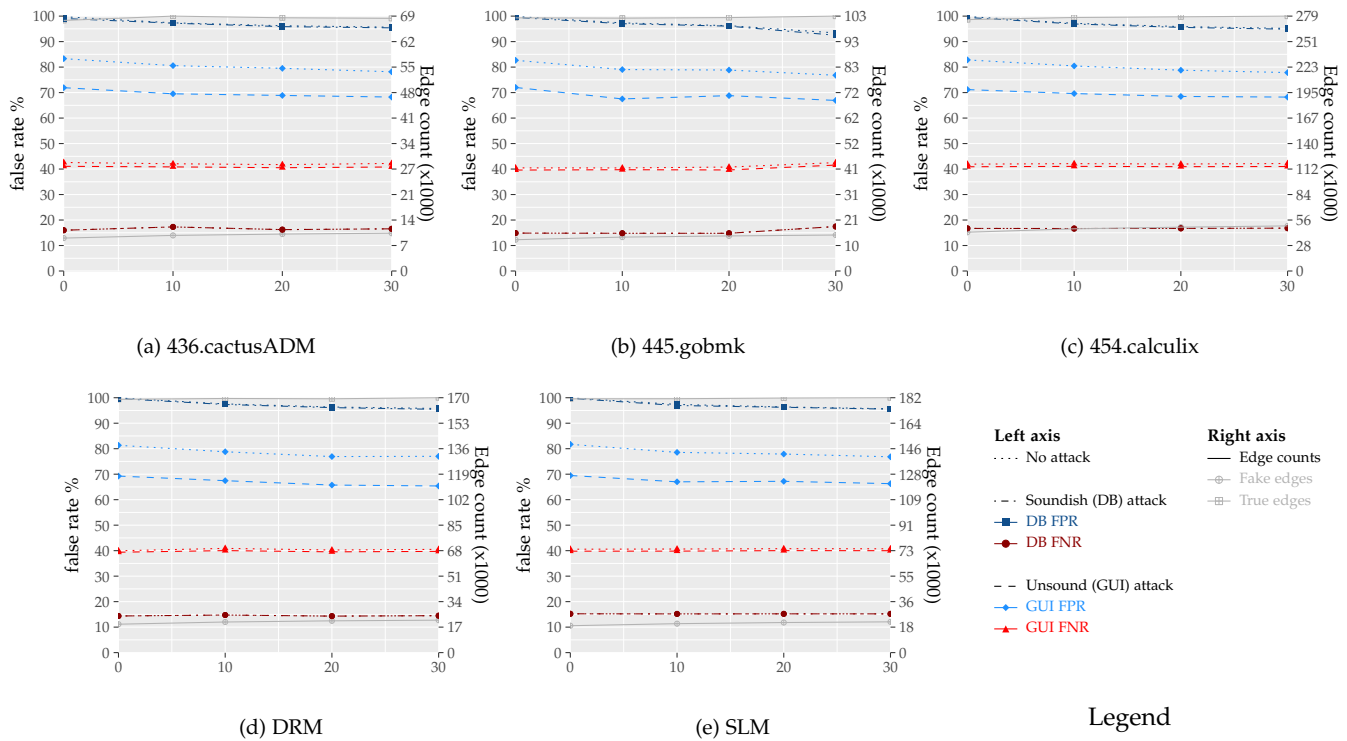

  \centering
  \subfloat[436.cactusADM]
    {\resizebox{0.33\linewidth}{!}{\input{sensitivity_436_fakeentry}}}
  \subfloat[445.gobmk]
    {\resizebox{0.33\linewidth}{!}{\input{sensitivity_445_fakeentry}}}
  \subfloat[454.calculix]
    {\resizebox{0.33\linewidth}{!}{\input{sensitivity_454_fakeentry}}}
  \\
  \subfloat[DRM]
    {\resizebox{0.33\linewidth}{!}{\input{sensitivity_DRM_fakeentry}}}
  \subfloat[SLM]
    {\resizebox{0.33\linewidth}{!}{\input{sensitivity_SLM_fakeentry}}}
  \subfloat{\stackunder[0em]{\resizebox{0.25\linewidth}{!}{\relax
\begin{tikzpicture}[x=1pt,y=1pt]

\newcommand{\Loffset}{20}
\newcommand{\Roffset}{140}
\newcommand{\length}{13}
\newcommand{\hlength}{6.5}
\newcommand{\hmarker}{2}
\newcommand{\plus}[2]{\the\numexpr #1 + #2}
\newcommand{\plusT}[3]{\the\numexpr #1 + #2 + #3}
\newcommand{\LToffset}{\plus{\plus{\Loffset}{\length}}{5}}
\newcommand{\RToffset}{\plus{\plus{\Roffset}{\length}}{5}}
\newcommand{\VToffset}[1]{\the\numexpr 150 - #1 * 12}
\newcommand{\VMoffset}[1]{\VToffset{#1} + 2}

\definecolor{fillColor}{RGB}{255,0,0}
\path[use as bounding box,fill=fillColor,fill opacity=0.00] (0,0) rectangle (190,180.67);

\begin{scope}
	\begin{scope}
		\definecolor{drawColor}{RGB}{0,0,0}
		\node[text=drawColor,anchor=base west,inner sep=0pt, outer sep=0pt, scale=  1.00] at (\Loffset,\VToffset{0}) {\textbf{Left axis}};
	\end{scope}

	\begin{scope}
		\definecolor{drawColor}{RGB}{0,0,0}
		\node[text=drawColor,anchor=base west,inner sep=0pt, outer sep=0pt, scale=  1.00] at (\Roffset,\VToffset{0}) {\textbf{Right axis}};
	\end{scope}
\end{scope}

\begin{scope}
	\begin{scope}
		\definecolor{drawColor}{RGB}{0,0,0}
		\path[draw=drawColor,line width= 0.6pt,dash pattern=on 1pt off 3pt ,line join=round] (\Loffset, \VMoffset{1}) -- (\plus{\Loffset}{\length}, \VMoffset{1});
	\end{scope}

	\begin{scope}
		\definecolor{drawColor}{RGB}{0,0,0}
		\node[text=drawColor,anchor=base west,inner sep=0pt, outer sep=0pt, scale=  0.88] at (\LToffset, \VToffset{1}) {No attack};
	\end{scope}
\end{scope}

\begin{scope}
	\begin{scope}
		\definecolor{drawColor}{RGB}{0,0,0}
		\path[draw=drawColor,line width= 0.6pt,dash pattern=on 1pt off 3pt on 4pt off 3pt ,line join=round] (\Loffset, \VMoffset{3}) -- (\plus{\Loffset}{\length}, \VMoffset{3});
	\end{scope}

	\begin{scope}
		\definecolor{drawColor}{RGB}{0,0,0}
			\node[text=drawColor,anchor=base west,inner sep=0pt, outer sep=0pt, scale=  0.88] at (\LToffset, \VToffset{3}) {Soundish (DB) attack};
	\end{scope}
\end{scope}

\begin{scope}
	\begin{scope}
		\definecolor{fillColor}{RGB}{16,78,139}
		\path[fill=fillColor] (\plusT{\Loffset}{\hlength}{-\hmarker},\plus{\VMoffset{4}}{-\hmarker}) --
			(\plusT{\Loffset}{\hlength}{\hmarker},\plus{\VMoffset{4}}{-\hmarker}) --
			(\plusT{\Loffset}{\hlength}{\hmarker},\plus{\VMoffset{4}}{\hmarker}) --
			(\plusT{\Loffset}{\hlength}{-\hmarker},\plus{\VMoffset{4}}{\hmarker}) --
			cycle;
	\end{scope}

	\begin{scope}
		\definecolor{drawColor}{RGB}{16,78,139}
		\path[draw=drawColor,line width= 0.6pt,line join=round] (\Loffset, \VMoffset{4}) -- (\plus{\Loffset}{\length}, \VMoffset{4});
	\end{scope}

	\begin{scope}
		\definecolor{drawColor}{RGB}{16,78,139}
		\node[text=drawColor,anchor=base west,inner sep=0pt, outer sep=0pt, scale=  0.88] at (\LToffset,\VToffset{4}) {DB FPR};
	\end{scope}
\end{scope}

\begin{scope}
	\begin{scope}
		\definecolor{drawColor}{RGB}{139,0,0}
		\definecolor{fillColor}{RGB}{139,0,0}
		\path[draw=drawColor,line width= 0.4pt,line join=round,line cap=round,fill=fillColor] (\plus{\Loffset}{\hlength},\VMoffset{5}) circle (\hmarker);
	\end{scope}

	\begin{scope}
		\definecolor{drawColor}{RGB}{139,0,0}
		\path[draw=drawColor,line width= 0.6pt,line join=round] (\Loffset,\VMoffset{5}) -- (\plus{\Loffset}{\length},\VMoffset{5});
	\end{scope}

	\begin{scope}
		\definecolor{drawColor}{RGB}{139,0,0}
		\node[text=drawColor,anchor=base west,inner sep=0pt, outer sep=0pt, scale=  0.88] at (\LToffset,\VToffset{5}) {DB FNR};
	\end{scope}
\end{scope}

\begin{scope}
	\begin{scope}
		\definecolor{drawColor}{RGB}{0,0,0}
		\path[draw=drawColor,line width= 0.6pt,dash pattern=on 4pt off 4pt ,line join=round] (\Loffset, \VMoffset{7}) -- (\plus{\Loffset}{\length}, \VMoffset{7});
	\end{scope}

	\begin{scope}
		\definecolor{drawColor}{RGB}{0,0,0}
			\node[text=drawColor,anchor=base west,inner sep=0pt, outer sep=0pt, scale=  0.88] at (\LToffset, \VToffset{7}) {Unsound (GUI) attack};
	\end{scope}
\end{scope}

\begin{scope}
	\begin{scope}
		\definecolor{fillColor}{RGB}{30,144,255}
		\path[fill=fillColor] (\plusT{\Loffset}{\hlength}{-\hmarker}, \VMoffset{8}) --
			(\plus{\Loffset}{\hlength},\plus{\VMoffset{8}}{\hmarker}) --
			(\plusT{\Loffset}{\hlength}{\hmarker}, \VMoffset{8}) --
			(\plus{\Loffset}{\hlength}, \plus{\VMoffset{8}}{-\hmarker}) --
			cycle;
	\end{scope}

	\begin{scope}
		\definecolor{drawColor}{RGB}{30,144,255}
		\path[draw=drawColor,line width= 0.6pt,line join=round] (\Loffset, \VMoffset{8}) -- (\plus{\Loffset}{\length}, \VMoffset{8});
	\end{scope}

	\begin{scope}
		\definecolor{drawColor}{RGB}{30,144,255}
		\node[text=drawColor,anchor=base west,inner sep=0pt, outer sep=0pt, scale=  0.88] at (\LToffset, \VToffset{8}) {GUI FPR};
	\end{scope}
\end{scope}

\begin{scope}
	\begin{scope}
		\definecolor{fillColor}{RGB}{255,0,0}
		\path[fill=fillColor] (\plus{\Loffset}{\hlength},\plus{\VMoffset{9}}{\hmarker}) --
			(\plusT{\Loffset}{\hlength}{\hmarker},\plus{\VMoffset{9}}{-\hmarker}) --
			(\plusT{\Loffset}{\hlength}{-\hmarker},\plus{\VMoffset{9}}{-\hmarker}) --
			cycle;
	\end{scope}

	\begin{scope}
		\definecolor{drawColor}{RGB}{255,0,0}
		\path[draw=drawColor,line width= 0.6pt,line join=round] (\Loffset,\VMoffset{9}) -- (\plus{\Loffset}{\length},\VMoffset{9});
	\end{scope}

	\begin{scope}
		\definecolor{drawColor}{RGB}{255,0,0}
		\node[text=drawColor,anchor=base west,inner sep=0pt, outer sep=0pt, scale=  0.88] at (\LToffset,\VToffset{9}) {GUI FNR};
	\end{scope}
\end{scope}

\begin{scope}
	\begin{scope}
		\definecolor{drawColor}{gray}{0.70}
		\path[draw=drawColor,line width= 0.4pt,line join=round,line cap=round] (\plus{\Roffset}{\hlength},\VMoffset{2}) circle (\hmarker);
		\path[draw=drawColor,line width= 0.4pt,line join=round,line cap=round] (\plusT{\Roffset}{\hlength}{-\hmarker},\VMoffset{2}) -- (\plusT{\Roffset}{\hlength}{\hmarker},\VMoffset{2});
		\path[draw=drawColor,line width= 0.4pt,line join=round,line cap=round] (\plus{\Roffset}{\hlength},\plus{\VMoffset{2}}{-\hmarker}) -- (\plus{\Roffset}{\hlength},\plus{\VMoffset{2}}{\hmarker});
	\end{scope}

	\begin{scope}
		\definecolor{drawColor}{gray}{0.70}
		\path[draw=drawColor,line width= 0.6pt,line join=round] (\Roffset,\VMoffset{2}) -- (\plus{\Roffset}{\length},\VMoffset{2});
	\end{scope}

	\begin{scope}
		\definecolor{drawColor}{gray}{0.70}
		\node[text=drawColor,anchor=base west,inner sep=0pt, outer sep=0pt, scale=  0.88] at (\RToffset,\VToffset{2}) {Fake edges};
	\end{scope}
\end{scope}

\begin{scope}
	\begin{scope}
		\definecolor{drawColor}{gray}{0.70}
		\path[draw=drawColor,line width= 0.4pt,line join=round,line cap=round] (\plusT{\Roffset}{\hlength}{-\hmarker},\VMoffset{3}) -- (\plusT{\Roffset}{\hlength}{\hmarker},\VMoffset{3});
		\path[draw=drawColor,line width= 0.4pt,line join=round,line cap=round] (\plus{\Roffset}{\hlength},\plus{\VMoffset{3}}{-\hmarker}) -- (\plus{\Roffset}{\hlength},\plus{\VMoffset{3}}{\hmarker});
		\path[draw=drawColor,line width= 0.4pt,line join=round,line cap=round] (\plusT{\Roffset}{\hlength}{-\hmarker},\plus{\VMoffset{3}}{-\hmarker}) rectangle (\plusT{\Roffset}{\hlength}{\hmarker},\plus{\VMoffset{3}}{\hmarker});
	\end{scope}

	\begin{scope}
		\definecolor{drawColor}{gray}{0.70}
		\path[draw=drawColor,line width= 0.6pt,line join=round] (\Roffset,\VMoffset{3}) -- (\plus{\Roffset}{\length},\VMoffset{3});
	\end{scope}

	\begin{scope}
		\definecolor{drawColor}{gray}{0.70}
		\node[text=drawColor,anchor=base west,inner sep=0pt, outer sep=0pt, scale=  0.88] at (\RToffset,\VToffset{3}) {True edges};
	\end{scope}
\end{scope}

\begin{scope}
	\begin{scope}
		\definecolor{drawColor}{RGB}{0,0,0}
		\path[draw=drawColor,line width= 0.6pt,line join=round] (\Roffset, \VMoffset{1}) -- (\plus{\Roffset}{\length}, \VMoffset{1});
	\end{scope}

	\begin{scope}
		\definecolor{drawColor}{RGB}{0,0,0}
		\node[text=drawColor,anchor=base west,inner sep=0pt, outer sep=0pt, scale=  0.88] at (\RToffset, \VToffset{1}) {Edge counts};
	\end{scope}
\end{scope}

\end{tikzpicture}}}{Legend}}

  \caption{Sensitivity analysis on the fake switch entry insertion probability (X-axis)}
  \label{fig:sense_fakeentry}
\end{figure*}

In Figure~\ref{fig:sense_fakeft}, we show the results of the sensitivity analysis for the probability to make fake edges fall-through when we set it to 0\%, 25\%, 50\%, 75\%, and 100\%. As the branch flipping transformation (controlled by parameter 11) randomly flips the condition of (opaque) conditional branches, we disabled that transformation here.
\begin{itemize}
  \item \textbf{Before attacking} the program, the \dbfpr{} remains constant at about 95\%, which is due to the default value for the sparseness parameter of 30\%. Similarly, the \dbfnr{} remains constant at about 17\%. The \guifpr{} increases from about 70\% to about 80\%, confirming our observation in Section~\ref{sec:opaque} that IDA Pro prefers drawing fall-through edges over branch-taken ones. For the same reasons as for the \dbfnr, the \guifnr{} remains constant at about 40\%.
  \item The \textbf{soundish attack} does not change the false rates. This is of course explained by the coupling of the opaque predicates, as discussed for Figure~\ref{fig:sense_cyclechance}. (The default value for that parameter is 100\%.)
  \item After the \textbf{unsound attack}, the \guifpr{} increases from about 55\% to about 65\%. The number of resolved opaque predicates slightly decreases with increasing probabilities, indicating that the parameter has only a small effect. The \guifnr{} does not change much compared to the unattacked program.
\end{itemize}
We conclude that the small impact of this parameter does not force a defender to pick a value within a small range. This ensures that attackers will not be able to create custom attacks that exploit a preferred direction of fake edges.
\begin{figure*}[t]
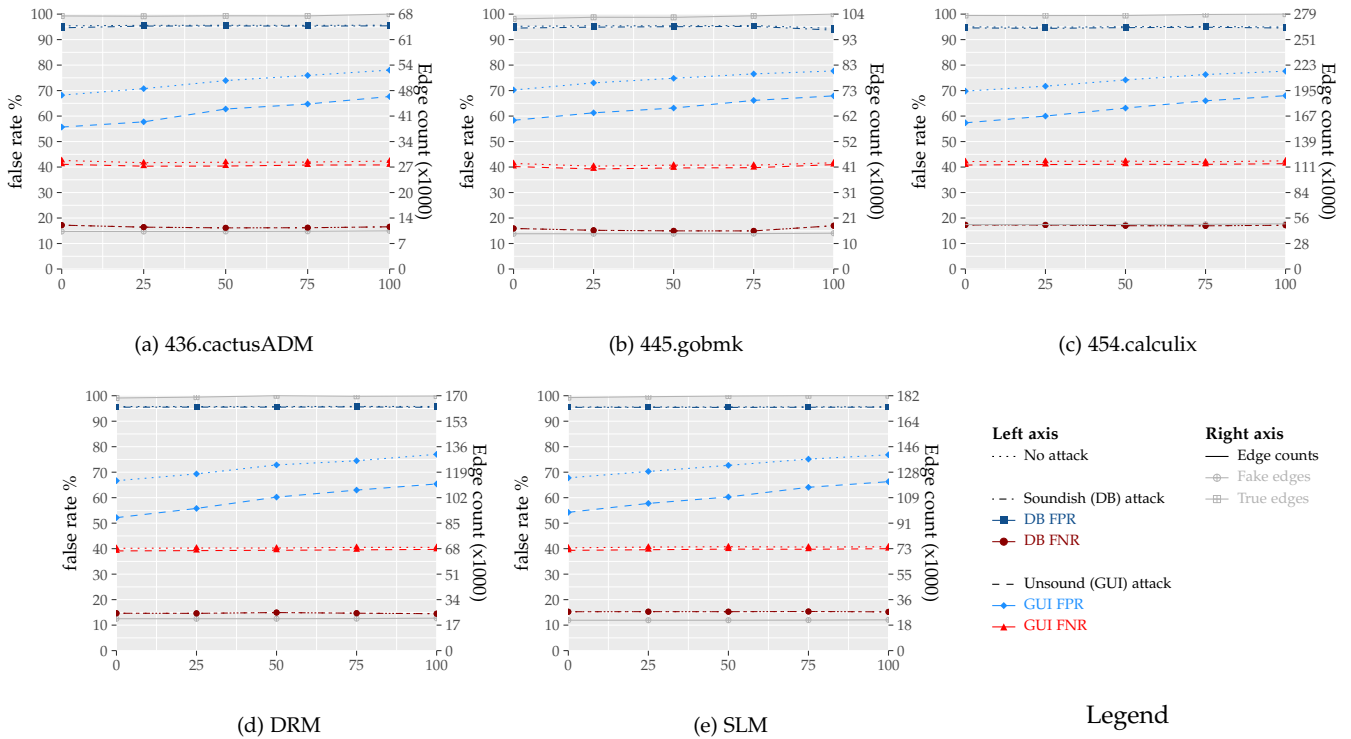

  \centering
  \subfloat[436.cactusADM]
    {\resizebox{0.33\linewidth}{!}{\input{sensitivity_436_fakeft}}}
  \subfloat[445.gobmk]
    {\resizebox{0.33\linewidth}{!}{\input{sensitivity_445_fakeft}}}
  \subfloat[454.calculix]
    {\resizebox{0.33\linewidth}{!}{\input{sensitivity_454_fakeft}}}
  \\
  \subfloat[DRM]
    {\resizebox{0.33\linewidth}{!}{\input{sensitivity_DRM_fakeft}}}
  \subfloat[SLM]
    {\resizebox{0.33\linewidth}{!}{\input{sensitivity_SLM_fakeft}}}
  \subfloat{\stackunder[0em]{\resizebox{0.25\linewidth}{!}{\relax
\begin{tikzpicture}[x=1pt,y=1pt]

\newcommand{\Loffset}{20}
\newcommand{\Roffset}{140}
\newcommand{\length}{13}
\newcommand{\hlength}{6.5}
\newcommand{\hmarker}{2}
\newcommand{\plus}[2]{\the\numexpr #1 + #2}
\newcommand{\plusT}[3]{\the\numexpr #1 + #2 + #3}
\newcommand{\LToffset}{\plus{\plus{\Loffset}{\length}}{5}}
\newcommand{\RToffset}{\plus{\plus{\Roffset}{\length}}{5}}
\newcommand{\VToffset}[1]{\the\numexpr 150 - #1 * 12}
\newcommand{\VMoffset}[1]{\VToffset{#1} + 2}

\definecolor{fillColor}{RGB}{255,0,0}
\path[use as bounding box,fill=fillColor,fill opacity=0.00] (0,0) rectangle (190,180.67);

\begin{scope}
	\begin{scope}
		\definecolor{drawColor}{RGB}{0,0,0}
		\node[text=drawColor,anchor=base west,inner sep=0pt, outer sep=0pt, scale=  1.00] at (\Loffset,\VToffset{0}) {\textbf{Left axis}};
	\end{scope}

	\begin{scope}
		\definecolor{drawColor}{RGB}{0,0,0}
		\node[text=drawColor,anchor=base west,inner sep=0pt, outer sep=0pt, scale=  1.00] at (\Roffset,\VToffset{0}) {\textbf{Right axis}};
	\end{scope}
\end{scope}

\begin{scope}
	\begin{scope}
		\definecolor{drawColor}{RGB}{0,0,0}
		\path[draw=drawColor,line width= 0.6pt,dash pattern=on 1pt off 3pt ,line join=round] (\Loffset, \VMoffset{1}) -- (\plus{\Loffset}{\length}, \VMoffset{1});
	\end{scope}

	\begin{scope}
		\definecolor{drawColor}{RGB}{0,0,0}
		\node[text=drawColor,anchor=base west,inner sep=0pt, outer sep=0pt, scale=  0.88] at (\LToffset, \VToffset{1}) {No attack};
	\end{scope}
\end{scope}

\begin{scope}
	\begin{scope}
		\definecolor{drawColor}{RGB}{0,0,0}
		\path[draw=drawColor,line width= 0.6pt,dash pattern=on 1pt off 3pt on 4pt off 3pt ,line join=round] (\Loffset, \VMoffset{3}) -- (\plus{\Loffset}{\length}, \VMoffset{3});
	\end{scope}

	\begin{scope}
		\definecolor{drawColor}{RGB}{0,0,0}
			\node[text=drawColor,anchor=base west,inner sep=0pt, outer sep=0pt, scale=  0.88] at (\LToffset, \VToffset{3}) {Soundish (DB) attack};
	\end{scope}
\end{scope}

\begin{scope}
	\begin{scope}
		\definecolor{fillColor}{RGB}{16,78,139}
		\path[fill=fillColor] (\plusT{\Loffset}{\hlength}{-\hmarker},\plus{\VMoffset{4}}{-\hmarker}) --
			(\plusT{\Loffset}{\hlength}{\hmarker},\plus{\VMoffset{4}}{-\hmarker}) --
			(\plusT{\Loffset}{\hlength}{\hmarker},\plus{\VMoffset{4}}{\hmarker}) --
			(\plusT{\Loffset}{\hlength}{-\hmarker},\plus{\VMoffset{4}}{\hmarker}) --
			cycle;
	\end{scope}

	\begin{scope}
		\definecolor{drawColor}{RGB}{16,78,139}
		\path[draw=drawColor,line width= 0.6pt,line join=round] (\Loffset, \VMoffset{4}) -- (\plus{\Loffset}{\length}, \VMoffset{4});
	\end{scope}

	\begin{scope}
		\definecolor{drawColor}{RGB}{16,78,139}
		\node[text=drawColor,anchor=base west,inner sep=0pt, outer sep=0pt, scale=  0.88] at (\LToffset,\VToffset{4}) {DB FPR};
	\end{scope}
\end{scope}

\begin{scope}
	\begin{scope}
		\definecolor{drawColor}{RGB}{139,0,0}
		\definecolor{fillColor}{RGB}{139,0,0}
		\path[draw=drawColor,line width= 0.4pt,line join=round,line cap=round,fill=fillColor] (\plus{\Loffset}{\hlength},\VMoffset{5}) circle (\hmarker);
	\end{scope}

	\begin{scope}
		\definecolor{drawColor}{RGB}{139,0,0}
		\path[draw=drawColor,line width= 0.6pt,line join=round] (\Loffset,\VMoffset{5}) -- (\plus{\Loffset}{\length},\VMoffset{5});
	\end{scope}

	\begin{scope}
		\definecolor{drawColor}{RGB}{139,0,0}
		\node[text=drawColor,anchor=base west,inner sep=0pt, outer sep=0pt, scale=  0.88] at (\LToffset,\VToffset{5}) {DB FNR};
	\end{scope}
\end{scope}

\begin{scope}
	\begin{scope}
		\definecolor{drawColor}{RGB}{0,0,0}
		\path[draw=drawColor,line width= 0.6pt,dash pattern=on 4pt off 4pt ,line join=round] (\Loffset, \VMoffset{7}) -- (\plus{\Loffset}{\length}, \VMoffset{7});
	\end{scope}

	\begin{scope}
		\definecolor{drawColor}{RGB}{0,0,0}
			\node[text=drawColor,anchor=base west,inner sep=0pt, outer sep=0pt, scale=  0.88] at (\LToffset, \VToffset{7}) {Unsound (GUI) attack};
	\end{scope}
\end{scope}

\begin{scope}
	\begin{scope}
		\definecolor{fillColor}{RGB}{30,144,255}
		\path[fill=fillColor] (\plusT{\Loffset}{\hlength}{-\hmarker}, \VMoffset{8}) --
			(\plus{\Loffset}{\hlength},\plus{\VMoffset{8}}{\hmarker}) --
			(\plusT{\Loffset}{\hlength}{\hmarker}, \VMoffset{8}) --
			(\plus{\Loffset}{\hlength}, \plus{\VMoffset{8}}{-\hmarker}) --
			cycle;
	\end{scope}

	\begin{scope}
		\definecolor{drawColor}{RGB}{30,144,255}
		\path[draw=drawColor,line width= 0.6pt,line join=round] (\Loffset, \VMoffset{8}) -- (\plus{\Loffset}{\length}, \VMoffset{8});
	\end{scope}

	\begin{scope}
		\definecolor{drawColor}{RGB}{30,144,255}
		\node[text=drawColor,anchor=base west,inner sep=0pt, outer sep=0pt, scale=  0.88] at (\LToffset, \VToffset{8}) {GUI FPR};
	\end{scope}
\end{scope}

\begin{scope}
	\begin{scope}
		\definecolor{fillColor}{RGB}{255,0,0}
		\path[fill=fillColor] (\plus{\Loffset}{\hlength},\plus{\VMoffset{9}}{\hmarker}) --
			(\plusT{\Loffset}{\hlength}{\hmarker},\plus{\VMoffset{9}}{-\hmarker}) --
			(\plusT{\Loffset}{\hlength}{-\hmarker},\plus{\VMoffset{9}}{-\hmarker}) --
			cycle;
	\end{scope}

	\begin{scope}
		\definecolor{drawColor}{RGB}{255,0,0}
		\path[draw=drawColor,line width= 0.6pt,line join=round] (\Loffset,\VMoffset{9}) -- (\plus{\Loffset}{\length},\VMoffset{9});
	\end{scope}

	\begin{scope}
		\definecolor{drawColor}{RGB}{255,0,0}
		\node[text=drawColor,anchor=base west,inner sep=0pt, outer sep=0pt, scale=  0.88] at (\LToffset,\VToffset{9}) {GUI FNR};
	\end{scope}
\end{scope}

\begin{scope}
	\begin{scope}
		\definecolor{drawColor}{gray}{0.70}
		\path[draw=drawColor,line width= 0.4pt,line join=round,line cap=round] (\plus{\Roffset}{\hlength},\VMoffset{2}) circle (\hmarker);
		\path[draw=drawColor,line width= 0.4pt,line join=round,line cap=round] (\plusT{\Roffset}{\hlength}{-\hmarker},\VMoffset{2}) -- (\plusT{\Roffset}{\hlength}{\hmarker},\VMoffset{2});
		\path[draw=drawColor,line width= 0.4pt,line join=round,line cap=round] (\plus{\Roffset}{\hlength},\plus{\VMoffset{2}}{-\hmarker}) -- (\plus{\Roffset}{\hlength},\plus{\VMoffset{2}}{\hmarker});
	\end{scope}

	\begin{scope}
		\definecolor{drawColor}{gray}{0.70}
		\path[draw=drawColor,line width= 0.6pt,line join=round] (\Roffset,\VMoffset{2}) -- (\plus{\Roffset}{\length},\VMoffset{2});
	\end{scope}

	\begin{scope}
		\definecolor{drawColor}{gray}{0.70}
		\node[text=drawColor,anchor=base west,inner sep=0pt, outer sep=0pt, scale=  0.88] at (\RToffset,\VToffset{2}) {Fake edges};
	\end{scope}
\end{scope}

\begin{scope}
	\begin{scope}
		\definecolor{drawColor}{gray}{0.70}
		\path[draw=drawColor,line width= 0.4pt,line join=round,line cap=round] (\plusT{\Roffset}{\hlength}{-\hmarker},\VMoffset{3}) -- (\plusT{\Roffset}{\hlength}{\hmarker},\VMoffset{3});
		\path[draw=drawColor,line width= 0.4pt,line join=round,line cap=round] (\plus{\Roffset}{\hlength},\plus{\VMoffset{3}}{-\hmarker}) -- (\plus{\Roffset}{\hlength},\plus{\VMoffset{3}}{\hmarker});
		\path[draw=drawColor,line width= 0.4pt,line join=round,line cap=round] (\plusT{\Roffset}{\hlength}{-\hmarker},\plus{\VMoffset{3}}{-\hmarker}) rectangle (\plusT{\Roffset}{\hlength}{\hmarker},\plus{\VMoffset{3}}{\hmarker});
	\end{scope}

	\begin{scope}
		\definecolor{drawColor}{gray}{0.70}
		\path[draw=drawColor,line width= 0.6pt,line join=round] (\Roffset,\VMoffset{3}) -- (\plus{\Roffset}{\length},\VMoffset{3});
	\end{scope}

	\begin{scope}
		\definecolor{drawColor}{gray}{0.70}
		\node[text=drawColor,anchor=base west,inner sep=0pt, outer sep=0pt, scale=  0.88] at (\RToffset,\VToffset{3}) {True edges};
	\end{scope}
\end{scope}

\begin{scope}
	\begin{scope}
		\definecolor{drawColor}{RGB}{0,0,0}
		\path[draw=drawColor,line width= 0.6pt,line join=round] (\Roffset, \VMoffset{1}) -- (\plus{\Roffset}{\length}, \VMoffset{1});
	\end{scope}

	\begin{scope}
		\definecolor{drawColor}{RGB}{0,0,0}
		\node[text=drawColor,anchor=base west,inner sep=0pt, outer sep=0pt, scale=  0.88] at (\RToffset, \VToffset{1}) {Edge counts};
	\end{scope}
\end{scope}

\end{tikzpicture}}}{Legend}}

  \caption{Sensitivity analysis on the probability that fake edges are fall-through (X-axis)}
  \label{fig:sense_fakeft}
\end{figure*}

Figure~\ref{fig:sense_opaque} shows the effect of changing the probability to insert opaque predicates by setting it to 0\%, 2\%, 5\%, 20\%, 50\%, and 100\%.
\begin{itemize}
\item \textbf{Before attacking} the program, the \dbfpr{} increases from about 90\% to about 100\%, while the \dbfnr{} decreases from about 20\% to about 10\%. When no opaque predicates are inserted, the only fake edges are those inserted by the dispatcher's sparse switch tables, which are not always analysed properly. Hence the 90\% starting value for the \dbfpr{}. As more opaque predicates are inserted, the fraction of fake edges contributed by the switch tables decreases. The trend for the \dbfnr{} can be explained similarly. The non-monotonic trends for \guifpr{} and \guifnr{} are due to the FRs for edges out of switch-based factoring dispatchers dominating the total FRs when no or little opaque predicates are inserted, whereas FRs for opaque predicate edges dominate the total FRs as more such predicates get inserted. FPRs (and FNRs) for both types of edges decrease (and increase) when more opaque predicates are inserted, but they start at different values for 0\% opaque predicate probability and converge to different values for 100\% probability.
  \item The \textbf{soundish attack} does not change the false rates significantly.
  \item After the \textbf{unsound attack}, the \guifpr{} is lower that in the unattacked case when opaque predicates are actually insered and monotonically decreasing due to the increasing number of resolved opaque predicates. The \guifnr{} does not change compared to the unattacked program.
\end{itemize}
\begin{figure*}[t]
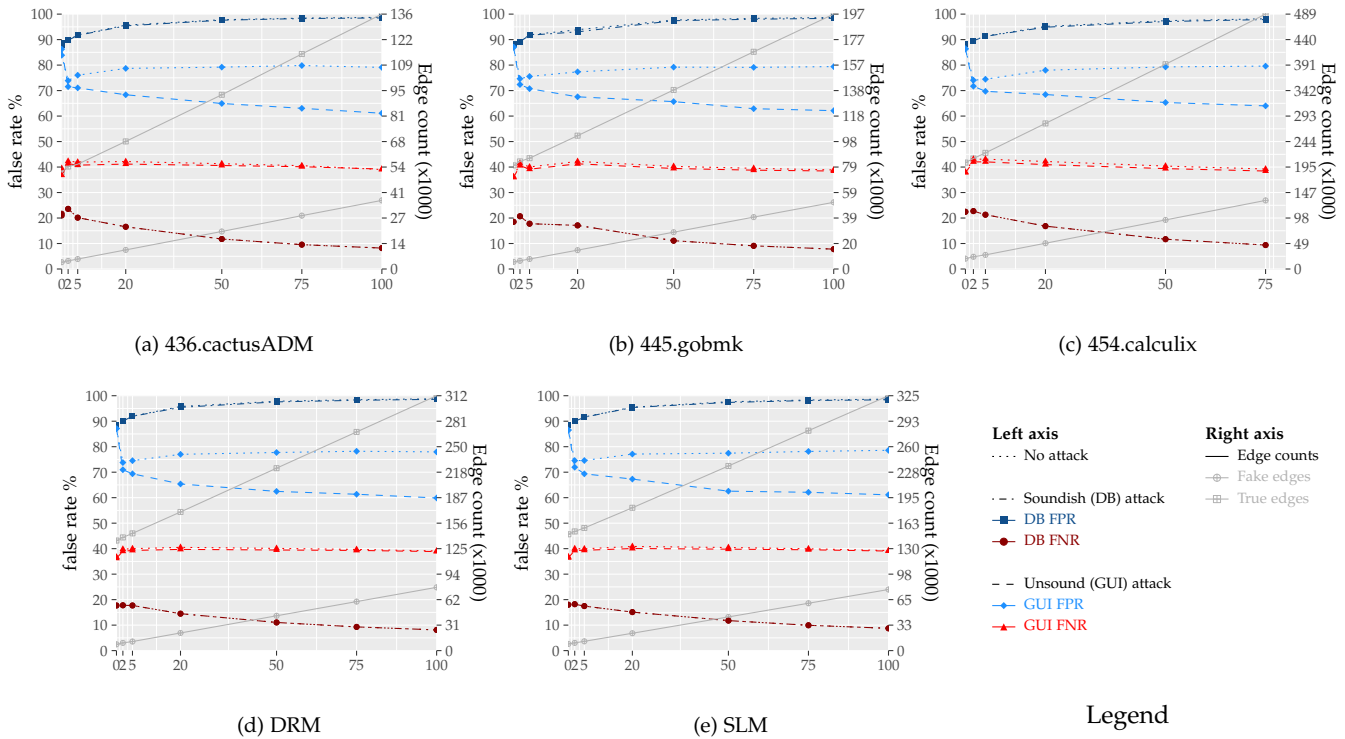

  \centering
  \subfloat[436.cactusADM]
    {\resizebox{0.33\linewidth}{!}{\input{sensitivity_436_opaque}}}
  \subfloat[445.gobmk]
    {\resizebox{0.33\linewidth}{!}{\input{sensitivity_445_opaque}}}
  \subfloat[454.calculix]
    {\resizebox{0.33\linewidth}{!}{\input{sensitivity_454_opaque}}}
  \\
  \subfloat[DRM]
    {\resizebox{0.33\linewidth}{!}{\input{sensitivity_DRM_opaque}}}
  \subfloat[SLM]
    {\resizebox{0.33\linewidth}{!}{\input{sensitivity_SLM_opaque}}}
  \subfloat{\stackunder[0em]{\resizebox{0.25\linewidth}{!}{\relax
\begin{tikzpicture}[x=1pt,y=1pt]

\newcommand{\Loffset}{20}
\newcommand{\Roffset}{140}
\newcommand{\length}{13}
\newcommand{\hlength}{6.5}
\newcommand{\hmarker}{2}
\newcommand{\plus}[2]{\the\numexpr #1 + #2}
\newcommand{\plusT}[3]{\the\numexpr #1 + #2 + #3}
\newcommand{\LToffset}{\plus{\plus{\Loffset}{\length}}{5}}
\newcommand{\RToffset}{\plus{\plus{\Roffset}{\length}}{5}}
\newcommand{\VToffset}[1]{\the\numexpr 150 - #1 * 12}
\newcommand{\VMoffset}[1]{\VToffset{#1} + 2}

\definecolor{fillColor}{RGB}{255,0,0}
\path[use as bounding box,fill=fillColor,fill opacity=0.00] (0,0) rectangle (190,180.67);

\begin{scope}
	\begin{scope}
		\definecolor{drawColor}{RGB}{0,0,0}
		\node[text=drawColor,anchor=base west,inner sep=0pt, outer sep=0pt, scale=  1.00] at (\Loffset,\VToffset{0}) {\textbf{Left axis}};
	\end{scope}

	\begin{scope}
		\definecolor{drawColor}{RGB}{0,0,0}
		\node[text=drawColor,anchor=base west,inner sep=0pt, outer sep=0pt, scale=  1.00] at (\Roffset,\VToffset{0}) {\textbf{Right axis}};
	\end{scope}
\end{scope}

\begin{scope}
	\begin{scope}
		\definecolor{drawColor}{RGB}{0,0,0}
		\path[draw=drawColor,line width= 0.6pt,dash pattern=on 1pt off 3pt ,line join=round] (\Loffset, \VMoffset{1}) -- (\plus{\Loffset}{\length}, \VMoffset{1});
	\end{scope}

	\begin{scope}
		\definecolor{drawColor}{RGB}{0,0,0}
		\node[text=drawColor,anchor=base west,inner sep=0pt, outer sep=0pt, scale=  0.88] at (\LToffset, \VToffset{1}) {No attack};
	\end{scope}
\end{scope}

\begin{scope}
	\begin{scope}
		\definecolor{drawColor}{RGB}{0,0,0}
		\path[draw=drawColor,line width= 0.6pt,dash pattern=on 1pt off 3pt on 4pt off 3pt ,line join=round] (\Loffset, \VMoffset{3}) -- (\plus{\Loffset}{\length}, \VMoffset{3});
	\end{scope}

	\begin{scope}
		\definecolor{drawColor}{RGB}{0,0,0}
			\node[text=drawColor,anchor=base west,inner sep=0pt, outer sep=0pt, scale=  0.88] at (\LToffset, \VToffset{3}) {Soundish (DB) attack};
	\end{scope}
\end{scope}

\begin{scope}
	\begin{scope}
		\definecolor{fillColor}{RGB}{16,78,139}
		\path[fill=fillColor] (\plusT{\Loffset}{\hlength}{-\hmarker},\plus{\VMoffset{4}}{-\hmarker}) --
			(\plusT{\Loffset}{\hlength}{\hmarker},\plus{\VMoffset{4}}{-\hmarker}) --
			(\plusT{\Loffset}{\hlength}{\hmarker},\plus{\VMoffset{4}}{\hmarker}) --
			(\plusT{\Loffset}{\hlength}{-\hmarker},\plus{\VMoffset{4}}{\hmarker}) --
			cycle;
	\end{scope}

	\begin{scope}
		\definecolor{drawColor}{RGB}{16,78,139}
		\path[draw=drawColor,line width= 0.6pt,line join=round] (\Loffset, \VMoffset{4}) -- (\plus{\Loffset}{\length}, \VMoffset{4});
	\end{scope}

	\begin{scope}
		\definecolor{drawColor}{RGB}{16,78,139}
		\node[text=drawColor,anchor=base west,inner sep=0pt, outer sep=0pt, scale=  0.88] at (\LToffset,\VToffset{4}) {DB FPR};
	\end{scope}
\end{scope}

\begin{scope}
	\begin{scope}
		\definecolor{drawColor}{RGB}{139,0,0}
		\definecolor{fillColor}{RGB}{139,0,0}
		\path[draw=drawColor,line width= 0.4pt,line join=round,line cap=round,fill=fillColor] (\plus{\Loffset}{\hlength},\VMoffset{5}) circle (\hmarker);
	\end{scope}

	\begin{scope}
		\definecolor{drawColor}{RGB}{139,0,0}
		\path[draw=drawColor,line width= 0.6pt,line join=round] (\Loffset,\VMoffset{5}) -- (\plus{\Loffset}{\length},\VMoffset{5});
	\end{scope}

	\begin{scope}
		\definecolor{drawColor}{RGB}{139,0,0}
		\node[text=drawColor,anchor=base west,inner sep=0pt, outer sep=0pt, scale=  0.88] at (\LToffset,\VToffset{5}) {DB FNR};
	\end{scope}
\end{scope}

\begin{scope}
	\begin{scope}
		\definecolor{drawColor}{RGB}{0,0,0}
		\path[draw=drawColor,line width= 0.6pt,dash pattern=on 4pt off 4pt ,line join=round] (\Loffset, \VMoffset{7}) -- (\plus{\Loffset}{\length}, \VMoffset{7});
	\end{scope}

	\begin{scope}
		\definecolor{drawColor}{RGB}{0,0,0}
			\node[text=drawColor,anchor=base west,inner sep=0pt, outer sep=0pt, scale=  0.88] at (\LToffset, \VToffset{7}) {Unsound (GUI) attack};
	\end{scope}
\end{scope}

\begin{scope}
	\begin{scope}
		\definecolor{fillColor}{RGB}{30,144,255}
		\path[fill=fillColor] (\plusT{\Loffset}{\hlength}{-\hmarker}, \VMoffset{8}) --
			(\plus{\Loffset}{\hlength},\plus{\VMoffset{8}}{\hmarker}) --
			(\plusT{\Loffset}{\hlength}{\hmarker}, \VMoffset{8}) --
			(\plus{\Loffset}{\hlength}, \plus{\VMoffset{8}}{-\hmarker}) --
			cycle;
	\end{scope}

	\begin{scope}
		\definecolor{drawColor}{RGB}{30,144,255}
		\path[draw=drawColor,line width= 0.6pt,line join=round] (\Loffset, \VMoffset{8}) -- (\plus{\Loffset}{\length}, \VMoffset{8});
	\end{scope}

	\begin{scope}
		\definecolor{drawColor}{RGB}{30,144,255}
		\node[text=drawColor,anchor=base west,inner sep=0pt, outer sep=0pt, scale=  0.88] at (\LToffset, \VToffset{8}) {GUI FPR};
	\end{scope}
\end{scope}

\begin{scope}
	\begin{scope}
		\definecolor{fillColor}{RGB}{255,0,0}
		\path[fill=fillColor] (\plus{\Loffset}{\hlength},\plus{\VMoffset{9}}{\hmarker}) --
			(\plusT{\Loffset}{\hlength}{\hmarker},\plus{\VMoffset{9}}{-\hmarker}) --
			(\plusT{\Loffset}{\hlength}{-\hmarker},\plus{\VMoffset{9}}{-\hmarker}) --
			cycle;
	\end{scope}

	\begin{scope}
		\definecolor{drawColor}{RGB}{255,0,0}
		\path[draw=drawColor,line width= 0.6pt,line join=round] (\Loffset,\VMoffset{9}) -- (\plus{\Loffset}{\length},\VMoffset{9});
	\end{scope}

	\begin{scope}
		\definecolor{drawColor}{RGB}{255,0,0}
		\node[text=drawColor,anchor=base west,inner sep=0pt, outer sep=0pt, scale=  0.88] at (\LToffset,\VToffset{9}) {GUI FNR};
	\end{scope}
\end{scope}

\begin{scope}
	\begin{scope}
		\definecolor{drawColor}{gray}{0.70}
		\path[draw=drawColor,line width= 0.4pt,line join=round,line cap=round] (\plus{\Roffset}{\hlength},\VMoffset{2}) circle (\hmarker);
		\path[draw=drawColor,line width= 0.4pt,line join=round,line cap=round] (\plusT{\Roffset}{\hlength}{-\hmarker},\VMoffset{2}) -- (\plusT{\Roffset}{\hlength}{\hmarker},\VMoffset{2});
		\path[draw=drawColor,line width= 0.4pt,line join=round,line cap=round] (\plus{\Roffset}{\hlength},\plus{\VMoffset{2}}{-\hmarker}) -- (\plus{\Roffset}{\hlength},\plus{\VMoffset{2}}{\hmarker});
	\end{scope}

	\begin{scope}
		\definecolor{drawColor}{gray}{0.70}
		\path[draw=drawColor,line width= 0.6pt,line join=round] (\Roffset,\VMoffset{2}) -- (\plus{\Roffset}{\length},\VMoffset{2});
	\end{scope}

	\begin{scope}
		\definecolor{drawColor}{gray}{0.70}
		\node[text=drawColor,anchor=base west,inner sep=0pt, outer sep=0pt, scale=  0.88] at (\RToffset,\VToffset{2}) {Fake edges};
	\end{scope}
\end{scope}

\begin{scope}
	\begin{scope}
		\definecolor{drawColor}{gray}{0.70}
		\path[draw=drawColor,line width= 0.4pt,line join=round,line cap=round] (\plusT{\Roffset}{\hlength}{-\hmarker},\VMoffset{3}) -- (\plusT{\Roffset}{\hlength}{\hmarker},\VMoffset{3});
		\path[draw=drawColor,line width= 0.4pt,line join=round,line cap=round] (\plus{\Roffset}{\hlength},\plus{\VMoffset{3}}{-\hmarker}) -- (\plus{\Roffset}{\hlength},\plus{\VMoffset{3}}{\hmarker});
		\path[draw=drawColor,line width= 0.4pt,line join=round,line cap=round] (\plusT{\Roffset}{\hlength}{-\hmarker},\plus{\VMoffset{3}}{-\hmarker}) rectangle (\plusT{\Roffset}{\hlength}{\hmarker},\plus{\VMoffset{3}}{\hmarker});
	\end{scope}

	\begin{scope}
		\definecolor{drawColor}{gray}{0.70}
		\path[draw=drawColor,line width= 0.6pt,line join=round] (\Roffset,\VMoffset{3}) -- (\plus{\Roffset}{\length},\VMoffset{3});
	\end{scope}

	\begin{scope}
		\definecolor{drawColor}{gray}{0.70}
		\node[text=drawColor,anchor=base west,inner sep=0pt, outer sep=0pt, scale=  0.88] at (\RToffset,\VToffset{3}) {True edges};
	\end{scope}
\end{scope}

\begin{scope}
	\begin{scope}
		\definecolor{drawColor}{RGB}{0,0,0}
		\path[draw=drawColor,line width= 0.6pt,line join=round] (\Roffset, \VMoffset{1}) -- (\plus{\Roffset}{\length}, \VMoffset{1});
	\end{scope}

	\begin{scope}
		\definecolor{drawColor}{RGB}{0,0,0}
		\node[text=drawColor,anchor=base west,inner sep=0pt, outer sep=0pt, scale=  0.88] at (\RToffset, \VToffset{1}) {Edge counts};
	\end{scope}
\end{scope}

\end{tikzpicture}}}{Legend}}

  \caption{Sensitivity analysis on the opaque predicate insertion probability (X-axis)}
  \label{fig:sense_opaque}
\end{figure*}

To conclude our sensitivity analysis, we report how well pairs of code fragments originating from the same function but split apart by factorisation are still put in the same function by IDA Pro with increasing amounts of factoring (1\%, 10\%, 25\%, 50\%, 75\%, and 100\%). Figure~\ref{fig:sense_links} shows how the fraction of such broken pairs (i.e., split pairs not put in the same function by the attacker-improved IDA Pro) evolves when more factoring is applied for when (1) \linknoop{} and (2) \linkop{} are inserted.
\begin{itemize}
  \item \textbf{Before attacking} the program, when \linknoop{} are inserted, the fraction of broken pairs starts at about 55\% when 1\% is factored and increases to about 75\% when all factoring is applied. This is evidence of the claim in Section~\ref{sec:potency} where we stated that applying more factoring will result in more broken pairs. When \linkop{} are inserted, the fraction starts at about 85\% when little factoring is applied, decreasing to about 80\% with increasing amounts of factoring.
  \item When \linknoop{} are inserted, the \textbf{soundish attack} is able to reunite about 20\% of the broken pairs when little factoring is applied. With increasing amounts of factoring, the fraction of broken pairs that get reunited shrinks due to the total number of pairs increasing. When \linkop{} are inserted, the effects are smaller yet similar.
  \item When \linknoop{} are inserted, the \textbf{unsound attack} does not yield better results than the soundish attack. When \linkop{} are inserted, some more broken pairs can be reunited, due to opaque predicates not being fully drawn in the GUI, but this advantage quickly diminishes when more factoring is applied.
\end{itemize}
When we compare the discussed results for 436.cactusADM with those for the other benchmarks, we observe that whereas the number of broken and fixed pairs varies, the conclusions remain the same.
\begin{figure*}[t]
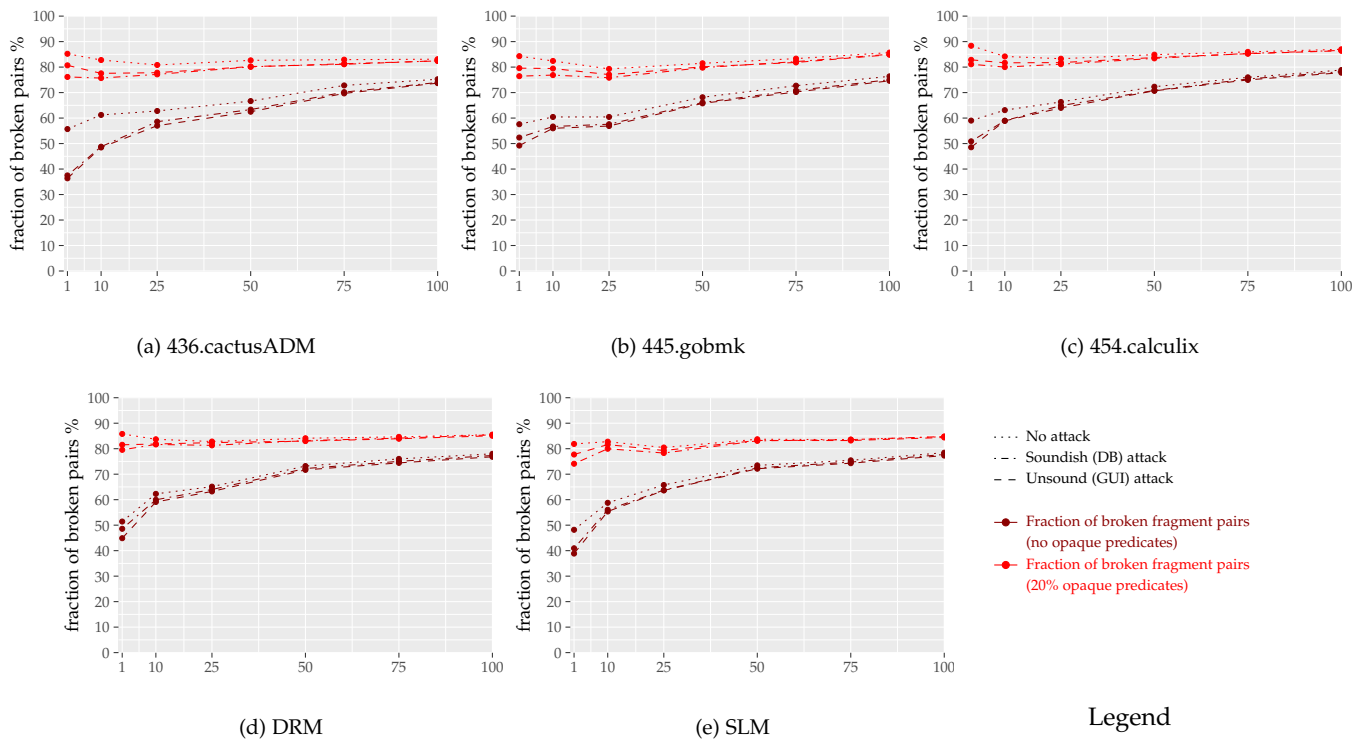

  \centering
  \subfloat[436.cactusADM]
    {\resizebox{0.33\linewidth}{!}{\input{sensitivity_436_links}}}
  \subfloat[445.gobmk]
    {\resizebox{0.33\linewidth}{!}{\input{sensitivity_445_links}}}
  \subfloat[454.calculix]
    {\resizebox{0.33\linewidth}{!}{\input{sensitivity_454_links}}}
  \\
  \subfloat[DRM]
    {\resizebox{0.33\linewidth}{!}{\input{sensitivity_DRM_links}}}
  \subfloat[SLM]
    {\resizebox{0.33\linewidth}{!}{\input{sensitivity_SLM_links}}}
  \subfloat{\stackunder[0em]{\resizebox{0.25\linewidth}{!}{\relax
\begin{tikzpicture}[x=1pt,y=1pt]

\newcommand{\Loffset}{20}
\newcommand{\Roffset}{140}
\newcommand{\length}{13}
\newcommand{\hlength}{6.5}
\newcommand{\hmarker}{2}
\newcommand{\plus}[2]{\the\numexpr #1 + #2}
\newcommand{\plusT}[3]{\the\numexpr #1 + #2 + #3}
\newcommand{\LToffset}{\plus{\plus{\Loffset}{\length}}{5}}
\newcommand{\RToffset}{\plus{\plus{\Roffset}{\length}}{5}}
\newcommand{\VToffset}[1]{\the\numexpr 150 - #1 * 12}
\newcommand{\VMoffset}[1]{\VToffset{#1} + 2}

\definecolor{fillColor}{RGB}{255,0,0}
\path[use as bounding box,fill=fillColor,fill opacity=0.00] (0,0) rectangle (190,180.67);

\begin{scope}
	\begin{scope}
		\definecolor{drawColor}{RGB}{0,0,0}
		\path[draw=drawColor,line width= 0.6pt,dash pattern=on 1pt off 3pt ,line join=round] (\Loffset, \VMoffset{0}) -- (\plus{\Loffset}{\length}, \VMoffset{0});
	\end{scope}

	\begin{scope}
		\definecolor{drawColor}{RGB}{0,0,0}
		\node[text=drawColor,anchor=base west,inner sep=0pt, outer sep=0pt, scale=  0.88] at (\LToffset, \VToffset{0}) {No attack};
	\end{scope}
\end{scope}

\begin{scope}
	\begin{scope}
		\definecolor{drawColor}{RGB}{0,0,0}
		\path[draw=drawColor,line width= 0.6pt,dash pattern=on 1pt off 3pt on 4pt off 3pt ,line join=round] (\Loffset, \VMoffset{1}) -- (\plus{\Loffset}{\length}, \VMoffset{1});
	\end{scope}

	\begin{scope}
		\definecolor{drawColor}{RGB}{0,0,0}
			\node[text=drawColor,anchor=base west,inner sep=0pt, outer sep=0pt, scale=  0.88] at (\LToffset, \VToffset{1}) {Soundish (DB) attack};
	\end{scope}
\end{scope}

\begin{scope}
	\begin{scope}
		\definecolor{drawColor}{RGB}{0,0,0}
		\path[draw=drawColor,line width= 0.6pt,dash pattern=on 4pt off 4pt ,line join=round] (\Loffset, \VMoffset{2}) -- (\plus{\Loffset}{\length}, \VMoffset{2});
	\end{scope}

	\begin{scope}
		\definecolor{drawColor}{RGB}{0,0,0}
			\node[text=drawColor,anchor=base west,inner sep=0pt, outer sep=0pt, scale=  0.88] at (\LToffset, \VToffset{2}) {Unsound (GUI) attack};
	\end{scope}
\end{scope}

\begin{scope}
	\begin{scope}
		\definecolor{drawColor}{RGB}{139,0,0}
		\definecolor{fillColor}{RGB}{139,0,0}
		\path[draw=drawColor,line width= 0.4pt,line join=round,line cap=round,fill=fillColor] (\plus{\Loffset}{\hlength},\VMoffset{4}) circle (\hmarker);
	\end{scope}

	\begin{scope}
		\definecolor{drawColor}{RGB}{139,0,0}
		\path[draw=drawColor,line width= 0.6pt,line join=round] (\Loffset,\VMoffset{4}) -- (\plus{\Loffset}{\length},\VMoffset{4});
	\end{scope}

	\begin{scope}
		\definecolor{drawColor}{RGB}{139,0,0}
		\node[text=drawColor,anchor=base west,inner sep=0pt, outer sep=0pt, scale=  0.88] at (\LToffset,\VToffset{4}) {Fraction of broken fragment pairs};
		\node[text=drawColor,anchor=base west,inner sep=0pt, outer sep=0pt, scale=  0.88] at (\LToffset,\VToffset{5}) {(no opaque predicates)};
	\end{scope}
\end{scope}

\begin{scope}
	\begin{scope}
		\definecolor{drawColor}{RGB}{255,0,0}
		\definecolor{fillColor}{RGB}{255,0,0}
		\path[draw=drawColor,line width= 0.4pt,line join=round,line cap=round,fill=fillColor] (\plus{\Loffset}{\hlength},\VMoffset{6}) circle (\hmarker);
	\end{scope}

	\begin{scope}
		\definecolor{drawColor}{RGB}{255,0,0}
		\path[draw=drawColor,line width= 0.6pt,line join=round] (\Loffset,\VMoffset{6}) -- (\plus{\Loffset}{\length},\VMoffset{6});
	\end{scope}

	\begin{scope}
		\definecolor{drawColor}{RGB}{255,0,0}
		\node[text=drawColor,anchor=base west,inner sep=0pt, outer sep=0pt, scale=  0.88] at (\LToffset,\VToffset{6}) {Fraction of broken fragment pairs};
		\node[text=drawColor,anchor=base west,inner sep=0pt, outer sep=0pt, scale=  0.88] at (\LToffset,\VToffset{7}) {(20\% opaque predicates)};
	\end{scope}
\end{scope}

\end{tikzpicture}}}{Legend}}

  \caption{Effect of the probability to apply code factoring (X-axis) on the fraction of broken pairs (Y-axis)}
  \label{fig:sense_links}
\end{figure*}
\fi

\if \techreport1
\subsection{Experimental setup}

We evaluated the considered benchmarks on two ARM development boards. One is an Arndale Board that runs Ubuntu 15.04, the other is a SABRE Lite i.MX6 board running Android 4.4.3 (KitKat). Table~\ref{tab:boards} lists the most important specifications of each board, along with the compiler versions used to compile the benchmarks with.
\begin{table*}[t]
  \centering
  \begin{tabular}{|l|c|c|}
    \cline{2-3}
    \multicolumn{1}{l|}{} & \textbf{Linux environment} & \textbf{Android environment}\\
    \hline
    \textbf{Board} & ArndaleBoard & Boundary Devices\\
          & 5250-A & BD-SL-i.MX6\\
    \hline
    \textbf{CPU}   & ARM Cortex-A15 & ARM Cortex-A9\\
          & (1.7 GHz) & (1 GHz)\\
    \hline
    \textbf{OS}    & Ubuntu 15.04 & Android 4.4.3 KitKat\\
          & (kernel 4.3.0) & (kernel 3.10.53)\\
    \hline
    \textbf{Compiler} & GNU GCC 4.8.1   & GNU GCC 4.8\\
    \textbf{} & binutils 2.23.2 & binutils 2.23.2\\
    \hline
  \end{tabular}
  \caption{Development board and software specifications}
  \label{tab:boards}
\end{table*}

For the SPEC benchmarks, we collected an execution profile with the \texttt{train} inputs. Of course, the overhead was measured on other inputs. For the 445.gobmk benchmark, we used the default \texttt{ref} inputs. We could not use them for the 436.cactusADM and 454.calculix benchmarks, as they are impractical to perform measurements on the relatively slow development boards. Instead, we slightly modified the 436.cactusADM \texttt{ref} input, and we used an alternative input for 454.calculix from its test suite (\texttt{beamnldyp.inp})~\cite{ccx215}.

For the SLM use case we used a fixed set of inputs to generate the execution profile and to measure the overhead. These inputs were carefully chosen to cover as much of the library as possible.

For the DRM use case, we collected an execution profile by logging in on the GUI application, after which we played an encrypted and an unencrypted movie. As discussed in Section~\ref{sec:overhead}, we did not measure the overhead for this use case due to its interactive nature.
\fi

\section{Related Work}
\label{sec:related}

\subsection{Code factoring}
Existing work on code factoring focused mainly on compaction, i.e., the removal of duplicate code to make binaries smaller. Production tool chains already include optimisation passes to factor identical procedures: Microsofts Visual C++ compiler~\cite{microsofticf}, GNU GCC~\cite{gccicfliska}\cite{gccicf}, Gold~\cite{goldicf} and LLVM~\cite{llvmicf}. In academic research, Debray et al.~\cite{debray2000compiler}, De Sutter et al.~\cite{siftingmud} and Von Koch et al.~\cite{edler2014exploiting} have developed code factoring techniques to factor almost identical code on the basic block level (the former two) and the procedural level (the latter two). Computation time is reduced by defining a fingerprint for each basic block and/or procedure and small differences between procedures are compensated for by parameterizing the factored code. Debray et al.\ and De Sutter et al.\ mitigated differences between basic blocks by using an ad-hoc register renaming algorithm and by canonicalising the instruction schedule. This was not an issue for Von Koch et al.\ because LLVM IR was used. Recently, Rocha et al.~\cite{rocha2019function} used a DNA sequence alignment algorithm from bioinformatics to identify factoring candidates and to compensate for differences between them. Similar to the work by Von Koch et al., they only support factoring on the procedural level but, as they implemented their technique on LLVM IR, they are not bound by register allocation schemes.

Inspired by the existing implementations, we factor (sub)blocks for obfuscation rather than compaction. Thus, we can give up on code size overhead to factor more code. Our technique is orthogonal and complementary to whole function merging, which by definition does not obfuscate function boundaries. Importantly, we rely only on intraprocedural control flow idioms rather than calls and returns.

\subsection{Obfuscations}
Many obfuscation transformations exist, each with its own strengths and weaknesses, as surveyed by Schrittwieser et al.~\cite{schrittwieser2016protecting}. Collberg et al.~\cite{collberg1997taxonomy} categorized obfuscation techniques into layout (e.g., code layout randomization), control flow and data transformations. One example of control flow obfuscations are opaque predicates. These can range from simple~\cite{myles2006software} to complex~\cite{majumdar2006manufacturing}. While easy to implement, the simple ones are not resilient against modern attacks such as symbolic or concolic execution~\cite{yadegari2015symbolic}. Recently, some alternatives were proposed to counter these advanced attacks. The range dividers of Banescu et al.~\cite{banescu2016code} introduce additional feasible code paths, exploding the analysis complexity. The bi-opaque predicates of Xu et al.~\cite{biopaquepreds} exploit the NP-hard problem of resolving symbolic memory. Zobernig et al.\ researched a technique to make opaque predicates indistinguishable from the program's predicates by hashing the calculation of each (opaque) predicate~\cite{zobernig2017,zobernig2019}. Attackers need to invert the hash function to prove the opaqueness of a predicate, a process that is known to be impossible but for brute-forcing. Our use of opaque predicates in this paper is orthogonal to mentioned work, as our work focuses on choosing the target of the fake edge, which needs to be done whatever the kind of computation is used to implement the opaque predicate that steers the conditional branch. 

Another example of control flow obfuscations are branch functions~\cite{linn2003obfuscation}, which replace direct with indirect branches to thwart code identification heuristics. Control flow flattening~\cite{wang2000software} is another obfuscation, which replaces direct with dispatcher-based control flow. Our factoring dispatchers resemble these obfuscations, but focus on thwarting function repartitioning heuristics, as we aim for attackers to identify fake inter-component control flow paths to confuse them even more and to hide the boundaries of components, rather than to obfuscate the components' internals themselves. Asghar et al.\ propose another way to obfuscate the control flow of a program by removing conditional branches~\cite{asghar2016}. Contrary to other techniques, they avoid the insertion of additional calculations but instead build on the increased complexity of the linearized calculations.

Obfuscations can be inserted at source level~\cite{collberg2015tigress}, by compilers~\cite{ieeespro2015JunodRWM} or by binary code rewriters~\cite{diablo2005}. As we want to hide the boundaries of linked-in components, we obviously opted for a post link-time rewriter.

\section{Conclusions and Future Work}
\label{sec:conclusions}
We presented a novel technique to apply code factoring across component boundaries with intraprocedural control flow idioms. We combined our technique with existing opaque predicates with which we also inject fake direct control flow accros component boundaries, and with fine-grained code layout randomization. In our evaluation with IDA Pro, a commonly used, state-of-the-art reverse engineering tool, we demonstrated that this thwarts disassembler and CFG reconstruction heuristics and that a program protected with our technique is more resilient to some known attacks. We can conclude that our technique increases the potency and resilience of protected applications against modern reverse engineering attacks.

In future research, approaches to generate more similar, and thus factorable code fragments can be investigated, rather that only identifying existing ones. Another research path can be the use of machine learning techniques to steer the insertion of fake control flow, so that attack tools are more purposively thwarted rather than stochastically.

\bibliographystyle{IEEEtran}
\bibliography{paper}

\begin{IEEEbiography} [{\includegraphics[width=1in,height=1.25in,clip,keepaspectratio]{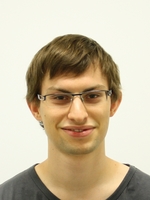}}]
  {Jens Van den Broeck} is a PhD student at Ghent University in the Computer
  Systems Lab. He obtained his MSc degree in Electrical Engineering from Ghent
  University's Faculty of Engineering and Architecture in 2013. His research
  focuses on software protection.
\end{IEEEbiography}

\begin{IEEEbiography} [{\includegraphics[width=1in,height=1.25in,clip,keepaspectratio]{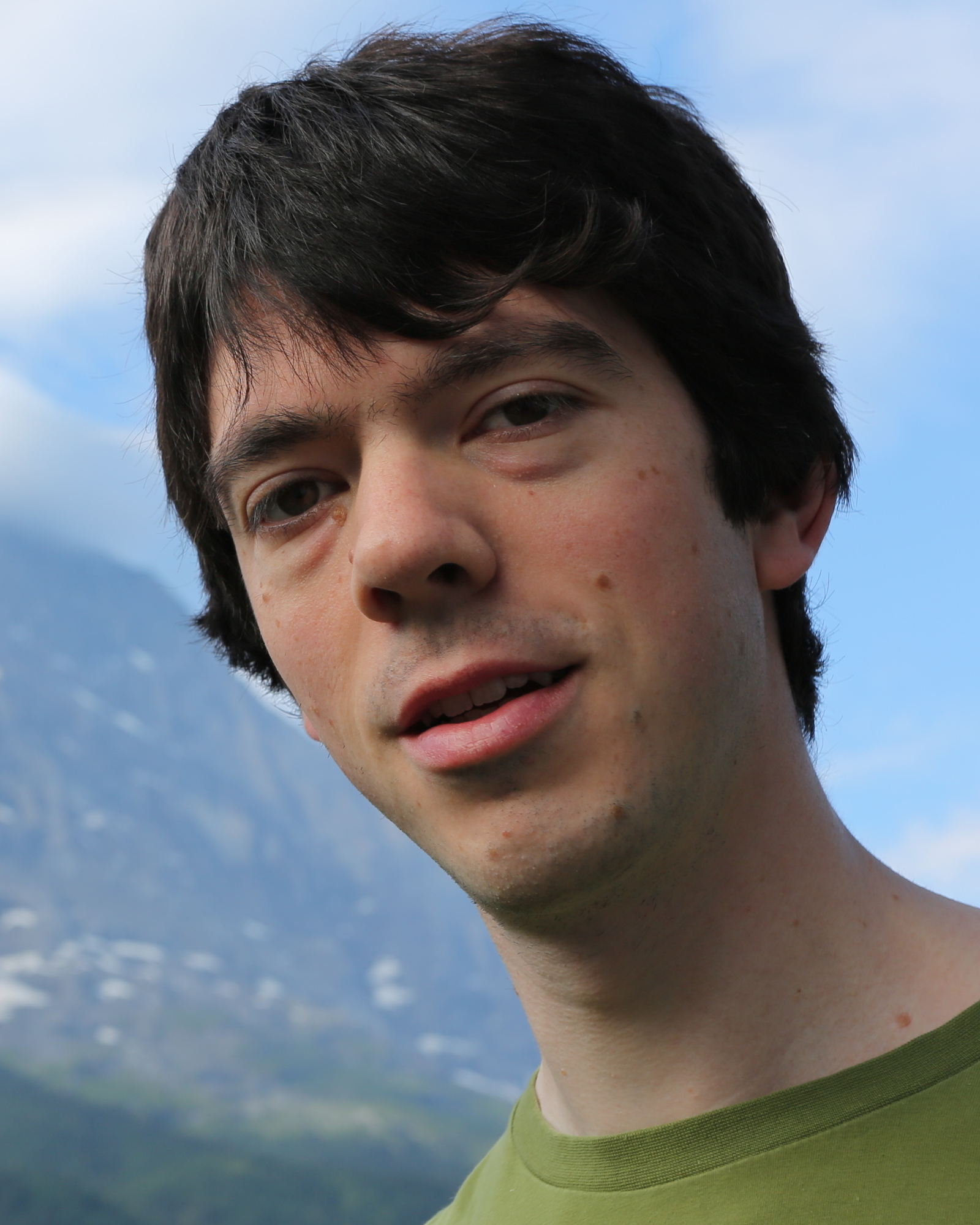}}]
  {Bart Coppens} is a postdoctoral researcher at Ghent University in the
  Computer Systems Lab. He received his PhD in Computer Science Engineering
  from the Faculty of Engineering and Architecture at Ghent University in 2013.
  His research focuses on protecting software against different forms of
  attacks using compiler-based techniques and run-time techniques.
\end{IEEEbiography}

\begin{IEEEbiography} [{\includegraphics[width=1in,height=1.25in,clip,keepaspectratio]{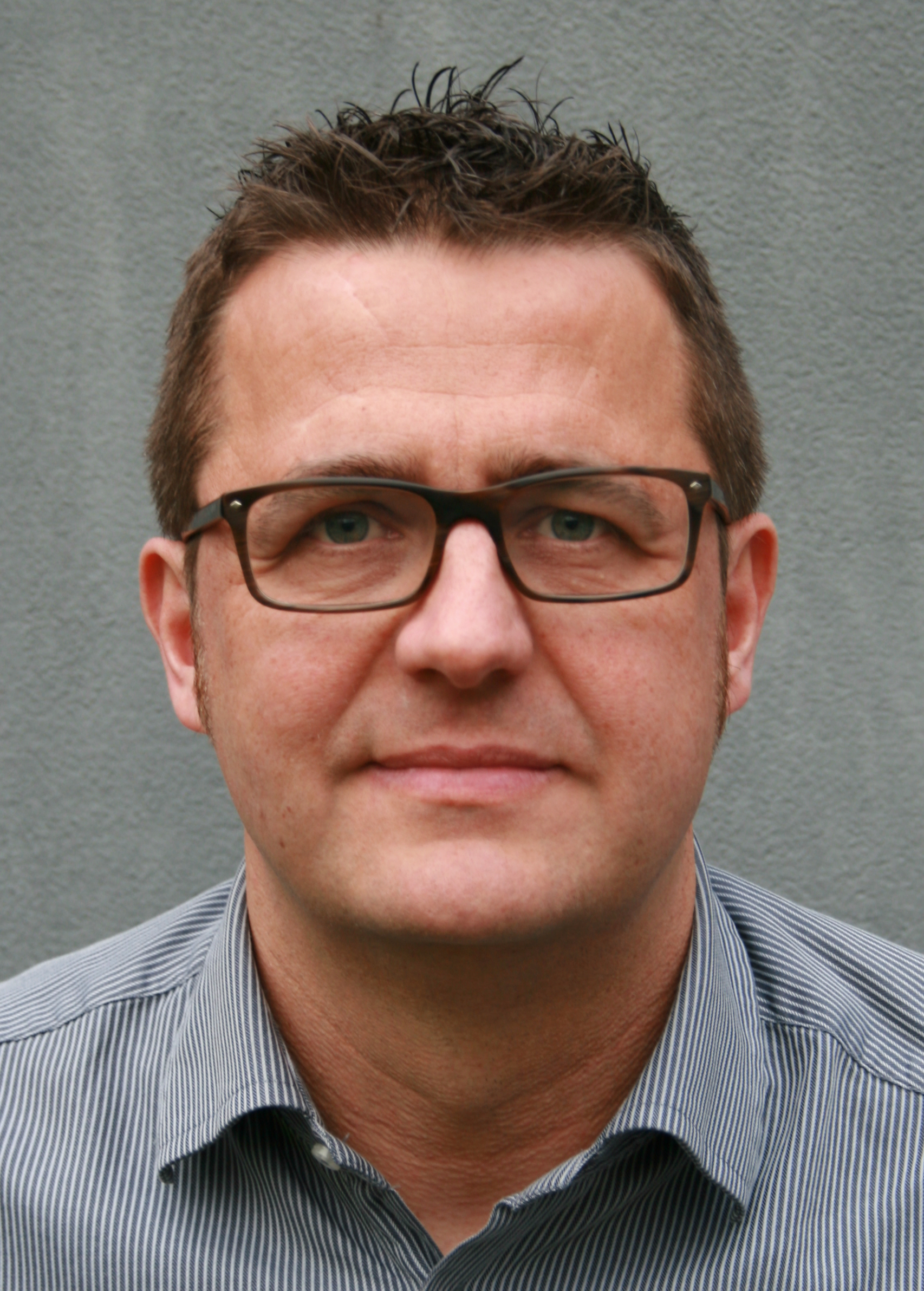}}]
  {Bjorn De Sutter} is professor at Ghent University in the Computer Systems
  Lab. He obtained his MSc and PhD degrees in Computer Science from the
  university's Faculty of Engineering in 1997 and 2002. His research focuses on
  techniques to aid programmers with non-functional aspects such as performance
  and software protection to mitigate reverse engineering, software tampering,
  code reuse attacks, fault injection, and side channel attacks. He co-authored
  over 80 papers and coordinated the ASPIRE project.
\end{IEEEbiography}

\end{document}